\documentclass{article}

\usepackage[eandd, final]{neurips_2026}

\usepackage[utf8]{inputenc} 
\usepackage[T1]{fontenc}    
\usepackage{hyperref}       
\usepackage{url}            
\usepackage{booktabs}       
\usepackage{amsfonts}       
\usepackage{nicefrac}       
\usepackage{microtype}      
\usepackage{xcolor}         

\definecolor{darkblue}{rgb}{0,0,0.6}
\hypersetup{colorlinks=true, linkcolor=darkblue, citecolor=darkblue, urlcolor=darkblue}

\usepackage[pdftex]{graphicx}
\usepackage{multirow}
\usepackage{enumitem}

\usepackage{fbox}
\usepackage{amsmath}
\usepackage{tabularx}
\usepackage{colortbl}
\usepackage{bm}
\usepackage[most]{tcolorbox}
\usepackage{xcolor}

\usepackage{longtable}
\usepackage{booktabs}

\usepackage{capt-of}

\title{TSNBench: Benchmarking LLM Proficiency in Time-Sensitive Networking}

\author{
\bfseries
Rubi~Debnath$^{1}$\thanks{Corresponding author. Email: \texttt{rubi.debnath@tum.de}} \quad
Daniel~Bujosa$^{2}$ \quad
Luxi~Zhao$^{3}$
\\
\bfseries
Silviu~S.~Craciunas$^{2,4}$ \quad
Paul~Pop$^{2}$ \quad
Sebastian~Steinhorst$^{1}$
\\
$^{1}$Technical University of Munich, Germany \quad
$^{2}$Technical University of Denmark, Denmark
\\
$^{3}$Beihang University, China \quad
$^{4}$NXP Semiconductors, Austria
}

\begin{document}

\maketitle

\begin{abstract}
We present TSNBench, the first benchmark for evaluating large language model (LLM) proficiency in Time-Sensitive Networking (TSN), a suite of IEEE 802.1 standards for deterministic communication with bounded latency in safety-critical domains such as autonomous vehicles, aviation, defense, and industrial automation. While LLMs have been extensively evaluated on general knowledge tasks, their capabilities in safety-critical networking domains remain largely unexplored. TSNBench comprises 939 expert-validated multiple-choice questions (MCQs) covering diverse TSN mechanisms, along with 100 open-ended Worst-Case Delay (WCD) computation tasks for each of Credit-Based Shaper (CBS) and Cyclic Queuing and Forwarding (CQF), across varying topologies and traffic flows. The ground-truth (GT) WCD values are computed using a verified Network Calculus (NC) solver for CBS and closed-form upper bounds for CQF. We evaluate 16 LLMs and find that although models achieve 67 to 95\% accuracy on MCQs, they fail substantially on open-ended WCD computation. For CBS, the best-performing model, GPT-5, deviates from the GT by 150.2~$\mu$s on average (mean absolute error, MAE), and its predicted WCD is within 50~$\mu$s of the GT for only 39.4\% of flows, while all other models exceed an MAE of 250~$\mu$s. Providing the WCD equations in the prompt lowers the error for 9 of 10 models on CQF but only 2 of 9 on CBS. Such errors are large relative to TSN latency budgets and can lead to violations of real-time constraints and unsafe configurations. TSNBench demonstrates that MCQ benchmarks may overestimate LLM capabilities in safety-critical networking. 
\end{abstract}

\section{Introduction}
Recent advances in large language models (LLMs) across different domains such as engineering~\citep{software_engineering_neurips, agi_eng_design_neurips}, medicine~\citep{tcmladder_medicine_neurips, CMExam_chinese_medical_exam_neurips, mediq_neurips}, clinical practice~\citep{EHRNoteQA_medicine_neurips}, computer networking~\citep{prosper_extract_protocol_spec}, telecommunications~\citep{teleqna, 6g_bench, 5g_neurips, tele_llm_poster_neurips_multi_agent}, and automation~\citep{taskbench_neurips} have shown groundbreaking performance in assisting engineers, practitioners, researchers~\citep{ai_research_agent_neurips, scieval_benchmark_scientific_research}, and doctors in solving real-world problems. System engineers are increasingly using LLMs to design and configure networks~\citep{netconfeval}, generate code, and analyze network logs. With this, they are entering new territory: safety-critical application domains such as autonomous vehicles, aerospace~\citep{spacecraft_tsn, microlaunchers_tsn}, defense~\citep{us_army_military}, and industrial communication~\citep{tsn_for_industrial}. In these contexts, the accuracy, reliability, and consistency of LLMs become far more than leaderboard metrics, as they become engineering requirements.

Time-Sensitive Networking (TSN)~\citep{8021Q}, standardized by the IEEE 802.1 Working Group (WG), is a layer-2 Ethernet technology that provides deterministic communication guarantees for safety-critical applications. TSN deployments typically separate traffic based on timing criticality. Safety-critical periodic communication with guaranteed latency and bounded jitter is categorized as time-triggered (TT) traffic and is served using the IEEE 802.1Qbv timed-gate mechanism. TT transmissions are controlled by a Gate-Control List (GCL), computed offline using exact methods such as SMT-based synthesis~\citep{silviu_tas} or heuristic approaches~\citep{PopIET16,8548553,bujosa2022hermes}. In contrast, periodic or sporadic communication requiring bounded end-to-end latency but less stringent jitter control is classified as Audio Video Bridging (AVB) stream traffic~\citep{electronics10202477,Bruckner}. Consequently, Worst-Case Delay (WCD) estimation errors of tens or hundreds of microseconds are significant, as they can consume timing margins, violate deadlines, or lead to infeasible TSN configurations. In mission-critical deployments, such errors can have severe consequences. A misconfigured TSN network can cause, for example, a robotic arm to miss a critical assembly step, a brake system to fail on a highway, an aircraft control system to respond incorrectly, a defense mechanism to collapse, or a spacecraft to miss a vital signal. These failures may result from sub-millisecond timing violations caused by a single misconfiguration. These risks highlight the importance of accurate analysis and configuration in TSN systems, especially as LLMs are increasingly integrated into network management workflows. Therefore, their domain proficiency must be rigorously evaluated. However, to the best of our knowledge, no existing benchmark evaluates LLM proficiency in TSN.

To fill this gap, we introduce TSNBench, the first benchmark for evaluating LLM proficiency in TSN, comprising two complementary evaluation components. The first is a 939-question expert-validated multiple-choice question and answer (MCQA) dataset, generated from 83 peer-reviewed research papers using three LLMs from distinct model families and rigorously reviewed by five domain experts, each with over eight years of TSN research experience. The second is a set of open-ended questions requiring multi-step WCD computation for two widely deployed TSN mechanisms, namely Credit-Based Shaper (CBS)~\citep{8021qav} and Cyclic Queuing and Forwarding (CQF)~\citep{8021Qch, itp_cqf}, across varying network topologies and traffic flows, with ground truth (GT) computed using a verified Network Calculus (NC) solver~\citep{luxi_avb} for CBS and closed-form mathematical upper bounds for CQF~\citep{jrsp_cqf}. These open-ended WCD questions are intended as a stress test of standalone model capability, not as a deployment workflow for free-text LLM timing outputs. Background on TSN, NC, CBS, and CQF is provided in Appendices~\ref{appendix:tsn} to \ref{appendix:cqf}. 

While general-purpose benchmarks such as MMLU~\citep{mmlu_iclr} and MMLU-Pro~\citep{mmlu_pro} evaluate broad subject knowledge spanning elementary mathematics, history, and law, they are fundamentally unsuited for safety-critical domain-specific evaluation. Answering an MCQ about elementary school history is categorically different from answering TSN terminology questions and correctly computing a WCD under NC constraints for a given network topology. Without a benchmark that captures this distinction, there is no principled way to measure LLM progress in deterministic networking domains. TSNBench is designed precisely to expose this gap.

We evaluate 16 LLMs comprising open-source and closed-source models, as well as general-purpose and reasoning-specialized architectures. Our results reveal a striking dissociation, where models achieve 67 to 95\% accuracy on MCQA yet fail substantially on open-ended WCD computation. The best-performing model on CBS, GPT-5, still deviates from the ground-truth WCD by 150.2~$\mu$s on average (mean absolute error, MAE), while all other models exceed 250~$\mu$s. This is concerning in a domain where timing violations of tens of microseconds, even 1\% of a 1000 $\mu$s deadline, may cause system failures. The dataset (CC-BY 4.0) is available at~\url{https://doi.org/10.34740/kaggle/ds/13019279} and the code (MIT) at~\url{https://github.com/tum-esi/TSNBench}. 

Our key contributions are:
\begin{enumerate}
    \item \textbf{First expert-validated TSN benchmark:} TSNBench evaluates LLM knowledge of TSN mechanisms through 939 expert-validated MCQs derived from peer-reviewed TSN literature.
    \item \textbf{Open-ended timing-analysis tasks:}
    TSNBench includes open-ended WCD computation tasks for CBS and CQF with GT computed using a verified NC solver for CBS and closed-form mathematical bounds for CQF.
    \item \textbf{Evaluation across 16 LLMs:} We evaluate both open-source and closed-source models, including general-purpose and reasoning-specialized models, and show that high MCQA accuracy does not reliably predict accurate WCD computation.
    \item \textbf{Analysis beyond closed-book prompting:} We test whether providing the WCD equations in the prompt (formula-in-context) lowers the error, how sensitive the error is to prompt wording and output format, and whether the findings hold on a real-world topology. 
\end{enumerate}

In summary, TSNBench gives the real-time networking and machine learning communities the first rigorous resource for evaluating LLM proficiency in TSN and the limits of LLMs in safety-critical, computationally demanding domains.

\section{Related Work}
\label{sec:related_work}
\textbf{General LLM Benchmarks:} Benchmarking and datasets are essential for measuring LLM progress and identifying key gaps and limitations~\citep{mmlu_iclr, mmlu_pro}. General knowledge benchmarks such as MMLU~\citep{mmlu_iclr} and MMLU-Pro~\citep{mmlu_pro} evaluate broad subject knowledge including elementary mathematics, history, computer science, and law, using multiple-choice questions. Domain-specific benchmarks have extended this paradigm to medicine~\citep{tcmladder_medicine_neurips, 
CMExam_chinese_medical_exam_neurips, mediq_neurips}, clinical practice~\citep{EHRNoteQA_medicine_neurips}, law~\citep{legalbench_neurips}, code generation~\citep{researchcodebench_neurips, effibench_neurips}, and scientific research~\citep{scieval_benchmark_scientific_research}. While these benchmarks have driven significant progress, they are not designed to evaluate safety-critical networking tasks. Most rely on multiple-choice evaluation, and none assess whether a model can perform the multi-step computational reasoning required in safety-critical networking domains. TSNBench addresses this gap by introducing MCQA and open-ended WCD computation questions with GT computed by a verified state-of-the-art NC solver and closed-form bounds, providing an evaluation of TSN that no existing general benchmark captures.

\textbf{Networking and Telecommunications Benchmarks:}
In the last few years, several benchmarks have evaluated LLM proficiency in networking and telecommunications domains. TeleQnA~\citep{teleqna} presents an MCQ dataset for telecommunications, generated from research documents and 3GPP standards and validated by domain experts. 6G-Bench~\citep{6g_bench} presents an MCQ-based dataset for 6G networks containing 3,722 difficult questions validated through automated filtering and expert human review. Beyond question-answering benchmarks, NetConfEval~\citep{netconfeval} evaluates LLMs on network configuration tasks and demonstrates that LLMs can simplify and automate complex network management tasks.

\textbf{LLMs for TSN and Real-Time Networks:}
The application of LLMs to TSN management and orchestration is still at a very early stage, with only limited initial studies available. \citet{netpilot} explored the use of LLMs for configuring hybrid 5G/TSN networks by assisting users with manual configuration tasks and suggesting configurations in a 5G-TSN network. However, this work remains preliminary and does not provide experimental results. Overall, prior work does not provide a systematic benchmark or rigorous evaluation of LLM proficiency across TSN mechanisms, nor does it assess computational reasoning capabilities for WCD analysis. TSNBench fills this gap by providing the first structured benchmark covering both declarative TSN knowledge through MCQA and computational reasoning through open-ended WCD evaluation.

\section{TSNBench}
\label{sub:overview_tsnbench}
Unlike established domains such as medicine~\citep{tcmladder_medicine_neurips}, 
5G~\citep{5g_neurips, teleqna}, general human knowledge~\citep{humanities_exam_2026, mmlu_iclr, mmlu_pro}, 
coding~\citep{researchcodebench_neurips, effibench_neurips}, and 
law~\citep{legalbench_neurips}, no open-source TSN dataset exists for LLM evaluation~\citep{tsn_for_industrial, survey_in_vehicle_tsn, wireless_tsn_survey, 5g_tsn_survey}. As highlighted in~\citep{CMExam_chinese_medical_exam_neurips}, the data source determines the reliability of a dataset, and generating a high-quality dataset is a crucial prerequisite for meaningful benchmarking. We describe the TSNBench construction pipeline below, with full details provided in Appendix~\ref{appendix:more_on_tsnbench}.

\begin{figure}
  \centering
  \includegraphics[width=0.8\linewidth, trim=0cm 0.1cm 0cm 0cm, clip]{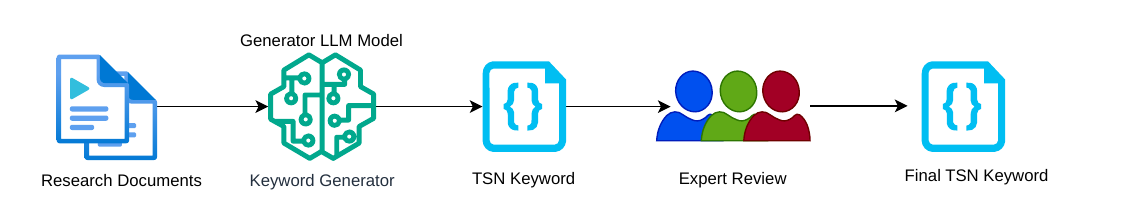}
  \caption{TSNBench keyword-generation pipeline. TSN keywords are extracted from research documents using an LLM, expert-verified, and used for MCQA generation as described in Section~\ref{sub:mcqa_gen}.}
  \label{fig:keyword_extractor}
\end{figure}

\subsection{Dataset Source Selection}
\label{sub:dataset_selection}
Published research papers and standards are among the most reliable sources for building domain-specific datasets~\citep{CMExam_chinese_medical_exam_neurips}. Since TSN knowledge originates primarily from peer-reviewed research and IEEE 802.1 TSN standards, we curate a collection of open-access research documents as our source corpus. To avoid copyright issues and exclude papers with incorrect results or flawed methodologies, we include only published or accepted papers that are openly available. For papers not available in open-access form, we use arXiv versions of papers that have been published or accepted, excluding unpublished preprints with unverified results. Where possible, we also collect author manuscript versions with proper attribution. To ensure quality, we prioritize highly cited papers from reputable venues while accounting for publication timeline, as recent papers naturally have fewer citations. In total, we collect 83 research papers covering a broad range of TSN mechanisms, including Time-Aware Shaper (TAS), CBS, CQF, NC-based schedulability analysis, performance evaluation, hardware experiments, combined shapers such as TAS+CBS~\citep{luxi_tnsm}, and Multi-CQF~\citep{paul_mcqf}. 

\subsection{Keyword and Acronym Extraction}
\label{sub:keyword_and_acronym_extract}
TSN employs specialized vocabulary, similar to other communication domains~\citep{5g, 6g, wifi}. A successful LLM that understands TSN should be able to reason correctly about TSN terminology. A model that cannot differentiate between TAS and CBS, or cannot correctly expand TSN-specific acronyms, cannot be considered proficient in TSN. To capture this dimension, we extract keywords and acronyms widely used in TSN literature and use them to guide MCQA generation. All terms are extracted from the 83 research documents using Claude Sonnet 4, as shown in Table~\ref{tab:gen_validator_model}, and stored in JSON format. Each document is preprocessed to remove non-relevant content, including author names, affiliations, figures, tables, URLs, and pseudocode. The model is instructed to extract only terms defined within the document, without relying on pretrained knowledge, and to provide each term’s acronym, full form, and one-to-two-sentence definition from the source. The extracted set is then reviewed by domain experts to resolve duplicates, retaining the longer definition in cases of conflict. Figure~\ref{fig:keyword_extractor} illustrates this pipeline. 

\subsection{MCQA Generation, Post-Processing, and Expert Review}
\label{sub:mcqa_gen}
\textbf{Raw MCQA Generation:} To optimize time and reduce manual effort, we use an LLM-based approach to generate MCQAs from research documents. The keyword file is provided alongside the research documents as additional input, serving as an independent source to complement research paper content during generation. We use three models from distinct families, namely Claude Sonnet 4, GPT-4o mini, and Llama 3.1 70B, as shown in Table~\ref{tab:gen_validator_model} (in Appendix~\ref{appendix:more_on_tsnbench}). These models are deliberately selected to ensure diverse styles and reasoning capabilities, thereby reducing generative bias. The same system prompt is used for all models, and each research paper is assigned to exactly one model in a round-robin manner. 

\textbf{Post-Processing:}
LLM-generated MCQAs cannot be used directly for benchmarking, as they may contain incorrectly formulated questions, incomplete options, or vague and incorrect answer choices. To address positional bias introduced by the generating model, answer options are shuffled randomly prior to human expert review, with the correct answer label updated to reflect the new ordering. 

\textbf{Human-Based Domain Expert Review:}
Given the safety-critical nature of TSN, rigorous human validation is essential. We engage five TSN domain experts: three senior professors with more than 15 years of research experience and two postdoctoral researchers with more than 8 years of expertise. Each question is independently evaluated with four outcomes: (i) \textit{accept} - correct and clear; (ii) \textit{revise} - requires modification for clarity or correctness; (iii) \textit{reject} - the question is incorrect, misleading, or irrelevant; or (iv) \textit{doubtful} - the expert is uncertain and passes it to remaining reviewers for consensus. Questions without consensus are discarded. Full review criteria are provided in Appendix~\ref{appendix:sub_human_review}. 

\begin{figure}[htbp]
  \centering
  \includegraphics[width=0.9\linewidth, trim=0cm 0.2cm 0cm 0cm, clip]{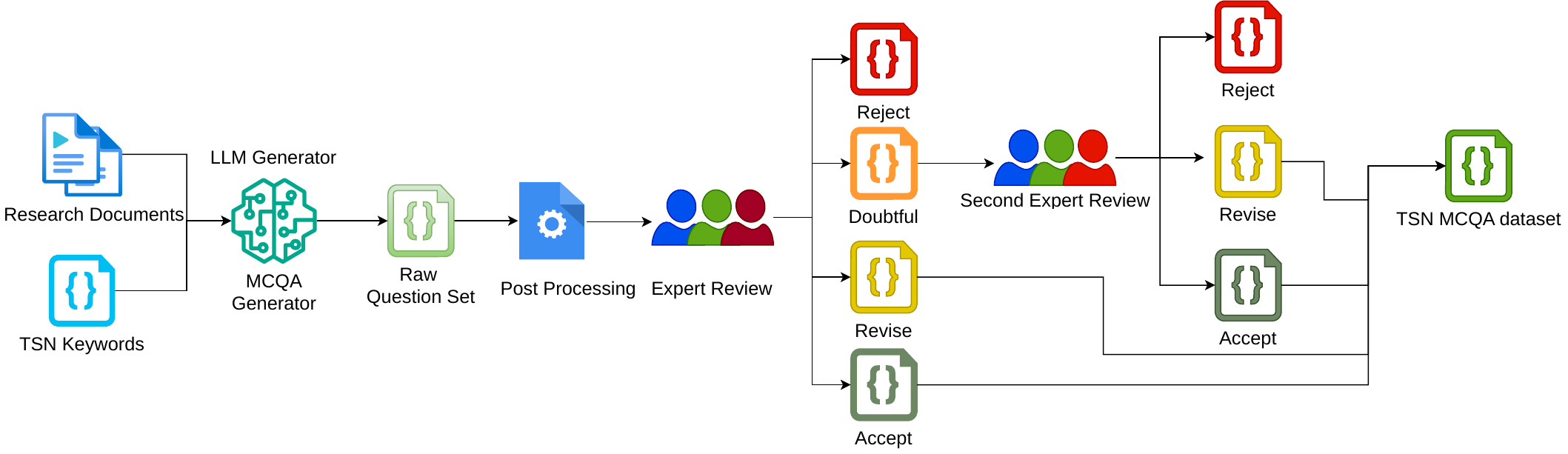}
  \caption{Pipeline of our TSNBench MCQA dataset generator, showing all steps from raw generation to the final validated dataset.}
  \label{fig:dataset_generator}
\end{figure}

\textbf{Mechanism Coverage:}
To check for under-representation, we categorize all 939 MCQs by TSN mechanism (Table~\ref{appendix:mcq_coverage} in the Appendix). TAS accounts for the largest share of the questions (36.1\%), reflecting its larger research base. Each TSN mechanism accounts for at least 5\% of the questions. Table~\ref{tab:mcqa_statistics} summarizes the dataset statistics and Figure~\ref{fig:dataset_generator} illustrates the full pipeline.

\begin{table}[htbp]
  \caption{TSNBench dataset construction statistics. Full generation details are in Appendix~\ref{appendix:more_on_tsnbench}.}
  \label{tab:mcqa_statistics}
  \centering
  \small
  \setlength{\tabcolsep}{3pt}
  \begin{tabular}{l|l|c}
    \toprule
    \textbf{Type} & \textbf{Category} & \textbf{Count} \\
    \midrule
    \multirow{3}{*}{\textbf{MCQA}}
      & Total raw questions generated by models & 1326 \\ 
      & Questions removed after expert review & 387 \\
      & Questions revised by domain experts & 185 \\
      & Questions in the final dataset (used for benchmarking) & \textbf{939} \\
    \midrule
    \multirow{3}{*}{\textbf{Open-ended questions}}
      & Credit-Based Shaper (CBS) & 100 \\
      & Cyclic Queuing and Forwarding (CQF) & 100 \\
    \midrule
    \multirow{3}{*}{\textbf{Orion Crew Exploration Vehicle (CEV)}}
      & CBS & 10 \\
      & CQF & 10 \\
    \bottomrule
  \end{tabular}
\end{table}

\subsection{Open-Ended Question Formulation}
\label{sub:opne_end_question}
\begin{figure}[t]
  \centering
  \includegraphics[width=0.5\linewidth, trim=0cm 0.3cm 0cm 0cm, clip]{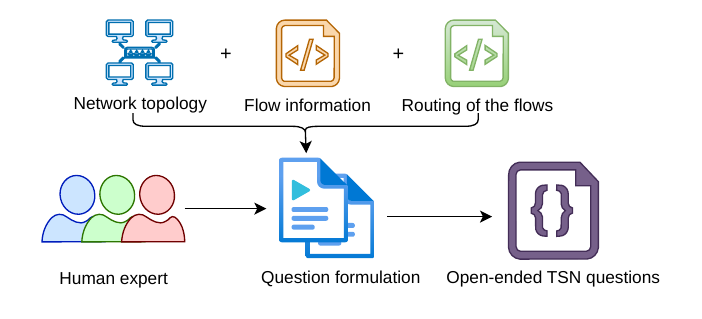}
  \caption{Pipeline for TSNBench open-ended question formulation by domain experts. Each question comprises three components: network topology, flow information, and flow routing.}
  \label{fig:open_end_question}
\end{figure}

While MCQA evaluates declarative TSN knowledge, open-ended questions assess whether LLMs can perform the multi-step mathematical reasoning required in real TSN deployment. We evaluate WCD computation, as WCD is a central key performance indicator (KPI) in TSN network design and directly determines whether a network meets its stringent timing requirements. We select two TSN mechanisms for this evaluation: CBS and CQF. CBS is widely deployed for audio-video traffic and requires NC-based analysis, making it mathematically demanding. CQF is a more recently standardized TSN mechanism whose WCD can be computed from a closed-form equation given routing and cycle duration ($T$), providing a complementary evaluation that isolates formula application from NC complexity. Together, these two mechanisms span a meaningful range of WCD computation difficulty. GT WCD values are computed using a verified state-of-the-art NC tool~\citep{luxi_avb} for CBS and closed-form mathematical upper bounds for CQF. We release all GT WCD values alongside the questions to support future open-source community evaluations. Each open-ended question is formulated by domain experts, as shown in Figure~\ref{fig:open_end_question}, and comprises three components: network topology, flow information, and flow routing. In TSNBench, three topologies are used to cover a broad range of scenarios: (i) one-switch topology (Figure~\ref{fig:one_switch_topo}), (ii) medium-mesh topology (Figure~\ref{fig:mm_topo}), and (iii) ring topology, representing industrial networks (Figure~\ref{fig:ring_topo}). We also evaluate a real-world topology (Figure~\ref{fig:orion_topo}). Each topology consists of end nodes and switches connected via Ethernet links, with unicast traffic flows transmitted from a sender to a single receiver. Flows consist of Ethernet frames whose maximum payload is bounded by the Maximum Transmission Unit (MTU). Further topology, flow, and routing details are provided in Appendix~\ref{appendix:open_end_question_details}. 

\subsection{Prompt Design}
For both MCQA and open-ended evaluations, each prompt defines the model’s role as a TSN expert. For MCQA, we use zero-shot prompting with no in-context examples, representing a conservative approach that measures inherent TSN proficiency, ensuring that the output performance reflects the model's domain knowledge rather than in-context pattern matching. For both question types, the model is asked to provide a confidence score alongside its answer.

For open-ended questions, we use two settings: (i) closed-book (CB), and (ii) formula-in-context (FC). In the CB setting, we use a zero-shot prompt, providing no example WCD calculations or NC or CQF equations, ensuring the model independently recalls and applies the correct computational methodology. In the FC setting, the prompt additionally provides the NC equations for CBS WCD analysis and the upper-bound equations of CQF WCD calculation, to test whether the errors persist when the model does not have to recall the methodology. The open-ended prompt comprises three variable components: network topology, flow parameters, and pre-computed shortest-path routes. The same prompt template is used across all 100 open-ended evaluation instances per mechanism and setting, with only these three components varying. Fixed network constants are maintained throughout to ensure comparability across models and instances. To measure the sensitivity to the prompt itself, the CB prompt is additionally evaluated in two variants: Variant A paraphrases the instructions, and Variant B requests the answer as plain text instead of JSON. A detailed discussion of the open-ended prompt design is provided in Appendix~\ref{appendix:prompt_open_end}.

\subsection{Model Scoring and Ground Truth}
\label{sec:ground_truth}
For the MCQA dataset, performance is measured as the percentage of questions answered correctly, reported as accuracy. For the open-ended questions, we evaluate the computational reasoning capability of each model by comparing its predicted WCD values against GT values. For CBS, GT WCD values are derived using NC-based Total Flow Analysis (TFA). Specifically, the worst-case delay upper bound $D_f^h$ for flow $f\in\mathcal{F}_{M_i}^h$ at $h$ equals the worst-case delay upper bound $D_{M_i}^{h}$ for all flows with the same priority $M_i$ aggregating at the output port $h$, 
\begin{equation}\label{g:aggDelay}
D_f^h\!=\!D_{M_i}^{h}\!=\!hDev(\alpha_{M_i}^h,\beta_{M_i}^h)=\!\sup_{t\geq0}\left\{\inf\left\{\tau\!\geq\!0\mid\alpha^h_{M_i}\!(t)\leq\beta^h_{M_i}\!(t\!+\!\tau)\right\}\right\},\\
\end{equation}
where $\alpha_{M_i}^h(t)$ represents the arrival curve of aggregate flows of priority $M_i$ passing through $h$, and $\beta_{M_i}^h(t)$ represents the service curve for these corresponding flows. The end-to-end WCD for a flow is obtained by summing per-port delay bounds along its route. Full NC methodology and proofs are provided in Appendix~\ref{appendix:network_calculus}.

For CQF, the worst-case end-to-end delay is given by the closed-form expression
\begin{equation}
    \mathrm{WCD} = f_i.\phi + (\mathrm{SW_{num}}+1) \cdot \mathrm{T} + \xi,
    \label{eq:cqf_delay}
\end{equation} 
where $f_i.\phi$ is the flow offset at the source node in $\mu$s, $\mathrm{SW_{num}}$ is the number of switches along the flow route, $\mathrm{T}$ is the cycle duration in $\mu$s, and $\xi$ denotes the network-specific delays including propagation delay, switching delay (per switch), and time synchronization error. Details and a description of this bound are provided in Appendix~\ref{appendix:cqf}. 

\begin{figure}[t]
\begin{minipage}[c]{0.64\textwidth}
\vspace{0pt}
\raggedright
\captionof{table}{TSNBench MCQA results across 16 models. Accuracy is the percentage of correct answers out of 939 questions, and consistency measures whether the model gives the same response across three runs. All models are evaluated at temperature 0.0 for deterministic performance. Models without temperature support use their default setting and are marked with $^\dagger$. Full model details are given in Table~\ref{tab:evaluation_models}, and extended results with temperature comparisons are provided in Table~\ref{tab:more_tsnbench_evaluation_appendix_deafult_temp_temp_0} in Appendix~\ref{appendix:more_on_evaluation}.}
\label{tab:tsnbench_evaluation_results}
\renewcommand{\arraystretch}{1.2}
\resizebox{\textwidth}{!}{%
\begin{tabular}{l|c|c|c|c|c|c|c}
\toprule
\textbf{Model} & 
\shortstack{\textbf{Accuracy} \\ \textbf{(\%)}} & 
\shortstack{\textbf{Avg.} \\ \textbf{Consistency}} & 
\shortstack{\textbf{Avg.} \\ \textbf{Latency (ms)}} & 
\shortstack{\textbf{Avg.} \\ \textbf{Conf.}} & 
\textbf{ECE}$\downarrow$ & 
\textbf{Brier}$\downarrow$ & 
\shortstack{\textbf{CW} \\ \textbf{Rate}$\downarrow$} \\
\midrule
Grok 4.1 Fast$^\dagger$ & 93.2 & 0.9858 & 6673 & 0.9509 & 0.0198 & 0.0599 & 99.0 \\
Grok 4.1 Fast (Non-Reasoning) & 91.7 & 0.9986 & 515 & 0.9760 & 0.0589 & 0.0764 & 100.0 \\
DeepSeek-V3.2 (Non-Thinking) & 94.0 & 0.9993 & 804 & 0.9312 & \cellcolor{green!20}\textbf{0.0104} & 0.0526 & 96.4 \\
GPT-4o & 91.8 & 0.9957 & 729 & 0.8782 & 0.0384 & 0.0765 & 99.2 \\
GPT-4o mini & 88.3 & 0.9950 & 799 & 0.9004 & 0.0171 & 0.0974 & 77.8 \\
Llama 3.3 70B & 88.9 & 0.9950 & 365 & 0.9082 & 0.0346 & 0.0918 & 100.0 \\
Mistral Medium 3.1 & 92.1 & 0.9965 & 653 & \cellcolor{green!20}\textbf{0.9779} & 0.0574 & 0.0750 & 100.0 \\
Mistral Large 3 & 92.8 & 0.9975 & 5498 & 0.9476 & 0.0194 & 0.0646 & 100.0 \\
Claude Sonnet 4.5 & \cellcolor{green!20}\textbf{95.3} & 0.9993 & 1842 & 0.9374 & 0.0178 & \cellcolor{green!20}\textbf{0.0429} & 86.6 \\
o3$^\dagger$ & 94.7 & 0.9840 & 3845 & \cellcolor{red!20}\textbf{0.7524} & \cellcolor{red!20}\textbf{0.1951} & 0.0852 & \cellcolor{green!20}\textbf{3.4} \\
GPT-5$^\dagger$ & 95.0 & 0.9886 & 5630 & 0.8773 & 0.0720 & 0.0475 & 51.7 \\
DeepSeek-V3.2 (Thinking)$^\dagger$ & 94.7 & 0.9819 & 4400 & 0.9202 & 0.0240 & 0.0487 & 78.1 \\
Gemini 2.5 Flash & 90.1 & 0.9847 & 6744 & 0.9674 & 0.0793 & 0.0942 & 95.4 \\
Llama 3.2 1B & \cellcolor{red!20}\textbf{67.4} & 1.0 & 669 & 0.8529 & 0.1840 & \cellcolor{red!20}\textbf{0.2544} & 99.0 \\
Qwen3 8B & 83.7 & 0.9897 & \cellcolor{red!20}\textbf{15103} & 0.8616 & 0.0360 & 0.1322 & 100.0 \\
Ministral 3 8B & 86.9 & 0.9954 & \cellcolor{green!20}\textbf{345} & 0.9649 & 0.0971 & 0.1230 & 100.0 \\
\bottomrule
\end{tabular}%
    }
    \smallskip
    {\scriptsize $^\dagger$ Temperature parameter not supported. 
    Evaluated with default settings. $\downarrow$ lower is better.}
\end{minipage}
\hfill
\begin{minipage}[c]{0.34\textwidth}
    \vspace{0pt}
    \centering
    \includegraphics[width=\textwidth, trim=0cm 0.3cm 0cm 0cm, clip]{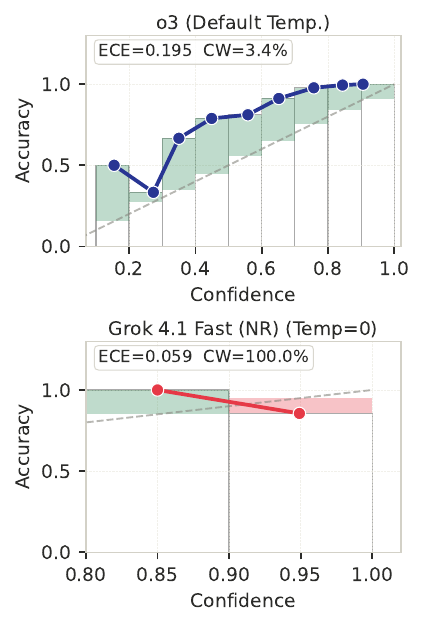}
    \caption{Reliability plot for o3 and Grok~4.1~Fast~(NR). Full reliability analysis is in Figure~\ref{fig:reliability_diagram}.}
    \label{fig:reliability_selected}
\end{minipage}
\end{figure}

\section{Experiments}
\label{sec:experiments}
We evaluate 16 state-of-the-art LLMs spanning open-source and closed-source models across general-purpose and reasoning-specialized architectures. Table~\ref{tab:evaluation_models} in Appendix~\ref{appendix:more_on_evaluation} 
provides the full list of models with their model IDs and organizations. All models are accessed via their vendor APIs with no fine-tuning applied: GPT (OpenAI API), DeepSeek (DeepSeek API), Mistral (Mistral AI API), Claude (Anthropic API), Gemini (Google AI API), Grok (xAI API), and Llama and Qwen (Hugging Face inference router). All client-side operations, including prompt construction, API handling, response parsing, and metric computation, are performed on a standard workstation. To assess repeatability, each MCQ is evaluated three times at temperature 0.0 and three times at 0.7, and each open-ended question in up to three runs at 0.0. Models that do not expose a temperature parameter use the vendor default configuration, as noted in Table~\ref{tab:tsnbench_evaluation_results}. Since TSN is widely used in safety-critical domains, deterministic responses are essential, as non-determinism would undermine the reliability of LLM-based TSN reasoning. Full cost and latency details are provided in Appendix~\ref{appendix:cost_latency}, Table~\ref{tab:more_tsnbench_evaluation_cost_and_latency}.

\subsection{MCQA Evaluation}
\label{sub:mcqa_evaluation_main}
\textbf{Contamination Analysis:}
Since the MCQs were generated using models from families included in the evaluation, as shown in Table~\ref{tab:gen_validator_model}, contamination is a potential concern. We therefore separate the evaluated models into generator families (Claude, GPT, Llama) and non-generator families (all remaining models) and compare their average MCQA accuracy. Generator-family models achieve an average accuracy of 88.8\%, whereas non-generator-family models achieve 91.0\%. The generator-family models do not perform better than the non-generator-family models, so we do not observe evidence of a systematic advantage. This analysis does not rule out all possible contamination pathways, but it addresses this specific concern. The open-ended timing tasks are less likely to be affected because their topology, flow, and routing inputs were constructed specifically for TSNBench.

\textbf{Evaluation Metrics:} Model performance on the MCQA dataset is measured using accuracy, defined as the percentage of correctly answered questions out of 939, averaged across three runs. We additionally report Expected Calibration Error (ECE)~\citep{ece} and Brier score~\citep{brier} to evaluate the alignment between the model's expressed confidence and its actual correctness. Calibration is particularly critical in safety-critical domains such as TSN, where high-confidence incorrect answers may lead to misleading configuration decisions, deadline violations, or network instability in industrial and automotive systems. We therefore also evaluate the Confidently Wrong (CW) rate to determine the fraction of incorrect answers where the model expresses high confidence ($\geq$0.8). All calibration metrics are computed on all 939 MCQs across three runs per model. 

\textbf{Results and Discussion:} 
Table~\ref{tab:tsnbench_evaluation_results} reports accuracy, average (avg.) consistency, calibration, and average latency for all 16 models. The top performers are Claude Sonnet 4.5 (95.3\%) and GPT-5 (95.0\%), with Claude Sonnet 4.5 also achieving the lowest Brier score (0.0429), indicating strong accuracy and calibration. Llama 3.2 1B achieves the lowest accuracy (67.4\%), consistent with its substantially smaller parameter count compared with the other models. 

A notable finding emerges from the reasoning models. Despite their stronger general reasoning capabilities, o3, GPT-5, and DeepSeek-V3.2 (Thinking) do not outperform the best non-reasoning models on MCQA, all scoring below Claude Sonnet 4.5. This suggests that TSN MCQA performance is primarily driven by domain knowledge rather than general reasoning, and that reasoning-specialized architectures offer limited advantage on declarative knowledge retrieval tasks. 

The calibration results reveal key differences across models. While most models are well-calibrated (ECE < 0.06), o3 has the highest ECE (0.1951) despite 94.7\% accuracy, yet achieves the lowest CW rate (3.4\%), rarely assigning high confidence to incorrect answers (refer to Figure~\ref{fig:reliability_selected}). In contrast, many non-reasoning models have CW rates of 100\%, assigning high confidence to incorrect answers. Mistral Medium 3.1 has the highest average confidence (0.9779) while maintaining 92.1\% accuracy. All models have zero refusal rate, indicating that the MCQA dataset does not trigger response refusals.

\subsection{Open-Ended Question Evaluation}
\label{sub:open_end_evaluation}
\textbf{Evaluation Metrics:} For the open-ended questions, we report the MAE, computed per test case (TC). Each ${TC}_x$ consists of $n_x$ flows, where $n_x$ is the number of flows in ${TC}_x$. Each flow is denoted as $f_i$, $i = 1, \ldots, n_x$. For each flow $f_i$, $\hat{y}_{{TC}_{x},f_i}$ denotes the WCD predicted by the model for ${TC}_x$ and $y_{{TC}_{x},f_i}$ denotes the GT WCD of flow $f_i$ for TC number $x$, computed using a verified NC solver for CBS and using Eq.~\ref{eq:cqf_delay} for CQF. Each TC is evaluated in up to three runs at one temperature per model (0.0, or the vendor default for models without temperature support), and $\hat{y}_{{TC}_{x},f_i}$ is the average of the predictions of flow $f_i$ over the valid runs. A run is discarded if it returns no WCD values, only zeros, or a flow identifier that does not exist in the TC (e.g., $f_0$ to $f_2$ instead of $f_1$ to $f_3$). Non-numeric answers such as ``unbounded'' or ``inf'' and values above $10^9~\mu$s are treated as missing. With $P_x$ denoting the flows of ${TC}_x$ that receive a predicted WCD, the MAE of a TC is defined as:
\begin{equation}
    \text{MAE}_{{TC}_{x}} = \frac{1}{|P_x|} \sum_{f_i \in P_x} \left|\hat{y}_{{TC}_{x},f_i} - y_{{TC}_{x},f_i}\right|,
    \label{eq:mae_per_tc}
\end{equation}
where $x \in \{1, \ldots, 100\}$ is the TC index. TCs in which fewer than 80\% of the flows are predicted are excluded. The overall MAE of a model is the average over its set $\mathcal{V}$ of valid TCs ($|\mathcal{V}| \leq 100$):
\begin{equation}
    \text{MAE} = \frac{1}{|\mathcal{V}|} \sum_{x \in \mathcal{V}} \text{MAE}_{{TC}_x},
\end{equation}
MAE is reported with its standard deviation across the valid TCs. A model is reported only if at least 50 valid TCs remain. We additionally report the median MAE across TCs as a robust measure against outlier TCs. We also report the percentage of flows whose absolute error falls below 50, 100, 500, and 1000~$\mu$s. We evaluate each model in the CB and FC settings and we report the relative change $\Delta = \left( \dfrac{\text{MAE}_{\text{CB}} - \text{MAE}_{\text{FC}}}{\text{MAE}_{\text{CB}}} \right) \cdot 100\, \%$, so a positive $\Delta$ means that providing the formula lowers the error. A worked example is given in Appendix~\ref{appendix:open_end_question_evaluation} (Table~\ref{tab:sample_mae_example}).

\begin{table*}[t]
\centering
\caption{Open-ended WCD estimation results for CBS and CQF across 100 test cases (TCs), reported as MAE $\pm$ standard deviation across valid TCs (\textdownarrow{} lower is better), under the closed-book (CB) and formula-in-context (FC) settings. $\Delta$ is the relative change of the MAE from CB to FC, with each MAE taken over the valid TCs of its setting. Positive values mean the formula lowers the error. Medians and full statistics are reported in Tables~\ref{appendix:cbs_mae_std_median} and \ref{appendix:cqf_mae_std_median}.}  
\label{tab:open_end_std_dev}
\resizebox{\textwidth}{!}{%
\begin{tabular}{l|ccc|ccc}
\toprule
\multirow{2}{*}{\textbf{Model}}
& \multicolumn{3}{c|}{\textbf{CBS WCD MAE}}
& \multicolumn{3}{c}{\textbf{CQF WCD MAE}} \\
\cmidrule(lr){2-4} \cmidrule(lr){5-7}
& \textbf{CB (µs)} $\downarrow$
& \textbf{FC (µs)} $\downarrow$
& \textbf{$\Delta$ (\%)}
& \textbf{CB (µs)} $\downarrow$
& \textbf{FC (µs)} $\downarrow$
& \textbf{$\Delta$ (\%)} \\
\midrule
Grok 4.1 Fast$^{\dagger,\P}$ & -- & -- & -- & $130.1 \pm 85.0$ & -- & -- \\
Grok 4.1 Fast (Non-Reasoning)$^{\P}$ & \cellcolor{red!20}\bm{$3637.3 \pm 3934.5$} & -- & -- & $163.9 \pm 37.8$ & -- & -- \\
DeepSeek-V3.2 (Non-Thinking)$^{\P}$ & -- & -- & -- & $168.1 \pm 66.3$ & -- & -- \\
GPT-4o$^\star$ & -- & -- & -- & $82.2 \pm 264.7$ & $11.3 \pm 15.8$ & $+86.3$ \\ 
GPT-4o mini & $378.5 \pm 189.7$ & $461.6 \pm 999.5$ & $-21.9$ & $180.5 \pm 82.0$ & \cellcolor{red!20}\bm{$1377.1 \pm 2818.7$} & \cellcolor{red!20}\bm{$-662.9$} \\
Llama 3.3 70B & $301.2 \pm 171.0$ & $333.5 \pm 171.3$ & $-10.7$ & $173.3 \pm 73.5$ & $5.1 \pm 15.0$ & $+97.1$ \\
Mistral Medium 3.1 & $339.6 \pm 224.9$ & $374.2 \pm 232.7$ & $-10.2$ & $150.3 \pm 42.2$ & $33.0 \pm 31.3$ & $+78.0$ \\
Mistral Large 3 & $270.6 \pm 180.2$ & $273.5 \pm 179.9$ & $-1.1$ & \cellcolor{green!20}\bm{$54.2 \pm 28.2$} & $19.3 \pm 17.8$ & $+64.3$ \\
Claude Sonnet 4.5 & $293.0 \pm 173.9$ & $307.7 \pm 178.3$ & $-5.0$ & $213.2 \pm 1062.5$ & $30.3 \pm 37.2$ & $+85.8$ \\
o3$^\dagger$ & $255.8 \pm 315.4$ & $501.9 \pm 2341.3$ & \cellcolor{red!20}\bm{$-96.2$} & $101.6 \pm 75.9$ & \cellcolor{green!20}\bm{$0.0 \pm 0.0$} & \cellcolor{green!20}\bm{$+100.0$} \\
GPT-5$^\dagger$ & \cellcolor{green!20}\bm{$150.2 \pm 198.2$} & \cellcolor{green!20}\bm{$161.4 \pm 188.0$} & $-7.5$ & $107.0 \pm 69.0$ & \cellcolor{green!20}\bm{$0.0 \pm 0.0$} & \cellcolor{green!20}\bm{$+100.0$} \\
Gemini 2.5 Flash & $554.6 \pm 1831.6$ & $258.0 \pm 231.2$ & \cellcolor{green!20}\bm{$+53.5$} & $112.0 \pm 90.1$ & $0.1 \pm 0.5$ & $+99.9$ \\
Qwen3 8B & -- & \cellcolor{red!20}\bm{$51855.3 \pm 460034.8$} & -- & -- & $2.0 \pm 6.2$ & -- \\
Ministral 3 8B & $1643.3 \pm 2618.4$ & $1551.7 \pm 2735.6$ & $+5.6$ & \cellcolor{red!20}\bm{$3057.4 \pm 4112.4$} & $1354.6 \pm 2166.2$ & $+55.7$ \\
\bottomrule
\end{tabular}%
}
\\[4pt]
\raggedright
\footnotesize
``--'' Not reported: fewer than 50 valid TCs, or not evaluated ($^\P$).\\
$^\dagger$ Temperature parameter not supported, evaluated with default settings.\\
$^\star$ GPT-4o returned all-zero WCD values (trivial failure) for 51 of the 100 CBS TCs, leaving 49 valid CBS TCs in the CB setting and 17 in the FC setting. It produced valid WCD estimates for all 100 CQF TCs.\\
$^\P$ Retired by the provider before the FC and prompt-sensitivity experiments, so only CB results are reported.\\
Not shown (fewer than 50 valid TCs in every setting): Llama 3.2 1B, DeepSeek-V3.2 (Thinking). Detailed failure modes are described in Appendix~\ref{appendix:failure_mode}.\\ 
\end{table*}

\begin{table}[htbp]
\centering
\caption{Percentage of flows with an absolute WCD error below 50, 100, 500, and 1000~$\mu$s in the CB setting for CBS and CQF. For each model, the percentage is computed per TC over the flows with a finite prediction and averaged over the valid TCs. FC results are given in Table~\ref{appendix:fc_error_thresholds}.}
\label{tab:error_thresholds}
\small
\resizebox{\textwidth}{!}{%
\begin{tabular}{l|cccc|cccc}
\toprule
\multirow{2}{*}{\textbf{Model}}
& \multicolumn{4}{c|}{\textbf{CBS (\%)}}
& \multicolumn{4}{c}{\textbf{CQF (\%)}} \\
\cmidrule(lr){2-5} \cmidrule(lr){6-9}
& \textbf{${<}50\mu$s} & \textbf{${<}100\mu$s} & \textbf{${<}500\mu$s} & \textbf{${<}1000\mu$s}
& \textbf{${<}50\mu$s} & \textbf{${<}100\mu$s} & \textbf{${<}500\mu$s} & \textbf{${<}1000\mu$s} \\
\midrule
Grok 4.1 Fast$^{\dagger,\P}$ & -- & -- & -- & -- & 28.2 & 50.0 & 99.5 & 100.0 \\
Grok 4.1 Fast (Non-Reasoning)$^\P$ & 2.6 & 5.5 & 23.2 & 31.1 & 0.0 & 16.7 & 100.0 & 100.0 \\
DeepSeek-V3.2 (Non-Thinking)$^{\P}$ & -- & -- & -- & -- & 3.8 & 23.4 & 99.4 & 99.8 \\
GPT-4o$^\star$ & -- & -- & -- & -- & 84.4 & 87.7 & 95.0 & 97.0 \\
GPT-4o mini & 1.0 & 6.8 & 74.3 & 96.6 & 1.7 & 18.9 & 98.1 & 99.6 \\
Llama 3.3 70B & 8.8 & 18.6 & 84.7 & 98.5 & 3.3 & 16.9 & 98.9 & 100.0 \\
Mistral Medium 3.1 & 10.7 & 21.6 & 77.1 & 97.0 & 1.5 & 23.2 & 99.8 & 100.0 \\
Mistral Large 3 & 14.8 & 25.3 & 86.8 & 98.0 & 48.6 & 92.6 & 100.0 & 100.0 \\
Claude Sonnet 4.5 & 7.2 & 19.3 & 85.6 & 98.1 & 49.1 & 71.3 & 98.0 & 98.0 \\
o3$^\dagger$ & 22.0 & 41.8 & 85.9 & 95.9 & 28.1 & 63.2 & 99.8 & 100.0 \\
GPT-5$^\dagger$ & 39.4 & 60.4 & 94.1 & 98.0 & 41.9 & 58.4 & 99.9 & 100.0 \\
Gemini 2.5 Flash & 16.6 & 29.3 & 80.5 & 94.2 & 42.0 & 60.4 & 98.5 & 100.0 \\
Ministral 3 8B & 1.9 & 4.1 & 21.4 & 67.8 & 2.6 & 6.4 & 14.2 & 46.3 \\
\bottomrule
\end{tabular}
}
\\[4pt]
\raggedright
\footnotesize
``--'', $^\dagger$, $^\star$, $^\P$ and the models not shown are explained in Table~\ref{tab:open_end_std_dev}. All thresholds are in $\mu$s (\textuparrow{} higher is better).
\end{table}

\textbf{Results and Discussion:} 
Table~\ref{tab:open_end_std_dev} presents the WCD computation results for both CBS and CQF across all 100 TCs under both CB and FC settings. The central finding is a striking dissociation between MCQA accuracy and computational reasoning performance. Models that achieve above 90\% accuracy on MCQA still fail substantially on open-ended WCD computation, with the best-performing model, GPT-5, achieving an MAE of 150.2~$\mu$s on CBS, which is concerning because industrial TSN traffic can have strict timing requirements~\citep{industrial_traffic}. Medians are given in Tables~\ref{appendix:cbs_mae_std_median} and~\ref{appendix:cqf_mae_std_median}, and per-TC results in Tables~\ref{tab:cbs_per_tc_mae_one_switch} to~\ref{tab:cqf_per_tc_mae_medium_mesh} in Appendix~\ref{appendix:open_end_question_evaluation}. 

For CBS, most models produce large errors, with all models except GPT-5 exceeding 250~$\mu$s MAE. Several models exhibit distinct failure modes. On CBS, Llama 3.2 1B returns WCD values for only 25 of its 61 evaluated TCs, all of them zero, and no parsable value for the others, so none of its TCs is valid. Grok 4.1 Fast returns truncated JSON, providing flow profile metadata but no WCD values, suggesting that the model hit an output length limit. DeepSeek-V3.2 (Thinking) returns empty responses for all CBS TCs and all but one CQF TC. The NC-based computation required for CBS is mathematically demanding and complex, and the zero-shot setting reveals that most models cannot independently recall or correctly apply the full NC methodology. Notably, the OpenAI reasoning models GPT-5 and o3 achieve the lowest CBS MAE of all models (150.2 and 255.8~$\mu$s), suggesting that multi-step mathematical reasoning capability provides an advantage for NC-based WCD computation even when it does not improve MCQA accuracy. 

For CQF, performance is more varied, with median MAE ranging from 1.2~$\mu$s (GPT-4o) to 1,046~$\mu$s (Ministral 3 8B). GPT-4o achieves this lowest median despite returning all-zero WCD values for 51 of the 100 CBS TCs, suggesting it can correctly apply the CQF closed-form equation, while Mistral Large 3 achieves the lowest mean MAE (54.2~$\mu$s). Llama 3.2 1B exhibits the most severe hallucination failure, listing up to 1,013 invented flows (F0 to F1012) instead of the flows of the TC and returning WCD~=~0 for all of them. Qwen3 8B has fewer than 50 valid TCs in the CB setting due to repeated API timeouts or responses with flow IDs that do not exist in the TC, and is reported only in the FC setting (Table~\ref{tab:open_end_std_dev}). Ministral 3 8B, despite its small size, produces valid responses for both mechanisms but with large errors (MAE 1,643.3~$\mu$s on CBS and 3,057.4~$\mu$s on CQF), showing that a valid answer format alone does not lead to correct WCD values.

\textbf{Effect of formula-in-context:} Providing the NC equations in the prompt does not help most models on CBS. CBS WCD computation is not a single formula but a multi-step derivation of arrival curves, the service curve, and the delay bound. Of the nine models with CBS results in both settings, FC lowers the mean MAE only for Gemini 2.5 Flash ($+53.5\%$) and Ministral 3 8B ($+5.6\%$) and raises it for the other seven, most strongly for o3 ($-96.2\%$, median from 138.8 to 187.4~$\mu$s). This suggests that formulas help models that lack a working strategy but can hurt models that already have one. For CQF, where the WCD reduces to a single closed-form upper bound, FC lowers the MAE for 9 of 10 models, by 56 to 100\%. o3 and GPT-5 reach an MAE of 0.0~$\mu$s, and Gemini 2.5 Flash and Llama 3.3 70B come close (0.1 and 5.1~$\mu$s). GPT-4o mini is the exception. Its CQF error rises from 180.5 to 1,377.1~$\mu$s and its median from 175.3 to 323.4~$\mu$s, so the increase is not caused only by a few outlier TCs. These results suggest that formulas help reliably when the task is substitution, but their effect depends on the model's own reasoning when the task requires a multi-step derivation.

\textbf{Error thresholds:} Error thresholds show how rarely the predictions are precise. In the CBS CB setting, even GPT-5 places only 39.4\% of flows within 50~$\mu$s of the GT, and all other models stay at or below 22\% (Table~\ref{tab:error_thresholds}, with FC results in Table~\ref{appendix:fc_error_thresholds}).

\textbf{Prompt sensitivity:} We repeat the CB evaluation with a paraphrased prompt (Variant A) and a prompt that requests plain text instead of JSON output (Variant B). On CBS, results are largely robust to paraphrasing for most models but sensitive to the output format (Table~\ref{tab:prompt_sensitivity_cbs_appendix} in the Appendix). Plain-text output raises the mean CBS MAE by more than two orders of magnitude for GPT-4o mini (316 times higher, Table~\ref{tab:prompt_sensitivity_cbs_appendix}). Claude Sonnet 4.5, GPT-5, and o3 remain stable and reach a lower MAE under both variants than with the original CB prompt. The most stable models are o3 (CV 1.2\%) and Claude Sonnet 4.5 (CV 2.0\%). Further results are given in Tables~\ref{appendix:prompt_sensitivity_cv_cqf} to~\ref{appendix:prompt_sensitivity_cqf_appendix} in the Appendix.

\textbf{Real-world topology case study:} To test whether the findings carry over to a real network topology, we evaluate CBS and CQF on the Orion Crew Exploration Vehicle (CEV) topology (Figure~\ref{fig:orion_topo}). For CBS, as on the synthetic topologies, FC lowers the mean MAE for only a minority of models (4 of 10), namely GPT-4o mini, o3, GPT-5, and Qwen3 8B. In the CB setting, o3 reaches the lowest MAE of 256.1~$\mu$s (Appendix Tables~\ref{tab:orion_cbs_summary}, \ref{appendix:orion_cbs_main}, and~\ref{appendix:orion_cbs_mae_std}). Per-TC results are given in Tables~\ref{tab:orion_cbs_per_tc_cb_appendix} and~\ref{tab:orion_cbs_per_tc_fc_appendix}. For CQF, FC lowers the mean MAE for 7 of 11 models (Appendix Tables~\ref{tab:orion_cqf_summary} and~\ref{appendix:orion_cqf_main} to~\ref{tab:orion_cqf_per_tc_fc_appendix}).

\section{Conclusion}
\label{sec:conclusion}
We present TSNBench, the first benchmark for evaluating LLM proficiency in Time-Sensitive Networking (TSN), comprising 939 expert-validated multiple-choice questions (MCQs) and 100 open-ended questions per mechanism for Credit-Based Shaper (CBS) and Cyclic Queuing and Forwarding (CQF). The ground truth (GT) WCD values are computed using a verified Network Calculus (NC) solver for CBS and closed-form mathematical upper bounds for CQF. We evaluate 16 LLMs and find that models achieve 67-95\% MCQA accuracy yet fail substantially on open-ended WCD computation, with the best model (GPT-5) still deviating from the GT WCD by 150.2~$\mu$s on average on CBS. Despite CBS being extensively researched and an older mechanism, models cannot correctly apply NC, whereas CQF, with its simpler closed-form equation, is handled more successfully, suggesting that WCD computation performance is governed by mathematical complexity rather than mechanism maturity. Moreover, providing the WCD equations in the prompt lowers the CQF error for 9 of 10 models but the CBS error for only 2 of 9. TSNBench demonstrates that MCQ benchmarks substantially overestimate LLM capability in safety-critical domains.

\textbf{Limitations and Future Directions:}
TSNBench has five main limitations. First, the MCQs are generated from open-access research papers, which limits coverage of recently standardized mechanisms such as ATS (Appendix~\ref{app:mcq-coverage}, Table~\ref{appendix:mcq_coverage}). Second, the MCQs were generated with models from the same families that we evaluate, so some contamination cannot be ruled out. Our contamination analysis finds no advantage for the generator families, and the open-ended tasks are constructed specifically for TSNBench and are therefore unlikely to be affected. Third, MCQA accuracy is high for the strongest models, which limits its ability to distinguish between them. We view this saturation as further evidence for our main claim that MCQA overestimates practical proficiency, and the open-ended tasks are designed to expose this gap. Fourth, the open-ended evaluation covers only CBS and CQF. Extending to TAS is a natural next step, though its NP-hard gate control list (GCL) synthesis problem poses additional challenges beyond CBS and CQF. Fifth, the open-ended tasks use three synthetic topologies and one real-world topology and evaluate standalone model behavior with and without formulas in the prompt. These results should not be interpreted as a recommended deployment workflow for safety-critical TSN systems. Evaluating LLMs in settings where they produce checkable artifacts verified by deterministic analysis tools is an important direction for future versions of TSNBench. 

\begin{ack}
\paragraph{Funding in direct support of this work:} This work received no external funding or financial support.
\paragraph{Competing interests:} The authors declare no competing interests.
\paragraph{Disclaimer:} Silviu S. Craciunas is employed by NXP Semiconductors and Technical University of Denmark (DTU). The views expressed in this paper are those of the authors and do not necessarily reflect those of NXP Semiconductors.
\end{ack}

{
\small
\bibliographystyle{plainnat}
\bibliography{references}

@misc{ademaj2019industrial, 
    title={{Industrial Automation Traffic Types and their Mapping to QoS/TSN Mechanisms}}, 
    url={https://www.ieee802.org/1/files/public/docs2018/60802-ademaj-traffic-type-characterization-1118-v01.pdf}, 
    journal={60802-ademaj-traffic-type-characterization-1118-v01}, 
    year={2018},
    howpublished = {IEEE 802.1 contribution, IEC/IEEE 60802},
    author={Ademaj, Astrit and Enzinger, Thomas}
}

@article{PopIET16,
  author = {Paul Pop and {Lander Raagaard}, Michael and Silviu S. Craciunas and Wilfried Steiner},
  title = {{Design Optimization of Cyber-Physical Distributed Systems using IEEE Time-sensitive Networks (TSN)}},
  journal = {IET-CPS},
  volume = {1},
  number = {1},
  year = {2016},
  pages = {86-94},
  doi = {10.1049/iet-cps.2016.0021},
  publisher = {IET}
}

@ARTICLE{8548553,
  author={Gavriluţ, Voica and Zhao, Luxi and Raagaard, Michael L. and Pop, Paul},
  journal={IEEE Access}, 
  title={{AVB-Aware Routing and Scheduling of Time-Triggered Traffic for TSN}}, 
  year={2018},
  volume={6},
  number={},
  pages={75229-75243},
  doi={10.1109/ACCESS.2018.2883644}}

@inproceedings{bujosa2022hermes,
  title={{HERMES: Heuristic multi-queue scheduler for TSN time-triggered traffic with zero reception jitter capabilities}},
  author={Bujosa, Daniel and Ashjaei, Mohammad and Papadopoulos, Alessandro V and Nolte, Thomas and Proenza, Juli{\'a}n},
  booktitle={Proceedings of the 30th International Conference on Real-Time Networks and Systems},
  doi = {10.1145/3534879.3534906},
  publisher = {Association for Computing Machinery},
  pages = {70-80},
  year={2022}
}

@Article{electronics10202477,
AUTHOR = {Böhm, Martin and Wermser, Diederich},
TITLE = {{Multi-Domain Time-Sensitive Networks—Control Plane Mechanisms for Dynamic Inter-Domain Stream Configuration}},
JOURNAL = {Electronics},
VOLUME = {10},
YEAR = {2021},
NUMBER = {20},
ARTICLE-NUMBER = {2477},
ISSN = {2079-9292},
DOI = {10.3390/electronics10202477}
}

@ARTICLE{Bruckner,
  author={Bruckner, Dietmar and Stănică, Marius-Petru and Blair, Richard and Schriegel, Sebastian and Kehrer, Stephan and Seewald, Maik and Sauter, Thilo},
  journal={Proceedings of the IEEE}, 
  title={{An Introduction to OPC UA TSN for Industrial Communication Systems}}, 
  year={2019},
  volume={107},
  number={6},
  pages={1121-1131},
  doi={10.1109/JPROC.2018.2888703}}

@inproceedings{
mmlu_iclr,
title={{Measuring Massive Multitask Language Understanding}},
author={Dan Hendrycks and Collin Burns and Steven Basart and Andy Zou and Mantas Mazeika and Dawn Song and Jacob Steinhardt},
booktitle={International Conference on Learning Representations},
year={2021},
url={https://openreview.net/forum?id=d7KBjmI3GmQ}
}

@inproceedings{
mmlu_pro,
title={{MMLU-Pro: A More Robust and Challenging Multi-Task Language Understanding Benchmark}},
author={Yubo Wang and Xueguang Ma and Ge Zhang and Yuansheng Ni and Abhranil Chandra and Shiguang Guo and Weiming Ren and Aaran Arulraj and Xuan He and Ziyan Jiang and Tianle Li and Max Ku and Kai Wang and Alex Zhuang and Rongqi Fan and Xiang Yue and Wenhu Chen},
booktitle={The Thirty-Eighth Conference on Neural Information Processing Systems Datasets and Benchmarks Track},
year={2024},
url={https://openreview.net/forum?id=y10DM6R2r3}
}

@inproceedings{5g_neurips,
title={{The LLM as a Network Operator: A Vision for Generative AI in the 6G Radio Access Network}},
author={Giwa Oluwaseyi and Michael Adewole and Tobi Awodumila and Pelumi Aderinto},
booktitle={NeurIPS 2025 Workshop: AI and ML for Next-Generation Wireless Communications and Networking},
year={2025},
url={https://openreview.net/forum?id=81mgAfsFJv}
}

@inproceedings{researchcodebench_neurips,
title={{ResearchCodeBench: Benchmarking LLMs on Implementing Novel Machine Learning Research Code}},
author={Tianyu Hua and Harper Hua and Violet Xiang and Benjamin Klieger and Sang T. Truong and Weixin Liang and Fan-Yun Sun and Nick Haber},
booktitle={The Thirty-ninth Annual Conference on Neural Information Processing Systems Datasets and Benchmarks Track},
year={2025},
url={https://openreview.net/forum?id=3k70Vt0YFS}
}

@inproceedings{effibench_neurips,
 author = {Huang, Dong and Qing, Yuhao and Shang, Weiyi and Cui, Heming and Zhang, Jie M.},
 booktitle = {Advances in Neural Information Processing Systems},
 doi = {10.52202/079017-0367},
 editor = {A. Globerson and L. Mackey and D. Belgrave and A. Fan and U. Paquet and J. Tomczak and C. Zhang},
 pages = {11506--11544},
 publisher = {Curran Associates, Inc.},
 title = {{EffiBench: Benchmarking the Efficiency of Automatically Generated Code}},
 url = {https://proceedings.neurips.cc/paper_files/paper/2024/file/15807b6e09d691fe5e96cdecde6d7b80-Paper-Datasets_and_Benchmarks_Track.pdf},
 volume = {37},
 year = {2024}
}

@inproceedings{
tcmladder_medicine_neurips,
title={{TCM-Ladder: A Benchmark for Multimodal Question Answering on Traditional Chinese Medicine}},
author={Jiacheng Xie and Yang Yu and Ziyang Zhang and Shuai Zeng and Jiaxuan He and Ayush Vasireddy and Xiaoting Tang and Congyu Guo and Lening Zhao and Congcong Jing and Guanghui An and Dong Xu},
booktitle={The Thirty-ninth Annual Conference on Neural Information Processing Systems Datasets and Benchmarks Track},
year={2025},
url={https://openreview.net/forum?id=ZDrT1eG54T}
}

@article{humanities_exam_2026,
   title={{A benchmark of expert-level academic questions to assess AI capabilities}},
   volume={649},
   ISSN={1476-4687},
   url={http://dx.doi.org/10.1038/s41586-025-09962-4},
   DOI={10.1038/s41586-025-09962-4},
   number={8099},
   journal={Nature},
   publisher={Springer Science and Business Media LLC},
   author={Phan, Long and Gatti, Alice and Li, Nathaniel and Khoja, Adam and Kim, Ryan and Ren, Richard and Hausenloy, Jason and Zhang, Oliver and Mazeika, Mantas and Hendrycks, Dan and others},
   year={2026},
   month=jan, pages={1139–1146} }

@inproceedings{legalbench_neurips,
 author = {Guha, Neel and Nyarko, Julian and Ho, Daniel and R\'{e}, Christopher and Chilton, Adam and K, Aditya and Chohlas-Wood, Alex and Peters, Austin and Waldon, Brandon and Rockmore, Daniel and Zambrano, Diego and Talisman, Dmitry and Hoque, Enam and Surani, Faiz and Fagan, Frank and Sarfaty, Galit and Dickinson, Gregory and Porat, Haggai and Hegland, Jason and Wu, Jessica and Nudell, Joe and Niklaus, Joel and Nay, John and Choi, Jonathan and Tobia, Kevin and Hagan, Margaret and Ma, Megan and Livermore, Michael and Rasumov-Rahe, Nikon and Holzenberger, Nils and Kolt, Noam and Henderson, Peter and Rehaag, Sean and Goel, Sharad and Gao, Shang and Williams, Spencer and Gandhi, Sunny and Zur, Tom and Iyer, Varun and Li, Zehua},
 booktitle = {Advances in Neural Information Processing Systems},
 editor = {A. Oh and T. Naumann and A. Globerson and K. Saenko and M. Hardt and S. Levine},
 pages = {44123--44279},
 publisher = {Curran Associates, Inc.},
 title = {{LegalBench: A Collaboratively Built Benchmark for Measuring Legal Reasoning in Large Language Models}},
 url = {https://proceedings.neurips.cc/paper_files/paper/2023/file/89e44582fd28ddfea1ea4dcb0ebbf4b0-Paper-Datasets_and_Benchmarks.pdf},
 volume = {36},
 year = {2023}
}

@ARTICLE{6g,
  author={Saad, Walid and Bennis, Mehdi and Chen, Mingzhe},
  journal={IEEE Network}, 
  title={{A Vision of 6G Wireless Systems: Applications, Trends, Technologies, and Open Research Problems}}, 
  year={2020},
  volume={34},
  number={3},
  pages={134-142},
  doi={10.1109/MNET.001.1900287}}

@article{wifi,
author = {Ma, Yongsen and Zhou, Gang and Wang, Shuangquan},
title = {{WiFi Sensing with Channel State Information: A Survey}},
year = {2019},
issue_date = {May 2020},
publisher = {Association for Computing Machinery},
address = {New York, NY, USA},
volume = {52},
number = {3},
issn = {0360-0300},
url = {https://doi.org/10.1145/3310194},
doi = {10.1145/3310194},
journal = {ACM Comput. Surv.},
month = jun,
articleno = {46},
numpages = {36}
}

@ARTICLE{5g,
  author={Andrews, Jeffrey G. and Buzzi, Stefano and Choi, Wan and Hanly, Stephen V. and Lozano, Angel and Soong, Anthony C. K. and Zhang, Jianzhong Charlie},
  journal={IEEE Journal on Selected Areas in Communications}, 
  title={{What Will 5G Be?}}, 
  year={2014},
  volume={32},
  number={6},
  pages={1065-1082},
  doi={10.1109/JSAC.2014.2328098}}

@ARTICLE{Zhao21,
  author={Zhao, Luxi and Pop, Paul and Zheng, Zhong and Daigmorte, Hugo and Boyer, Marc},
  journal={IEEE Transactions on Industrial Electronics}, 
  title={{Latency Analysis of Multiple Classes of AVB Traffic in TSN With Standard Credit Behavior Using Network Calculus}}, 
  year={2021},
  volume={68},
  number={10},
  pages={10291-10302},
  doi={10.1109/TIE.2020.3021638}}

@inproceedings{EHRNoteQA_medicine_neurips,
 author = {Kweon, Sunjun and Kim, Jiyoun and Kwak, Heeyoung and Cha, Dongchul and Yoon, Hangyul and Kim, Kwanghyun and Yang, Jeewon and Won, Seunghyun and Choi, Edward},
 booktitle = {Advances in Neural Information Processing Systems},
 doi = {10.52202/079017-3958},
 editor = {A. Globerson and L. Mackey and D. Belgrave and A. Fan and U. Paquet and J. Tomczak and C. Zhang},
 pages = {124575--124611},
 publisher = {Curran Associates, Inc.},
 title = {{EHRNoteQA: An LLM Benchmark for Real-World Clinical Practice Using Discharge Summaries}},
 url = {https://proceedings.neurips.cc/paper_files/paper/2024/file/e15c4afff22f12c4986c1fcb4e941e03-Paper-Datasets_and_Benchmarks_Track.pdf},
 volume = {37},
 year = {2024}
}

@inproceedings{CMExam_chinese_medical_exam_neurips,
 author = {Liu, Junling and Zhou, Peilin and Hua, Yining and Chong, Dading and Tian, Zhongyu and Liu, Andrew and Wang, Helin and You, Chenyu and Guo, Zhenhua and Zhu, Lei and Li, Michael Lingzhi},
 booktitle = {Advances in Neural Information Processing Systems},
 editor = {A. Oh and T. Naumann and A. Globerson and K. Saenko and M. Hardt and S. Levine},
 pages = {52430--52452},
 publisher = {Curran Associates, Inc.},
 title = {{Benchmarking Large Language Models on CMExam - A comprehensive Chinese Medical Exam Dataset}},
 url = {https://proceedings.neurips.cc/paper_files/paper/2023/file/a48ad12d588c597f4725a8b84af647b5-Paper-Datasets_and_Benchmarks.pdf},
 volume = {36},
 year = {2023}
}

@inproceedings{mediq_neurips,
 author = {Li, Shuyue Stella and Balachandran, Vidhisha and Feng, Shangbin and Ilgen, Jonathan S. and Pierson, Emma and Koh, Pang Wei and Tsvetkov, Yulia},
 booktitle = {Advances in Neural Information Processing Systems},
 doi = {10.52202/079017-0908},
 editor = {A. Globerson and L. Mackey and D. Belgrave and A. Fan and U. Paquet and J. Tomczak and C. Zhang},
 pages = {28858--28888},
 publisher = {Curran Associates, Inc.},
 title = {{MediQ: Question-Asking LLMs and a Benchmark for Reliable Interactive Clinical Reasoning}},
 url = {https://proceedings.neurips.cc/paper_files/paper/2024/file/32b80425554e081204e5988ab1c97e9a-Paper-Conference.pdf},
 volume = {37},
 year = {2024}
}

@inproceedings{taskbench_neurips,
 author = {Shen, Yongliang and Song, Kaitao and Tan, Xu and Zhang, Wenqi and Ren, Kan and Yuan, Siyu and Lu, Weiming and Li, Dongsheng and Zhuang, Yueting},
 booktitle = {Advances in Neural Information Processing Systems},
 doi = {10.52202/079017-0148},
 editor = {A. Globerson and L. Mackey and D. Belgrave and A. Fan and U. Paquet and J. Tomczak and C. Zhang},
 pages = {4540--4574},
 publisher = {Curran Associates, Inc.},
 title = {{TaskBench: Benchmarking Large Language Models for Task Automation}},
 url = {https://proceedings.neurips.cc/paper_files/paper/2024/file/085185ea97db31ae6dcac7497616fd3e-Paper-Datasets_and_Benchmarks_Track.pdf},
 volume = {37},
 year = {2024}
}

@inproceedings{ai_research_agent_neurips,
title={{Benchmarking Large Language Models as AI Research Agents}},
author={Qian Huang and Jian Vora and Percy Liang and Jure Leskovec},
booktitle={NeurIPS 2023 Foundation Models for Decision Making Workshop},
year={2023},
url={https://openreview.net/forum?id=kXlTY0BmK3}
}

@inproceedings{software_engineering_neurips,
title={{Predicting Emergent Software Engineering Capabilities by Fine-tuning}},
author={Jason J Jackson and Terry Huang and Henry Velasquez and Kevin Zhu and Sunishchal Dev},
booktitle={NeurIPS 2025 Workshop on Evaluating the Evolving LLM Lifecycle: Benchmarks, Emergent Abilities, and Scaling},
year={2025},
url={https://openreview.net/forum?id=EwchHtwavV}
}

@inproceedings{agi_eng_design_neurips,
title={{Toward Engineering AGI: Benchmarking the Engineering Design Capabilities of LLMs}},
author={Xingang Guo and Yaxin Li and XiangYi Kong and Yilan Jiang and Xiayu Zhao and Zhihua Gong and Yufan Zhang and Daixuan Li and Tianle Sang and Beixiao Zhu and Gregory Jun and Yingbing Huang and Yiqi Liu and Yuqi Xue and Rahul Dev Kundu and Qi Jian Lim and Yizhou Zhao and Luke Alexander Granger and Mohamed Badr Younis and Darioush Keivan and Nippun Sabharwal and Shreyanka Sinha and Prakhar Agarwal and Kojo Vandyck and Hanlin Mai and Zichen Wang and Aditya Venkatesh and Ayush Barik and Jiankun Yang and Chongying Yue and Jingjie He and Libin Wang and Licheng Xu and Hao Chen and Jinwen Wang and Liujun Xu and Rushabh Shetty and Ziheng Guo and Dahui Song and Manvi Jha and Weijie Liang and Weiman Yan and Bryan Zhang and Sahil Bhandary Karnoor and Jialiang Zhang and Rutva Pandya and Xinyi Gong and Mithesh Ballae Ganesh and Feize Shi and Ruiling Xu and Yifan Zhang and Yanfeng Ouyang and Lianhui Qin and Elyse Rosenbaum and Corey Snyder and Peter Seiler and Geir Dullerud and Xiaojia Shelly Zhang and Zuofu Cheng and Pavan Kumar Hanumolu and Jian Huang and Mayank Kulkarni and Mahdi Namazifar and Huan Zhang and Bin Hu},
booktitle={The Thirty-ninth Annual Conference on Neural Information Processing Systems Datasets and Benchmarks Track},
year={2025},
url={https://openreview.net/forum?id=Wmsnx7EPel}
}

@ARTICLE{6g_bench,
  author={Ferrag, Mohamed Amine and Lakas, Abderrahmane and Debbah, Mérouane},
  journal={IEEE Open Journal of the Communications Society}, 
  title={{6G-Bench: An Open Benchmark for Semantic Communication and Network-Level Reasoning With Foundation Models in AI-Native 6G Networks}}, 
  year={2026},
  volume={7},
  number={},
  pages={3305-3330},
  doi={10.1109/OJCOMS.2026.3680457}}

@ARTICLE{teleqna,
  author={Maatouk, Ali and Ayed, Fadhel and Piovesan, Nicola and Domenico, Antonio De and Debbah, Merouane and Luo, Zhi-Quan},
  journal={IEEE Network}, 
  title={{TeleQnA: A Benchmark Dataset to Assess Large Language Models Telecommunications Knowledge}}, 
  year={2026},
  volume={40},
  number={2},
  pages={253-260},
  doi={10.1109/MNET.2025.3576035}}

@inproceedings{tele_llm_poster_neurips_multi_agent,
title={{Tele-LLM-Hub: Building Context-Aware Multi-Agent LLM Systems for Telecom Networks}},
author={Pranshav Gajjar and Cong Shen and Vijay K Shah},
booktitle={NeurIPS 2025 Workshop: AI and ML for Next-Generation Wireless Communications and Networking},
year={2025},
url={https://openreview.net/forum?id=AencYkmJtl}
}

@inproceedings{prosper_extract_protocol_spec,
author = {Sharma, Prakhar and Yegneswaran, Vinod},
title = {{PROSPER: Extracting Protocol Specifications Using Large Language Models}},
year = {2023},
isbn = {9798400704154},
publisher = {Association for Computing Machinery},
address = {New York, NY, USA},
url = {https://doi.org/10.1145/3626111.3628205},
doi = {10.1145/3626111.3628205},
booktitle = {Proceedings of the 22nd ACM Workshop on Hot Topics in Networks},
pages = {41-47},
numpages = {7},
location = {Cambridge, MA, USA},
series = {HotNets '23}
}

@inproceedings{scieval_benchmark_scientific_research,
author = {Sun, Liangtai and Han, Yang and Zhao, Zihan and Ma, Da and Shen, Zhennan and Chen, Baocai and Chen, Lu and Yu, Kai},
title = {{SciEval: a multi-level large language model evaluation benchmark for scientific research}},
year = {2024},
isbn = {978-1-57735-887-9},
publisher = {AAAI Press},
url = {https://doi.org/10.1609/aaai.v38i17.29872},
doi = {10.1609/aaai.v38i17.29872},
booktitle = {Proceedings of the Thirty-Eighth AAAI Conference on Artificial Intelligence and Thirty-Sixth Conference on Innovative Applications of Artificial Intelligence and Fourteenth Symposium on Educational Advances in Artificial Intelligence},
articleno = {2124},
numpages = {9},
series = {AAAI'24/IAAI'24/EAAI'24}
}

@ARTICLE{8021Q,
  author  = {{IEEE}},
  journal={IEEE Std 802.1Q-2018 (Revision of IEEE Std 802.1Q-2014)}, 
  title={{IEEE Standard for Local and Metropolitan Area Network--Bridges and Bridged Networks}}, 
  year={2018},
  volume={},
  number={},
  pages={1-1993},
  doi={10.1109/IEEESTD.2018.8403927}}

@inproceedings{prompt_as_program,
author = {Reynolds, Laria and McDonell, Kyle},
title = {{Prompt Programming for Large Language Models: Beyond the Few-Shot Paradigm}},
year = {2021},
isbn = {9781450380959},
publisher = {Association for Computing Machinery},
address = {New York, NY, USA},
url = {https://doi.org/10.1145/3411763.3451760},
doi = {10.1145/3411763.3451760},
booktitle = {Extended Abstracts of the 2021 CHI Conference on Human Factors in Computing Systems},
articleno = {314},
numpages = {7},
series = {CHI EA '21}
}

@article{structeval,
title={{StructEval: Benchmarking LLMs' Capabilities to Generate Structural Outputs}},
author={Jialin Yang and Dongfu Jiang and Tony He and Sherman Siu and Yuxuan Zhang and Disen Liao and Zhuofeng Li and Huaye Zeng and Yiming Jia and Haozhe Wang and Benjamin Schneider and Chi Ruan and Wentao Ma and Zhiheng Lyu and Yifei Wang and Yi Lu and Quy Duc Do and Ziyan Jiang and Ping Nie and Wenhu Chen},
journal={Transactions on Machine Learning Research},
issn={2835-8856},
year={2026},
url={https://openreview.net/forum?id=buDwV7LUA7},
note={J2C Certification}
}

@ARTICLE{intro_tsn,
  author={Finn, Norman},
  journal={IEEE Communications Standards Magazine}, 
  title={{Introduction to Time-Sensitive Networking}}, 
  year={2018},
  volume={2},
  number={2},
  pages={22-28},
  doi={10.1109/MCOMSTD.2018.1700076}}

@inproceedings{silviu_tas,
author = {Craciunas, Silviu S. and Oliver, Ramon Serna and Chmel\'{\i}k, Martin and Steiner, Wilfried},
title = {{Scheduling Real-Time Communication in IEEE 802.1Qbv Time Sensitive Networks}},
year = {2016},
isbn = {9781450347877},
publisher = {Association for Computing Machinery},
address = {New York, NY, USA},
url = {https://doi.org/10.1145/2997465.2997470},
doi = {10.1145/2997465.2997470},
booktitle = {Proceedings of the 24th International Conference on Real-Time Networks and Systems},
pages = {183-192},
numpages = {10},
location = {Brest, France},
series = {RTNS '16}
}

@INPROCEEDINGS{silviu_gcl_array,
  author={Serna Oliver, Ramon and Craciunas, Silviu S. and Steiner, Wilfried},
  booktitle={2018 IEEE Real-Time and Embedded Technology and Applications Symposium (RTAS)}, 
  title={{IEEE 802.1Qbv Gate Control List Synthesis Using Array Theory Encoding}}, 
  year={2018},
  volume={},
  number={},
  pages={13-24},
  doi={10.1109/RTAS.2018.00008}}

@INPROCEEDINGS{ubs,
  author={Specht, Johannes and Samii, Soheil},
  booktitle={2016 28th Euromicro Conference on Real-Time Systems (ECRTS)}, 
  title={{Urgency-Based Scheduler for Time-Sensitive Switched Ethernet Networks}}, 
  year={2016},
  volume={},
  number={},
  pages={75-85},
  doi={10.1109/ECRTS.2016.27}}

@ARTICLE{ats_tas_ahmed,
  author={Nasrallah, Ahmed and Thyagaturu, Akhilesh S. and Alharbi, Ziyad and Wang, Cuixiang and Shao, Xing and Reisslein, Martin and Elbakoury, Hesham},
  journal={IEEE Access}, 
  title={{Performance Comparison of IEEE 802.1 TSN Time Aware Shaper (TAS) and Asynchronous Traffic Shaper (ATS)}}, 
  year={2019},
  volume={7},
  number={},
  pages={44165-44181},
  doi={10.1109/ACCESS.2019.2908613}}

@ARTICLE{jrsp_cqf,
  author={Wang, Xiaolong and Yao, Haipeng and Mai, Tianle and Xiong, Zehui and Wang, Fu and Liu, Yunjie},
  journal={IEEE Transactions on Vehicular Technology}, 
  title={{Joint Routing and Scheduling With Cyclic Queuing and Forwarding for Time-Sensitive Networks}}, 
  year={2023},
  volume={72},
  number={3},
  pages={3793-3804},
  doi={10.1109/TVT.2022.3216958}}

@ARTICLE{rubi_iotj,
  author={Debnath, Rubi and Barzegaran, Mohammadreza and Steinhorst, Sebastian},
  journal={IEEE Internet of Things Journal}, 
  title={{Toward an Optimized Multi-Cyclic Queuing and Forwarding in Time-Sensitive Networking With Time Injection}}, 
  year={2025},
  volume={12},
  number={20},
  pages={43034-43051},
  doi={10.1109/JIOT.2025.3597560}}

@INPROCEEDINGS{itp_cqf,
  author={Yan, Jinli and Quan, Wei and Jiang, Xuyan and Sun, Zhigang},
  booktitle={IEEE INFOCOM 2020 - IEEE Conference on Computer Communications}, 
  title={{Injection Time Planning: Making CQF Practical in Time-Sensitive Networking}}, 
  year={2020},
  volume={},
  number={},
  pages={616-625},
  doi={10.1109/INFOCOM41043.2020.9155434}}

@INPROCEEDINGS{luxi_avb,
  author={Zhao, Luxi and Pop, Paul and Zheng, Zhong and Li, Qiao},
  booktitle={2018 IEEE Real-Time and Embedded Technology and Applications Symposium (RTAS)}, 
  title={{Timing Analysis of AVB Traffic in TSN Networks Using Network Calculus}}, 
  year={2018},
  volume={},
  number={},
  pages={25-36},
  doi={10.1109/RTAS.2018.00009}}

@article{netconfeval,
author = {Wang, Changjie and Scazzariello, Mariano and Farshin, Alireza and Ferlin, Simone and Kosti\'{c}, Dejan and Chiesa, Marco},
title = {{NetConfEval: Can LLMs Facilitate Network Configuration?}},
year = {2024},
issue_date = {June 2024},
publisher = {Association for Computing Machinery},
address = {New York, NY, USA},
volume = {2},
number = {CoNEXT2},
url = {https://doi.org/10.1145/3656296},
doi = {10.1145/3656296},
journal = {Proc. ACM Netw.},
month = jun,
articleno = {7},
numpages = {25}
}

@ARTICLE{spacecraft_tsn,
  author={Fiori, Tiziana and Lavacca, Francesco Giacinto and Valente, Francesco and Eramo, Vincenzo},
  journal={IEEE Access}, 
  title={{Proposal and Investigation of a Lite Time Sensitive Networking Solution for the Support of Real Time Services in Space Launcher Networks}}, 
  year={2024},
  volume={12},
  number={},
  pages={10664-10680},
  doi={10.1109/ACCESS.2024.3353466}}

@ARTICLE{microlaunchers_tsn,
  author={Sanchez-Garrido, Jorge and Aparicio, Beatriz and Ramírez, José Gabriel and Rodriguez, Rafael and Melara, Mariasole and Cercós, Lorenzo and Ros, Eduardo and Diaz, Javier},
  journal={IEEE Transactions on Aerospace and Electronic Systems}, 
  title={{Implementation of a Time-Sensitive Networking (TSN) Ethernet Bus for Microlaunchers}}, 
  year={2021},
  volume={57},
  number={5},
  pages={2743-2758},
  doi={10.1109/TAES.2021.3061806}}

@inproceedings{us_army_military,
  author    = {Elliott, Leonard},
  title     = {{Time-Sensitive Networking (TSN) in Military Ground Vehicle Architectures}},
  booktitle = {2024 NDIA Michigan Chapter Ground Vehicle Systems Engineering and Technology Symposium},
  address   = {Novi, Michigan, United States},
  publisher = {National Defense Industrial Association Michigan Chapter},
  month     = aug,
  year      = {2024},
  doi       = {10.4271/2024-01-4122}
}

@article{tsn_for_industrial,
author = {Zhang, Tianyu and Wang, Gang and Xue, Chuanyu and Wang, Jiachen and Nixon, Mark and Han, Song},
title = {{Time-Sensitive Networking (TSN) for Industrial Automation: Current Advances and Future Directions}},
year = {2024},
issue_date = {February 2025},
publisher = {Association for Computing Machinery},
address = {New York, NY, USA},
volume = {57},
number = {2},
issn = {0360-0300},
url = {https://doi.org/10.1145/3695248},
doi = {10.1145/3695248},
journal = {ACM Comput. Surv.},
month = oct,
articleno = {30},
numpages = {38}
}

@phdthesis{mateu2024improved,
  title  = {{Improved Configuration and Analysis Solutions for Time-Sensitive Networks with Support for Legacy Systems}},
  author = {{Bujosa Mateu}, Daniel},
  school = {M{\"a}lardalen University},
  address = {V{\"a}ster{\aa}s, Sweden},
  year   = {2024}
}

@ARTICLE{luxi_tii,
  author={Zhao, Luxi and Yan, Yida and Zhou, Xuan},
  journal={IEEE Transactions on Industrial Informatics}, 
  title={{Minimum Bandwidth Reservation for CBS in TSN With Real-Time QoS Guarantees}}, 
  year={2024},
  volume={20},
  number={4},
  pages={6187-6198},
  doi={10.1109/TII.2023.3342466}}

@INPROCEEDINGS{rubi_icc,
  author={Debnath, Rubi and Zhao, Luxi and Steinhorst, Sebastian},
  booktitle={ICC 2025 - IEEE International Conference on Communications}, 
  title={{Learning-Based Traffic Classification for Mixed-Critical Flows in Time-Sensitive Networking}}, 
  year={2025},
  volume={},
  number={},
  pages={5926-5932},
  doi={10.1109/ICC52391.2025.11161468}}

@INPROCEEDINGS{rubi_rtcsa,
  author={Debnath, Rubi and Hortig, Philipp and Zhao, Luxi and Steinhorst, Sebastian},
  booktitle={2023 IEEE 29th International Conference on Embedded and Real-Time Computing Systems and Applications (RTCSA)}, 
  title={{Advanced Modeling and Analysis of Individual and Combined TSN Shapers in OMNeT++}}, 
  year={2023},
  volume={},
  number={},
  pages={176-185},
  doi={10.1109/RTCSA58653.2023.00029}}

@article{voica_tta,
author = {Gavrilu\c{t}, Voica and Pop, Paul},
title = {{Traffic-type Assignment for TSN-based Mixed-criticality Cyber-physical Systems}},
year = {2020},
journal   = {ACM Transactions on Cyber-Physical Systems},
issue_date = {April 2020},
publisher = {Association for Computing Machinery},
address = {New York, NY, USA},
volume = {4},
number = {2},
issn = {2378-962X},
doi = {10.1145/3371708},
month = jan,
articleno = {23},
numpages = {27}
}

@ARTICLE{luxi_tnsm,
  author={Zhao, Luxi and Pop, Paul and Steinhorst, Sebastian},
  journal={IEEE Transactions on Network and Service Management}, 
  title={{Quantitative Performance Comparison of Various Traffic Shapers in Time-Sensitive Networking}}, 
  year={2022},
  volume={19},
  number={3},
  pages={2899-2928},
  doi={10.1109/TNSM.2022.3180160}}

@inproceedings{ece,
title={{Understanding Model Calibration - A gentle introduction and visual exploration of calibration and the expected calibration error ({ECE})}},
author={Maja Pavlovic},
booktitle={The Fourth Blogpost Track at ICLR 2025},
year={2025},
url={https://openreview.net/forum?id=BxBeCjQd2y}
}

@article{brier,
  title   = {{On misconceptions about the Brier score in binary prediction models}},
  author  = {Hoessly, Linard},
  journal = {Glob. Epidemiol.},
  volume  = {11},
  pages   = {100242},
  month   = jun,
  year    = {2026},
  publisher = {Elsevier}
}

@ARTICLE{8021qav,
  author  = {{IEEE}},
  journal={IEEE Std 802.1Qav-2009 (Amendment to IEEE Std 802.1Q-2005)}, 
  title={{IEEE Standard for Local and Metropolitan Area Networks - Virtual Bridged Local Area Networks Amendment 12: Forwarding and Queuing Enhancements for Time-Sensitive Streams}}, 
  year={2010},
  volume={},
  number={},
  pages={C1-72},
  doi={10.1109/IEEESTD.2009.5375704}}

@ARTICLE{8021Qch,
  author  = {{IEEE}},
  journal={IEEE 802.1Qch-2017 (Amendment to IEEE Std 802.1Q-2014 as amended by IEEE Std 802.1Qca-2015, IEEE Std 802.1Qcd(TM)-2015, IEEE Std 802.1Q-2014/Cor 1-2015, IEEE Std 802.1Qbv-2015, IEEE Std 802.1Qbu-2016, IEEE Std 802.1Qbz-2016, and IEEE Std 802.1Qci-2017)}, 
  title={{IEEE Standard for Local and metropolitan area networks--Bridges and Bridged Networks--Amendment 29: Cyclic Queuing and Forwarding}}, 
  year={2017},
  volume={},
  number={},
  pages={1-30},
  doi={10.1109/IEEESTD.2017.7961303}}

@INPROCEEDINGS{rubi_ccnc,
  author={Debnath, Rubi and Zhao, Luxi and Barzegaran, Mohammadreza and Steinhorst, Sebastian},
  booktitle={2025 IEEE 22nd Consumer Communications \& Networking Conference (CCNC)}, 
  title={{CyclicSim: Comprehensive Evaluation of Cyclic Shapers in Time-Sensitive Networking}}, 
  year={2025},
  volume={},
  number={},
  pages={01-09},
  doi={10.1109/CCNC54725.2025.10975975}}

@ARTICLE{survey_in_vehicle_tsn,
  author={Peng, Yifei and Shi, Boxin and Jiang, Tigang and Tu, Xiaodong and Xu, Du and Hua, Kun},
  journal={IEEE Internet of Things Journal}, 
  title={{A Survey on In-Vehicle Time-Sensitive Networking}}, 
  year={2023},
  volume={10},
  number={16},
  pages={14375-14396},
  doi={10.1109/JIOT.2023.3264909}}

@ARTICLE{wireless_tsn_survey,
  author={Zanbouri, Kouros and Noor-A-Rahim, Md. and John, Jobish and Sreenan, Cormac J. and Vincent Poor, H. and Pesch, Dirk},
  journal={IEEE Communications Surveys \& Tutorials}, 
  title={{A Comprehensive Survey of Wireless Time-Sensitive Networking (TSN): Architecture, Technologies, Applications, and Open Issues}}, 
  year={2025},
  volume={27},
  number={4},
  pages={2129-2155},
  doi={10.1109/COMST.2024.3486618}}

@ARTICLE{5g_tsn_survey,
  author={Adil, Muhammad and Qiu, Tie and Zhou, Xiaobo and Javeed, Danish and Cao, Zhenrui and Oliver Wu, Dapeng},
  journal={IEEE Communications Surveys \& Tutorials}, 
  title={{Integrated 5G and Time Sensitive Networking for Emerging Applications: A Survey of Advancements, Challenges, and Future Directions}}, 
  year={2026},
  volume={28},
  number={},
  pages={4016-4050},
  doi={10.1109/COMST.2025.3632286}}

@ARTICLE{paul_mcqf,
  author={Alexandris, Konstantinos and Pop, Paul and Wang, Tongtong},
  journal={IEEE Access}, 
  title={{Configuration and Evaluation of Multi-CQF Shapers in IEEE 802.1 Time-Sensitive Networking (TSN)}}, 
  year={2022},
  volume={10},
  number={},
  pages={109068-109081},
  doi={10.1109/ACCESS.2022.3214007}}

@INPROCEEDINGS{industrial_traffic,
  author={Ekrad, Kasra and Vadillo, Inés Alvarez and Johansson, Bjarne and Mubeen, Saad and Ashjaei, Mohammad},
  booktitle={2025 IEEE 30th International Conference on Emerging Technologies and Factory Automation (ETFA)}, 
  title={{A Methodology to Map Industrial Automation Traffic to TSN Traffic Classes}}, 
  year={2025},
  volume={},
  number={},
  pages={1-8},
  doi={10.1109/ETFA65518.2025.11205571}}

@INPROCEEDINGS{netpilot,
  author={Windmann, Stefan and Albrecht, Janis and Friesen, Maxim and Jasperneite, Jürgen},
  booktitle={2025 IEEE 30th International Conference on Emerging Technologies and Factory Automation (ETFA)}, 
  title={{NetPilot - Towards LLM-Assisted Configuration of Hybrid TSN/5G Networks}}, 
  year={2025},
  volume={},
  number={},
  pages={1-4},
  doi={10.1109/ETFA65518.2025.11205555}}

@INPROCEEDINGS{rubi_vtc,
  author={Debnath, Rubi and Akinci, Mustafa Selman and Ajith, Devika and Steinhorst, Sebastian},
  booktitle={2023 IEEE 98th Vehicular Technology Conference (VTC2023-Fall)}, 
  title={{5GTQ: QoS-Aware 5G-TSN Simulation Framework}}, 
  year={2023},
  volume={},
  number={},
  pages={1-7},
  doi={10.1109/VTC2023-Fall60731.2023.10333533}}

@INPROCEEDINGS{rubi_noms,
  author={Debnath, Rubi and Hortig, Philipp and Zhao, Luxi and Steinhorst, Sebastian},
  booktitle={NOMS 2024-2024 IEEE Network Operations and Management Symposium}, 
  title={{Quantifying the Impact of Frame Preemption on Combined TSN Shapers}}, 
  year={2024},
  volume={},
  number={},
  pages={1-9},
  doi={10.1109/NOMS59830.2024.10575564}}
}


\newpage
\section*{NeurIPS Paper Checklist}

\begin{enumerate}
\item {\bf Claims}
    \item[] Question: Do the main claims made in the abstract and introduction accurately reflect the paper's contributions and scope?
    \item[] Answer: \answerYes{} 
    \item[] Justification: We present TSNBench, the first benchmark for evaluating LLM proficiency in safety-critical Time-Sensitive Networking. TSNBench covers TSN knowledge through expert-validated MCQA and computational reasoning through open-ended WCD computation for CBS and CQF. Ground-truth values are computed using a verified state-of-the-art Network Calculus solver for CBS and closed-form mathematical upper bounds for CQF. Dataset construction is detailed in Section~\ref{sub:overview_tsnbench}, and evaluation results are presented in Section~\ref{sec:experiments}. A detailed description of TSNBench is also given in Appendix~\ref{appendix:more_on_tsnbench}.

\item {\bf Limitations}
    \item[] Question: Does the paper discuss the limitations of the work performed by the authors?
    \item[] Answer: \answerYes{} 
    \item[] Justification: We discuss the limitations of TSNBench in detail in Section~\ref{sec:conclusion}. A more detailed explanation is provided in Appendix~\ref{appendix:limitations}. 

\item {\bf Theory assumptions and proofs}
    \item[] Question: For each theoretical result, does the paper provide the full set of assumptions and a complete (and correct) proof?
    \item[] Answer: \answerYes{} 
    \item[] Justification: The Network Calculus background and methodology underlying the CBS ground truth are provided in Appendix~\ref{appendix:network_calculus}. The closed-form WCD equation and its proof for CQF are given in Section~\ref{sec:ground_truth} and Appendix~\ref{appendix:cqf}, respectively. All network constants, parameters, and assumptions used in the open-ended evaluation are detailed in Appendix~\ref{appendix:prompt_open_end}.
    \item {\bf Experimental result reproducibility}
    \item[] Question: Does the paper fully disclose all the information needed to reproduce the main experimental results of the paper to the extent that it affects the main claims and/or conclusions of the paper (regardless of whether the code and data are provided or not)?
    \item[] Answer: \answerYes{} 
    \item[] Justification: Full details of dataset generation, evaluation framework, and results are provided in Section~\ref{sub:overview_tsnbench}, Section~\ref{sec:experiments}, and Appendix~\ref{appendix:more_on_tsnbench}. The TSNBench dataset, comprising MCQs and open-ended WCD test cases with GT values, is publicly released on Kaggle under the CC-BY 4.0 license at~\url{https://doi.org/10.34740/kaggle/ds/13019279}. The code, prompts, and evaluation scripts are open-sourced on GitHub under the MIT license at~\url{https://github.com/tum-esi/TSNBench}.

\item {\bf Open access to data and code}
    \item[] Question: Does the paper provide open access to the data and code, with sufficient instructions to faithfully reproduce the main experimental results, as described in supplemental material?
    \item[] Answer: \answerYes{} 
    \item[] Justification: The TSNBench dataset is publicly available on Kaggle at~\url{https://doi.org/10.34740/kaggle/ds/13019279} under the CC-BY 4.0 license, with Croissant metadata. It contains the complete MCQA dataset with answer keys and all open-ended test cases with network topologies, flow parameters, and routing configurations. Ground-truth WCD values are provided for all test cases, computed via Network Calculus for CBS and via closed-form mathematical upper bounds for CQF, enabling direct result verification. The full TSNBench codebase is available at~\url{https://github.com/tum-esi/TSNBench} under the MIT license, including a README with step-by-step environment setup and reproduction instructions. 

\item {\bf Experimental setting/details}
    \item[] Question: Does the paper specify all the training and test details (e.g., data splits, hyperparameters, how they were chosen, type of optimizer) necessary to understand the results?
    \item[] Answer: \answerYes{} 
    \item[] Justification: All experimental details including research document selection, MCQA generation models, expert review criteria, evaluation models, temperature configurations, and prompt design are provided in Section~\ref{sub:overview_tsnbench} and Section~\ref{sec:experiments}. Dataset construction is summarized in Table~\ref{tab:mcqa_statistics} and illustrated in Figure~\ref{fig:dataset_generator} for MCQA and Figure~\ref{fig:open_end_question} for open-ended questions. A more detailed explanation is given in Appendices~\ref{appendix:more_on_tsnbench} and \ref{appendix:sub_human_review}.

\item {\bf Experiment statistical significance}
    \item[] Question: Does the paper report error bars suitably and correctly defined or other appropriate information about the statistical significance of the experiments?
    \item[] Answer: \answerYes{} 
    \item[] Justification: For the open-ended questions, Tables~\ref{tab:open_end_std_dev}, \ref{appendix:cbs_mae_std_median}, and \ref{appendix:cqf_mae_std_median} report the MAE of each model as mean $\pm$ standard deviation, and Tables~\ref{appendix:cbs_mae_std_median} and~\ref{appendix:cqf_mae_std_median} also give the median MAE. The standard deviation is the population standard deviation of the per-TC MAE across the valid TCs ($n$) of a model. The prediction of each flow is averaged over up to three valid runs at one temperature per model before the MAE is computed, so the standard deviation captures the variation across TCs, not across runs. For MCQA, accuracy is averaged over three runs, and the consistency of answers across runs is reported in Table~\ref{tab:more_tsnbench_evaluation_appendix_deafult_temp_temp_0}. We do not run formal significance tests. More results are given in  Appendix~\ref{appendix:more_on_evaluation}.

\item {\bf Experiments compute resources}
    \item[] Question: For each experiment, does the paper provide sufficient information on the computer resources (type of compute workers, memory, time of execution) needed to reproduce the experiments?
    \item[] Answer: \answerYes{} 
    \item[] Justification: All experiments are conducted via vendor APIs on a standard workstation with no specialized compute required (details are given in Section~\ref{sec:experiments}). Per-model API latency and cost for both MCQA and the closed-book open-ended evaluations are reported in Table~\ref{tab:more_tsnbench_evaluation_cost_and_latency} in the Appendix. 
    
\item {\bf Code of ethics}
    \item[] Question: Does the research conducted in the paper conform, in every respect, with the NeurIPS Code of Ethics \url{https://neurips.cc/public/EthicsGuidelines}?
    \item[] Answer: \answerYes{} 
    \item[] Justification: The research conforms to the NeurIPS Code of Ethics in every respect. 

\item {\bf Broader impacts}
    \item[] Question: Does the paper discuss both potential positive societal impacts and negative societal impacts of the work performed?
    \item[] Answer: \answerYes{} 
    \item[] Justification: We provide the broader positive and negative impact of TSNBench in Appendix~\ref{appendix:broader_impact} and \ref{appendix:negative_impact} respectively. 
    
\item {\bf Safeguards}
    \item[] Question: Does the paper describe safeguards that have been put in place for responsible release of data or models that have a high risk for misuse (e.g., pre-trained language models, image generators, or scraped datasets)?
    \item[] Answer: \answerNA{} 
    \item[] Justification: TSNBench does not pose any such risk. 

\item {\bf Licenses for existing assets}
    \item[] Question: Are the creators or original owners of assets (e.g., code, data, models), used in the paper, properly credited and are the license and terms of use explicitly mentioned and properly respected?
    \item[] Answer: \answerYes{} 
    \item[] Justification: The TSNBench code and dataset are owned and released by the authors under the MIT and CC-BY 4.0 licenses, respectively. All research documents used for dataset construction are open-access publications, arXiv versions of published or accepted papers, or author manuscript versions. The authors of all directly used manuscript versions are properly credited, and full acknowledgments are disclosed in the paper. 

\item {\bf New assets}
    \item[] Question: Are new assets introduced in the paper well documented and is the documentation provided alongside the assets?
    \item[] Answer: \answerYes{} 
    \item[] Justification: TSNBench introduces two new assets, the 939-question expert-validated MCQA dataset and 100 open-ended WCD computation questions per mechanism on synthetic topologies plus 10 open-ended questions per mechanism on the real-world Orion CEV topology, all with verified ground-truth WCD values. Both are publicly released on Kaggle at~\url{https://doi.org/10.34740/kaggle/ds/13019279} under the CC-BY 4.0 license. The code and evaluation scripts are released on GitHub at~\url{https://github.com/tum-esi/TSNBench} under the MIT license, with a README that provides complete instructions for evaluation setup and result reproduction. 

\item {\bf Crowdsourcing and research with human subjects}
    \item[] Question: For crowdsourcing experiments and research with human subjects, does the paper include the full text of instructions given to participants and screenshots, if applicable, as well as details about compensation (if any)? 
    \item[] Answer: \answerNA{} 
    \item[] Justification: In this paper, we do not have any involvement with crowdsourcing or research with human subjects. 

\item {\bf Institutional review board (IRB) approvals or equivalent for research with human subjects}
    \item[] Question: Does the paper describe potential risks incurred by study participants, whether such risks were disclosed to the subjects, and whether Institutional Review Board (IRB) approvals (or an equivalent approval/review based on the requirements of your country or institution) were obtained?
    \item[] Answer: \answerNA{} 
    \item[] Justification: NA. 

\item {\bf Declaration of LLM usage}
    \item[] Question: Does the paper describe the usage of LLMs if it is an important, original, or non-standard component of the core methods in this research? Note that if the LLM is used only for writing, editing, or formatting purposes and does \emph{not} impact the core methodology, scientific rigor, or originality of the research, declaration is not required.
    \item[] Answer: \answerYes{} 
    \item[] Justification: LLMs are an integral part of the TSNBench framework, used for keyword extraction and MCQA generation as described in Section~\ref{sub:overview_tsnbench}. All extracted keywords and generated questions are reviewed by five domain experts with extensive TSN research experience, ensuring that LLM-generated content does not compromise the scientific rigor or correctness of the benchmark.

\end{enumerate}

\newpage
\appendix
\section{Limitations and Broader Impact}
\subsection{Limitations}
\label{appendix:limitations}
While TSNBench fills a significant research gap and proposes a step forward towards evaluating TSN capabilities in LLMs, it has several limitations:

\textbf{Dataset scope:} TSNBench currently only covers CBS and CQF in open-ended questions. Evaluating other TSN mechanisms is necessary to fully cover the entire TSN mechanisms.

\textbf{MCQA Scope:} MCQs are solely developed using published research papers and the IEEE standards are not used to generate the MCQs. Solving the license issue and utilizing standards to include MCQs using IEEE 802.1 standard will enhance the entire MCQA dataset.

\textbf{Topology coverage:} TSNBench open-ended questions currently cover four different topologies: one-switch, medium-mesh, ring topology, and Orion CEV topology. Covering diverse topologies and flow parameters will present a comprehensive evaluation.

\subsection{Improvement Strategies}
To address the limitations of TSNBench, we propose the following additions and improvements in future versions of TSNBench.
\begin{enumerate}
    \item \textbf{Larger and more diverse dataset:} Our current TSNBench dataset covers 100 TCs per mechanism across three synthetic topology types, and 10 TCs on the real-world Orion CEV topology. In future versions, we will include larger and more complex topologies with higher flow counts. As model performance improves, more complex open-ended evaluations should be integrated with complex topologies and combined TSN mechanisms.
    \item \textbf{Additional scheduling mechanisms:} TSNBench currently evaluates CBS and CQF. Future versions should extend to TAS and ATS to cover a broader range of the TSN standard suite.
    \item \textbf{Updated MCQA:} Our MCQA dataset was developed using published or accepted research documents that are openly available. In future work, we will update the dataset with MCQAs formulated directly from TSN standards.
    \item \textbf{Fine-tuned and domain-adapted models.} TSNBench currently evaluates general-purpose LLMs without any TSN-specific fine-tuning. Future versions should benchmark domain-adapted models trained on TSN standards and network calculus literature.
\end{enumerate}

\subsection{Broader Impact}
\label{appendix:broader_impact}
TSNBench enables the real-time systems community and the machine learning community to objectively measure LLM performance and readiness for management and deployment assistance in safety-critical deterministic networks. By highlighting the critical aspects of TSN and the performance gap of the models between MCQA and computational reasoning, TSNBench exposes weaknesses of current models which may lead to misconfigurations and safety-critical issues. This benchmark provides a concrete direction to improve LLMs for deterministic networking. TSNBench further highlights the potential benefits of using LLMs to automate the management and deployment of TSN networks. Moreover, open-sourced GT WCD values computed by NC solvers for CBS provide a reliable resource for the entire community to further evaluate different benchmarking datasets.

\subsection{Negative Impacts}
\label{appendix:negative_impact}
While TSNBench is intended to advance research on LLM proficiency in TSN, we acknowledge the following potential negative impacts.

\textbf{Overreliance on model outputs:} Models trained on the open-access dataset provided by TSNBench may achieve high accuracy on WCD analysis tasks, which could lead practitioners to deploy such models directly in real-world deployments without independent verification. Any WCD values or network configuration decisions produced by an LLM should be verified using formally verified solvers and NC tools before real-world deployment. 

\textbf{False confidence from MCQA performance:} Our results demonstrate that strong MCQA performance does not transfer to open-ended WCD estimation. A practitioner or system engineer who evaluates an LLM solely on MCQA benchmarks may incorrectly conclude that the model is suitable for TSN configuration tasks, leading to unsafe deployments in systems where timing guarantees are required.

\textbf{Data contamination and benchmark overfitting:} As TSNBench is released as an open-access dataset, future models may be trained directly on the benchmark questions, leading to inflated performance that does not reflect genuine TSN reasoning capability. We recommend that researchers introduce randomization in the test cases to prevent bias in results. Researchers should be cautious when interpreting results from models whose training data may overlap with the TSNBench dataset.

\textbf{Misuse of the dataset:} The dataset can be used to train models to configure TSN networks. Owing to the safety-critical nature of TSN applications, such models could potentially be exploited by attackers to manipulate network configurations, introduce timing violations, or deliberately cause deadline misses in industrial and automotive systems.

\section{Time-Sensitive Networking}
\label{appendix:tsn}
\begin{figure}[htbp]
  \centering
  \includegraphics[scale=0.35]{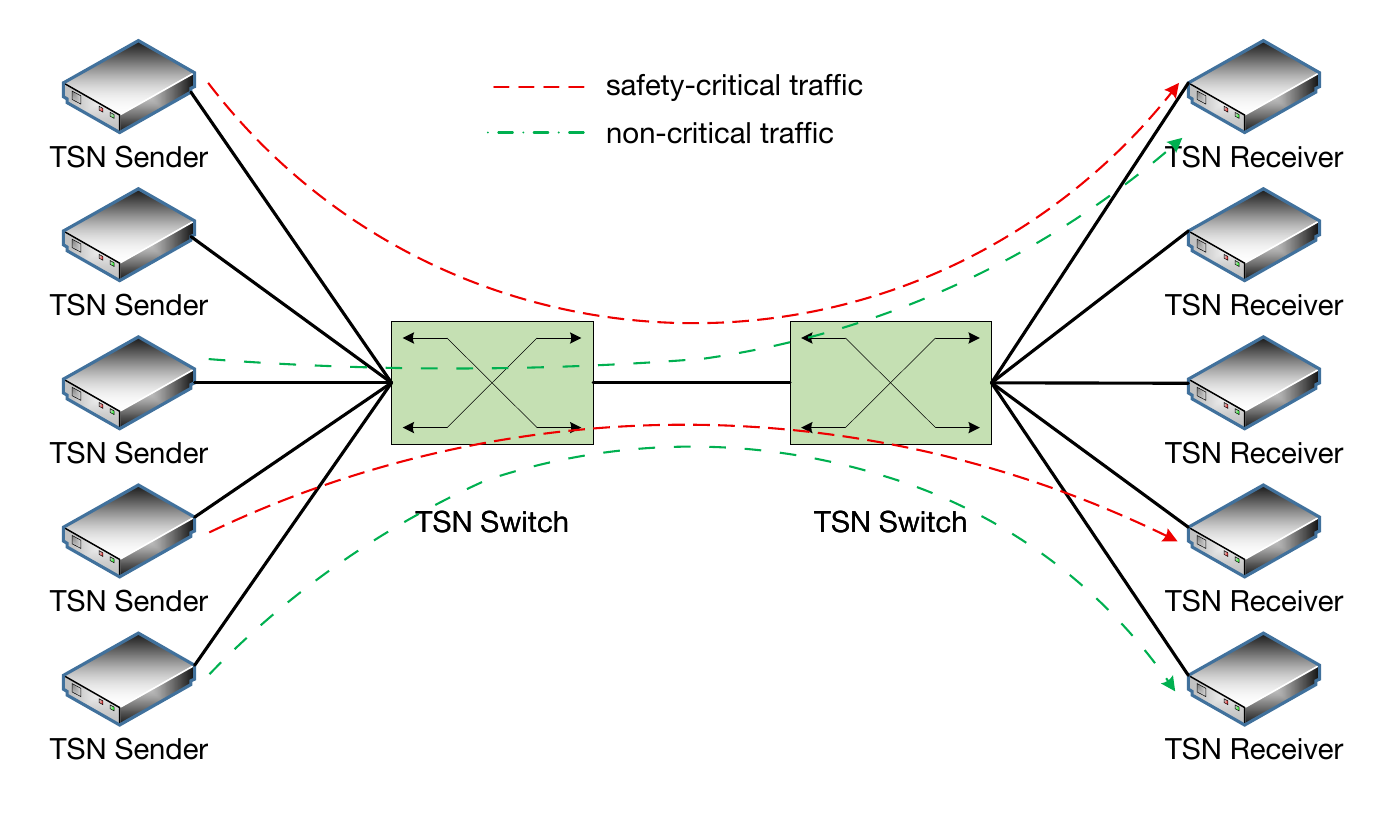}
  \caption{A sample TSN network with TSN senders, receivers, and TSN switches in the network. TSN senders are sending mixed-criticality traffic, including safety-critical and non-critical traffic to the TSN receivers.}
  \label{fig:tsn_network_example_with_flows}
\end{figure}

Time-Sensitive Networking (TSN)~\citep{intro_tsn} is a set of amendments and additions to the IEEE 802.1 standards that, since its inception in 2012, has become one of the most relevant technologies for enabling deterministic and real-time communications over Ethernet networks. TSN extends standard Ethernet by introducing mechanisms for bounded latency, low jitter, and high reliability, making it suitable for applications such as industrial automation, automotive systems, and professional audio-video networks. Figure~\ref{fig:tsn_network_example_with_flows} showcases a simple TSN network with flows.

\begin{figure}[htbp]
  \centering
  \includegraphics[scale=0.35]{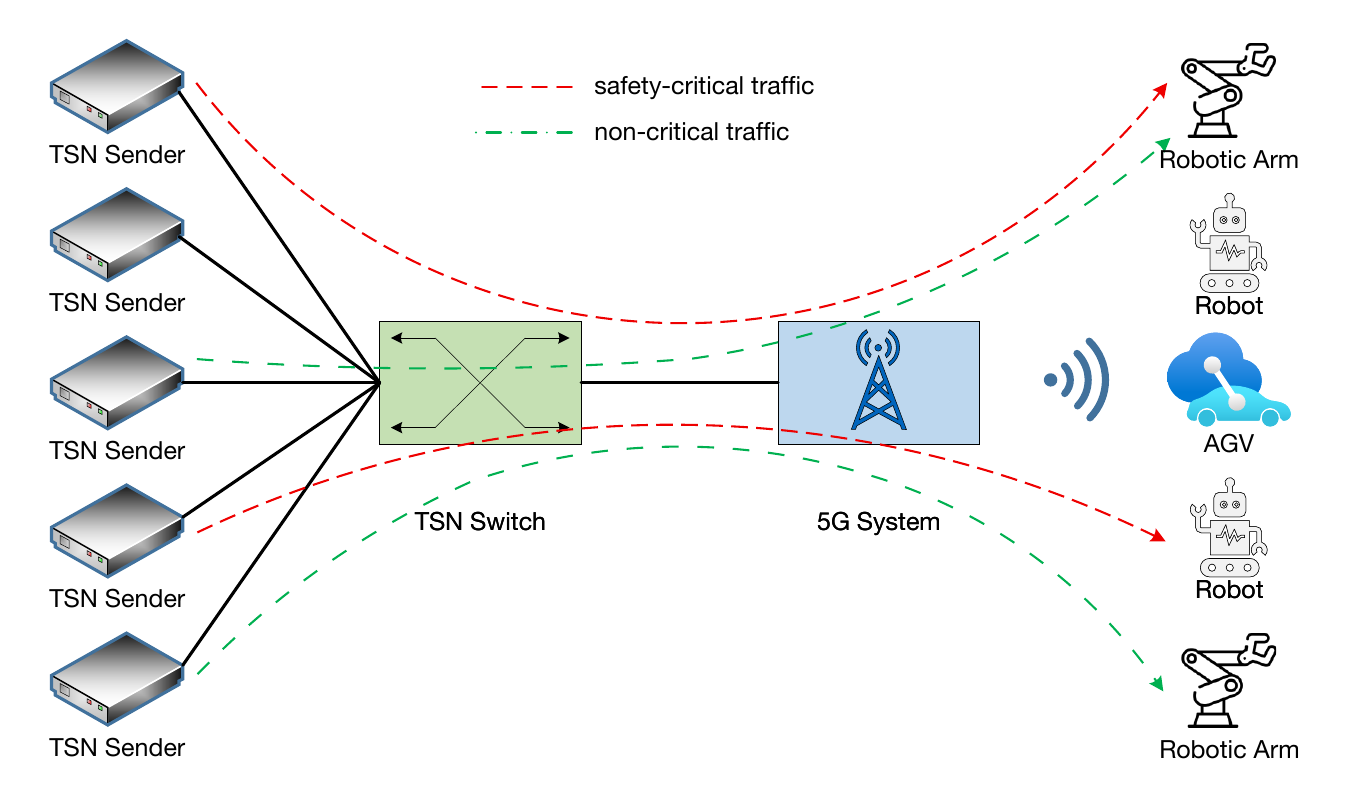}
  \caption{A sample wireless-TSN network with TSN senders, wireless receivers (such as robotic arm and automated guided vehicles (AGVs)), TSN switches, and 5G system in the network. TSN senders are sending mixed-critical including safety-critical and non-critical traffic to the wireless receivers.}
  \label{fig:5g_tsn_network_example_with_flows}
\end{figure}

In TSN, communication between end-stations is based on the transmission of Ethernet frames across a network of interconnected Ethernet links and TSN switches. These switches, as well as the output ports of end-stations, implement a queuing architecture with up to eight First-In-First-Out (FIFO) queues, each associated with one of the eight traffic priorities defined in IEEE 802.1Q~\citep{8021Q}. TSN is not just limited to the wired domain. The growing necessity of deterministic communication has extended to wireless domain gaining a significant interest in wireless-TSN networks. Although TSN is fundamentally an IEEE 802.1 bridged Ethernet technology, wireless and 5G-TSN~\citep{rubi_vtc} integration requires additional adaptation or translation functions, together with time-synchronization mechanisms that preserve deterministic latency guarantees across heterogeneous network segments.
We showcase a 5G-TSN system in Figure~\ref{fig:5g_tsn_network_example_with_flows}, where TSN senders are sending mixed criticality traffic types to wireless receiver nodes over a TSN switch and 5G system in the network. Some of the most commonly used abbreviations in TSN are given in Table~\ref{tab:abbreviations}.

Frames are classified into traffic classes and assigned to egress queues based on their priority, with transmission selection typically governed by strict priority. Industrial TSN traffic is commonly categorized into traffic types such as isochronous traffic, cyclic-synchronous traffic, cyclic-asynchronous traffic, network-control traffic, alarms and events, configuration and diagnostics, and best-effort traffic~\citep{ademaj2019industrial}. These traffic types require different timing guarantees: safety-critical isochronous traffic is typically mapped to time-triggered (TT) traffic, requiring guaranteed latency and bounded jitter, and is commonly handled by time-triggered mechanisms such as the Time-Aware Shaper (TAS)~\citep{silviu_tas, silviu_gcl_array}. In contrast, cyclic-synchronous or cyclic-asynchronous traffic that requires bounded end-to-end latency but less stringent jitter control is commonly mapped to AVB stream traffic and is often supported by the Credit-Based Shaper (CBS)~\citep{luxi_avb}. TSN also defines mechanisms such as Asynchronous Traffic Shaper (ATS)~\citep{ubs, rubi_rtcsa, ats_tas_ahmed}, Frame preemption (FP)~\citep{rubi_noms}, and Cyclic Queuing and Forwarding (CQF)~\citep{jrsp_cqf, rubi_iotj, itp_cqf} to provide deterministic communication under different traffic and deployment assumptions.

These mechanisms regulate when and how frames are transmitted, allowing the network to provide guarantees such as bounded delay, jitter, and controlled bandwidth allocation. In the MCQA dataset of TSNBench, we covered the basics of different TSN mechanisms, including TAS, CBS, ATS, and CQF. The MCQAs are theoretical in nature and cover the basic understanding of the mechanisms without going into their mathematical or analytical details. In contrast, for the open-ended questions, we evaluate the capability of the models to perform numerical analysis, formulate mathematical equations, and find the WCD values for the flows in the network. For this, we selected two TSN mechanisms: CBS and CQF. The WCD values of the flows using the CBS mechanism are calculated using NC analysis, which is mathematically complex. Therefore, we also evaluate the CQF mechanism as a simpler mechanism. The WCD values of the flows using the CQF mechanism can be directly calculated using the routing of the flow and the cycle duration. The working mechanism and architecture of CBS and CQF are described in detail in Appendices~\ref{appendix:cbs} and \ref{appendix:cqf}. The theory of NC is further explained along with the mathematical equations in Appendix~\ref{appendix:network_calculus}.

\begin{table}
\centering
\caption{Abbreviations and mechanisms used in TSNBench.}
\label{tab:abbreviations}
\begin{tabular}{lc}
\toprule
\textbf{Abbreviation} & \textbf{Full form}\\
\midrule
TSN & Time-Sensitive Networking \\
TAS & Time-Aware Shaper \\
CBS & Credit-Based Shaper \\
ATS & Asynchronous Traffic Shaper \\
CQF & Cyclic Queuing and Forwarding \\
NC & Network Calculus \\
WCD & Worst-Case Delay \\
AVB & Audio Video Bridging \\
TT & Time-Triggered \\
GCL & Gate Control List \\
BE & Best Effort \\
\bottomrule
\end{tabular}
\end{table}

\section{Network Calculus Theory}
\label{appendix:network_calculus}
Network Calculus (NC) is a theory for calculating worst-case bounds in communication networks based on min-plus algebra. Its basic paradigm involves two operators: convolution $\otimes$ 

\begin{equation}\label{eq:Conv}
	(f\!\otimes\!g)(t)\!=\!\inf_{0\leq s\leq t}\{\!f(t\!-\!s)\!+\!g(s)\!\},
\end{equation}
and deconvolution $\oslash$,

\begin{equation}\label{eq:Deconv}
	(f\!\oslash\!g)(t)\!=\!\sup_{s\ge 0}\{f(t\!+\!s)\!-\!g(s)\!\}. 
\end{equation}

Based on this algebra, the arrival curve and the service curve are constructed to describe the maximum arrival traffic data and the minimum service capability over any time interval, respectively. In the hybrid TSN/TAS+CBS architecture, the service for ET traffic is constrained not only by the bandwidth reservation, but also by high-priority TT traffic. We adopt the state-of-the-art network calculus model \citep{Zhao21, luxi_tii} to ensure deadline guarantees for ET flows with an arbitrary number of SR classes in the TSN/TAS+CBS architecture. Since our open-ended CBS questions do not have any TAS mechanism, we use the TSN/TAS+CBS architecture without the TAS mechanism in it with only CBS mechanism for the AVB flows in the network.

As described in~\citep{luxi_tii}, the service curve $\beta(t)$ is for constraining the minimum service capabilities, satisfying
\begin{equation}\label{g:SerCur}
	\mathcal{R}^{*}(t)\geq \left(\mathcal{R}\otimes\beta\right)(t).
\end{equation}
The function $\mathcal{R}(t)$ (resp. $\mathcal{R}^*(t)$) is the input (resp. output) cumulative function counting the total data bits of the flow that arrive at (resp. depart from) the server up to time $t$. A typical example of a service curve is the rate-latency form,
\begin{equation}\label{g:SerRateLatency}
	\beta_{R,T}(t)=R[t-T]^+
\end{equation}
with the service rate $R$ and latency $T$. The notation $[x]^+$ equals $x$ if $x\geq 0$, and 0 otherwise.

In the hybrid TSN/TAS+CBS architecture, the CBS service curve~\citep{Zhao21} for the arbitrary SR Class $M_i$ ($i\in[1,N_{SR}]$) with the impact of TT traffic at the output port $h$ is,
\begin{equation}\label{g:aggSerCBS}
	\beta^{h}_{M_i}(t)=idSl^h_{M_i}\left[t-\frac{\alpha_{TAS}^h(t)}{C}-\frac{c_{M_i}^{h,\max}}{idSl^h_{M_i}}\right]^+_\uparrow,
\end{equation}
where $c_{M_i}^{h,\max}$ is the credit upper bound for SR Class $M_i$, $idSl$ denotes the idleSlope,
\begin{equation}\label{g:creditMaxF}
	c_{M_i}^{h,\max}=idSl^h_{M_i}\cdot \frac{\sum_{j=1}^{i-1}c_{M_j}^{h,\min}-l^{h,\max}_{>i}}{\sum_{j=1}^{i-1}idSl^h_{M_j}-C},
\end{equation}
where 
$l^{h,\max}_{>i}=\max_{j>i}\{l^{h,\max}_{M_{j}}, l^{h,\max}_{BE}\}$ is the maximum frame size with priority lower than Class $M_i$ at $h$, $l^{h,\max}_{M_{j}}$ is the maximum frame size of Class $M_i$ at $h$, and $c_{M_i}^{h,\min}$ is the lower credit bound of SR Class $M_i$,
\begin{equation}\label{g:creditMin}
	c_{M_i}^{h,\min}=sdSl^h_{M_i}\cdot \frac{l^{h,\max}_{M_i}}{C},
\end{equation}
where $sdSl$ denotes the sendSlope.

$\alpha_{TAS}^h(t)$ in Eq.~(\ref{g:aggSerCBS}) is the arrival curve of TT traffic scheduled by GCL.

The arrival curve $\alpha(t)$ is for constraining the arrival process of the flow, satisfying 
\begin{equation}\label{g:ArrCur}
	\mathcal{R}(t)\leq\left(\mathcal{R}\otimes\alpha\right)(t).
\end{equation}
A typical example of an arrival curve is the burst-rate form,

\begin{equation}\label{g:ArrBurstRate}
	\alpha(t)=b+\rho\cdot t,
\end{equation}

\noindent
for $t>0$ and 0 otherwise, with the parameters $b$ as the maximum burst tolerance and $\rho$ as the long-term rate of the flow. 

For each ET flow $f$ at its source ES $h_0$, the arrival curve can be modeled as,

\begin{equation}\label{g:ArrAVB1}
	\alpha_f^{h_0}(t)=b_f^{h_0}+\rho_f^{h_0}t,
\end{equation}
where $b_f^{h_0}=l_f$, and $\rho_f^{h_0}=l_f/P_f$.
The arrival curve of flow $f$ at intermediate node $h$ is the output arrival curve of $f$ departing from the server $h^-$, 

\begin{equation}\label{g:outputArr2}
	\alpha_f^{h}(t)=\alpha_f^{h^-}\oslash\delta_{D_f^{h^-}}(t),
\end{equation}

\noindent
where $D_f^{h^-}$ is the latency upper bound of flow $f$ queuing at server $h^-$, and $\delta_{D}(t)$ is the pure-delay function. 

The aggregate arrival curve for ET flows of SR Class $M_i$ at $h$ is obtained by summing the arrival curves of individual flows. It also incorporates the link shaping curve and the CBS shaping curve to improve the tightness of the analysis results.
\begin{equation}\label{g:AggArr}
		\alpha^h_{M_i}\!(t)\!=\!\!\sum_{h^-\in\mathcal{H}}\sum_{f\in \mathcal{F}_{M_i}^{h^-\!,h}}\!\!\!\!\!\! \alpha^{h}_f(t)\!\wedge\!\sigma_{link}^{h^-\!,h}(t)\!\wedge\!\sigma_{M_i}^{h^-\!,h}(t),
\end{equation}

\noindent
where $x\wedge y=\min\{x,y\}$, $\sigma_{link}^{h^-\!,h}(t)$ is the link shaping curve from the preceding output $h^-$ to the current output port $h$:
\begin{equation}\label{g:LinkShapingCur}
	\sigma_{link}^{h^-\!,h}(t)=Ct+l_{M_i}^{h^-\!,h,\max},
\end{equation}
considering the packetization impact of the maximum frame size $l_{M_i}^{h^-\!,h,\max}$ of flows with Class $M_i$ from $h^-$ to $h$. $\sigma_{M_i}^{h^-\!,h}(t)$ is the CBS shaping curve of Class $M_i$ from $h^-$ to $h$:
\begin{equation}\label{g:CBSShapingCur}
	\sigma_{M_i}^{h^-\!,h}\!(t)\!=\!idSl^{h^-}_{M_i}\!\left[\!t\!-\!\frac{\beta^{h^-}_{TAS}(t)}{C}\!+\!\frac{c_{M_i}^{h^-\!,\max}\!-\!c_{M_i}^{h^-\!,\min}}{idSl^{h^-}_{M_i}}\!\right]\!+\!l_{M_i}^{h^-\!,h,\max},
\end{equation}
$\beta^{h}_{TAS}(t)$ represents the minimum service supplied to TT traffic on the output port $h$.

With NC-based Total Flow Analysis (TFA), the worst-case delay upper bound $D_f^h$ for flow $f\in\mathcal{F}_{M_i}^h$ at $h$ equals the worst-case delay upper bound $D_{M_i}^{h}$ for all flows with the same priority $M_i$ aggregating at $h$,
\begin{equation}\label{g:aggDelay}
D_f^h\!=\!D_{M_i}^{h}\!=\!hDev(\alpha_{M_i}^h,\beta_{M_i}^h)=\!\sup_{t\geq0}\left\{\inf\left\{\tau\!\geq\!0\mid\alpha^h_{M_i}\!(t)\leq\beta^h_{M_i}\!(t\!+\!\tau)\right\}\right\}\\
\end{equation}

\noindent
where $\alpha^h_{M_i}(t)$ is the arrival curve of aggregate flows of Class $M_i$ from Eq.~(\ref{g:AggArr}), and $\beta^h_{M_i}(t)$ is the service curve for Class $M_i$ from Eq.~(\ref{g:aggSerCBS}). The upper bound of the worst-case end-to-end delay for the flow $f$ is then obtained by summing the per-port latency bounds along its route.

\section{Credit-Based Shaper}
\label{appendix:cbs}
Credit-Based Shaper (CBS) is a TSN mechanism designed to prevent starvation of lower-priority traffic while guaranteeing a reserved portion of bandwidth for higher-priority queues, thereby providing reliability through bounded end-to-end delays. Traffic assigned to queues using CBS is typically referred to as Audio Video Bridging (AVB) traffic.

\begin{figure}[htbp]
  \centering
  \includegraphics[scale=0.2]{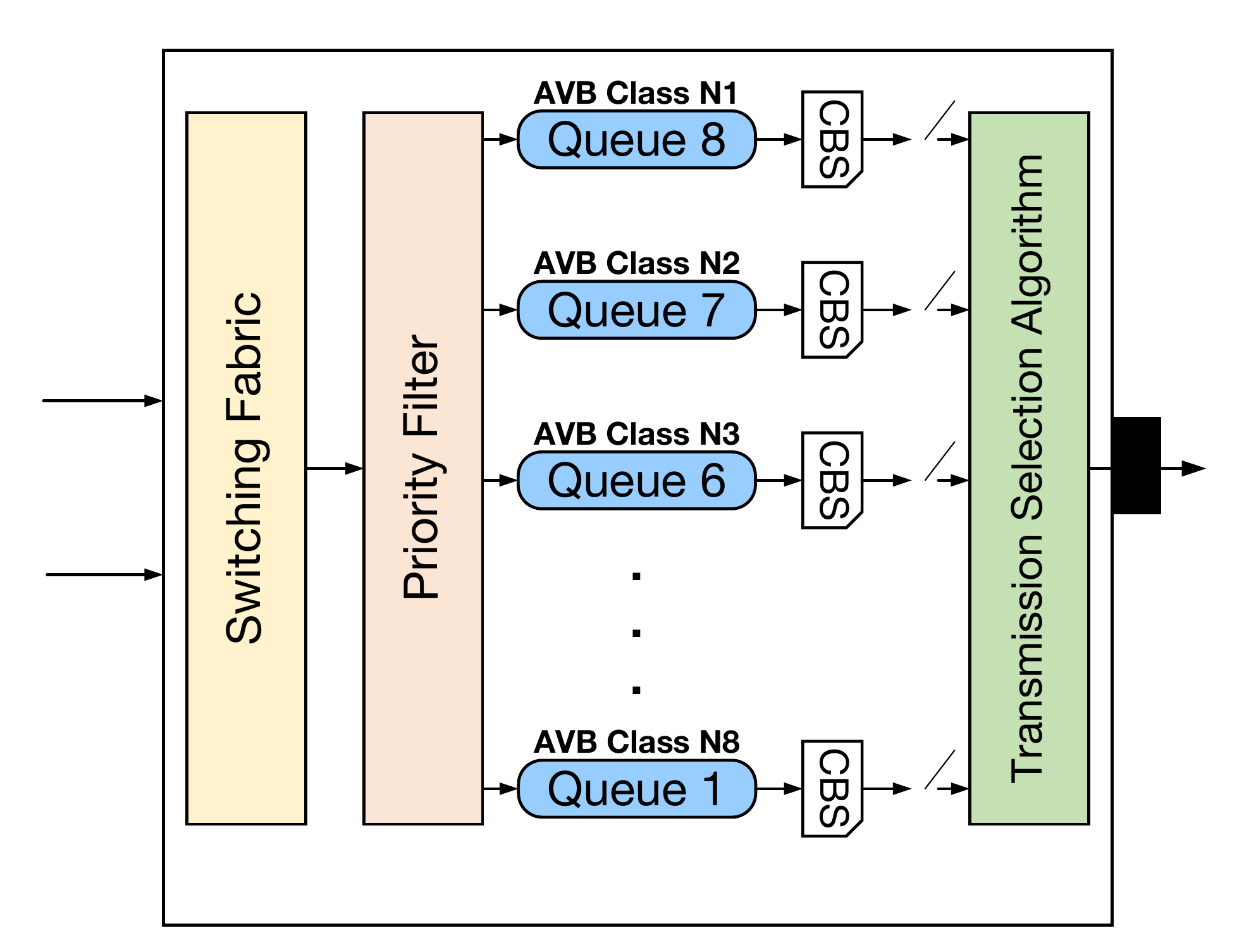}
  \caption{A simple CBS mechanism with eight queues in the egress port of the switch with different AVB classes mapped to different queues.}
  \label{fig:cbs}
\end{figure}

Here, we build on the description from~\citep{mateu2024improved}. In CBS, each AVB queue is associated with a credit value. This credit increases over time when a frame is waiting to be transmitted or when the credit is negative, and decreases while a frame is being transmitted. Moreover, if the credit is positive and there are no AVB frames waiting to be transmitted, the credit is immediately reset to 0. The rates at which credit is increased and decreased are defined by the parameters \textit{idleSlope} and \textit{sendSlope}, respectively. Each queue implementing CBS is configured with its own \textit{idleSlope} and \textit{sendSlope} values, which determine its allocated bandwidth share. In particular, the bandwidth reserved for a queue is expressed as Eq.~(\ref{eq: RBW}). A queue is eligible for transmission only when its credit is zero or positive.

\begin{equation} \label{eq: RBW}
    \mathrm{Reserved \; BW} = \frac{\textit{idleSlope}}{\textit{idleSlope} + \textit{sendSlope}} \cdot BW
\end{equation} 

\begin{figure}[htbp]
    \centering
    \includegraphics[width=\linewidth]{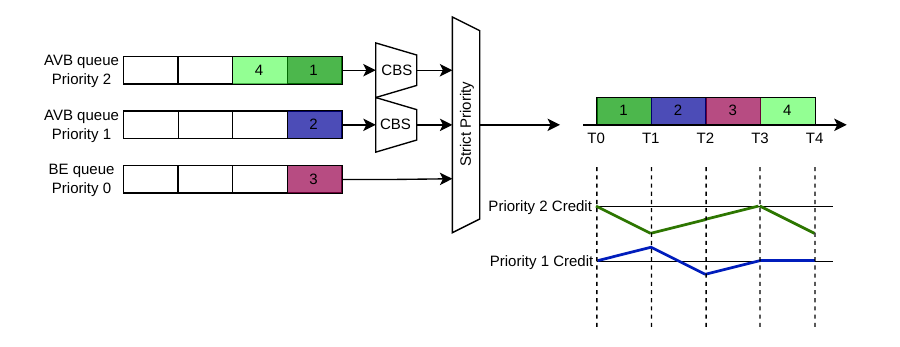}
    \caption{TSN output port with two AVB queues employing CBS and one BE queue.}
    \label{fig:TSNPort}
\end{figure}

Consider the example illustrated in Figure~\ref{fig:TSNPort}, which includes two AVB queues and one Best Effort (BE) queue. Frames 1 and 4 are assigned to the higher-priority AVB queue, while frames 2 and 3 belong to the lower-priority AVB queue and the BE queue, respectively.

At time T0, both AVB queues are eligible for transmission. Due to strict priority scheduling, the higher-priority AVB queue (priority 2) is selected, and frame 1 is transmitted. During this transmission, its credit decreases, while the credit of the lower-priority AVB queue increases because it is waiting.

At time T1, the higher-priority AVB queue has accumulated negative credit and is therefore no longer eligible for transmission. As a result, the lower-priority AVB queue is selected, and frame 2 is transmitted, even though a higher-priority frame (frame 4) is waiting. During this time, the lower-priority queue's credit decreases, while the higher-priority queue's credit recovers.

By time T2, both AVB queues have negative credit, making them ineligible for transmission. Consequently, the BE queue is selected, and frame 3 is transmitted, despite the presence of a higher-priority AVB frame waiting.

Finally, at time T3, the credit of the higher-priority AVB queue has recovered to zero, making it eligible again. Therefore, frame 4 is transmitted.

\section{Cyclic Queuing and Forwarding (CQF)}
\label{appendix:cqf}
Cyclic Queuing and Forwarding (CQF)~\citep{8021Qch, rubi_ccnc} is a TSN shaping mechanism which uses a single cycle duration, denoted as $T$, across the entire network. $T$ is the minimum scheduling unit where we put the TSN flows. Furthermore, $T$ defines the granularity of the end-to-end delay of the flows in the network. The unit of $T$ is in $\mu$s in TSNBench. In a TSN switch, every egress port in the network has eight queues. TSN flows are stored in the queues depending on their priority. In CQF, for each egress port, two queues are used: an \texttt{even queue} and an \texttt{odd queue}. Figure~\ref{fig:cqf} shows the basic working diagram of CQF with two queues (\texttt{even} and \texttt{odd}). As shown in Figure~\ref{fig:cqf}, CQF works by employing two queues, let's say, $Q_{8}$ and $Q_{7}$ for TT flows by operating them in a ping-pong manner where $Q_{7}$ receives and $Q_{8}$ transmits at the first cycle slot ($T_1$). During the second cycle slot ($T_2$), $Q_{8}$ receives and $Q_{7}$ transmits. Selecting or allotting a cycle slot for a flow means selecting the cycle slot number (within the hypercycle $H$) and the queue for the flow.

\begin{figure}[htbp]
  \centering
  \includegraphics[scale=0.2]{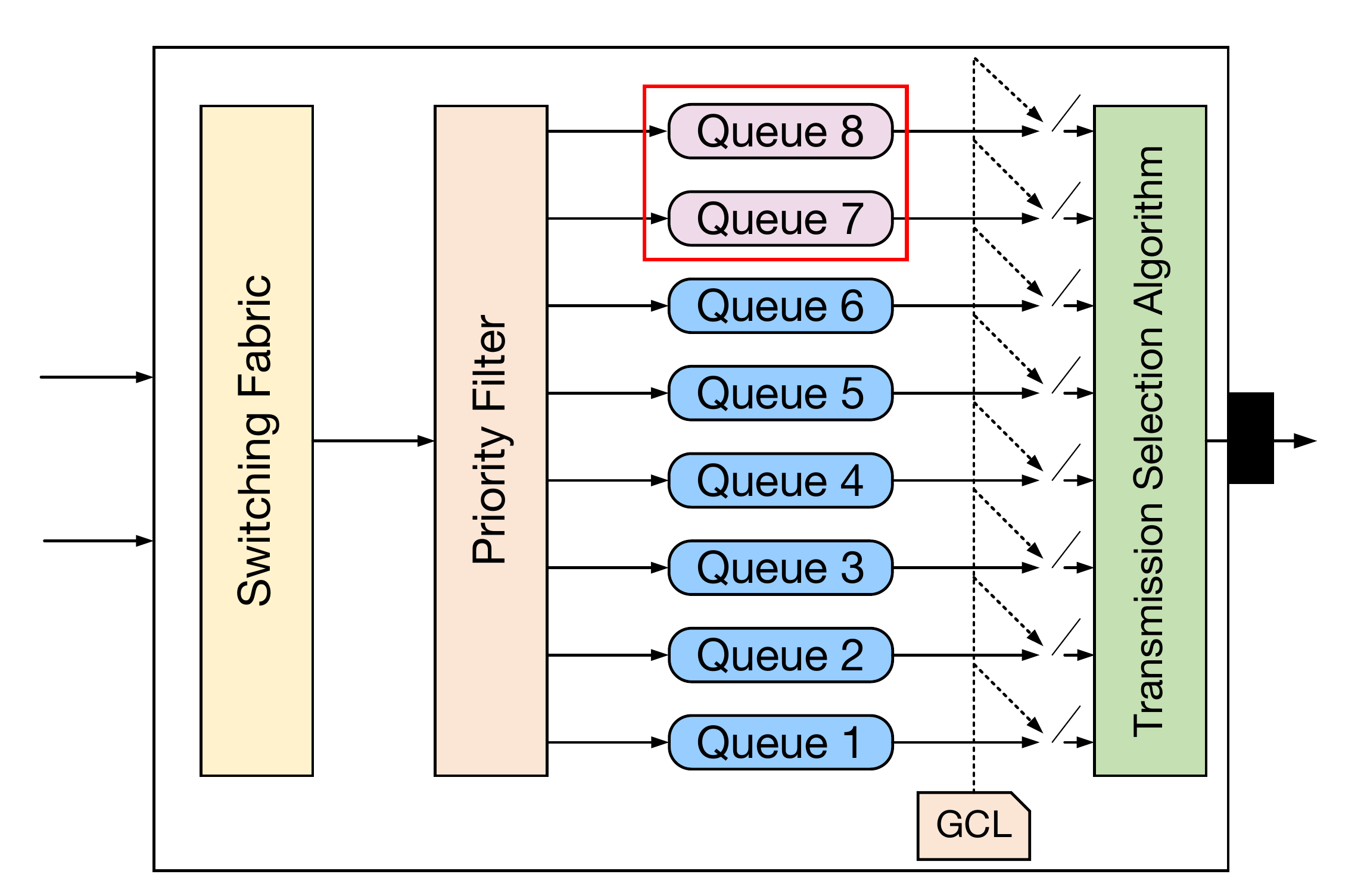}
  \caption{A simple CQF mechanism with eight queues in the egress port of the switch with two queues (Queue 8 and 7) operating as even and odd queue as shown in red.}
  \label{fig:cqf}
\end{figure}

\begin{figure}[htbp]
  \centering
  \includegraphics[scale=0.18]{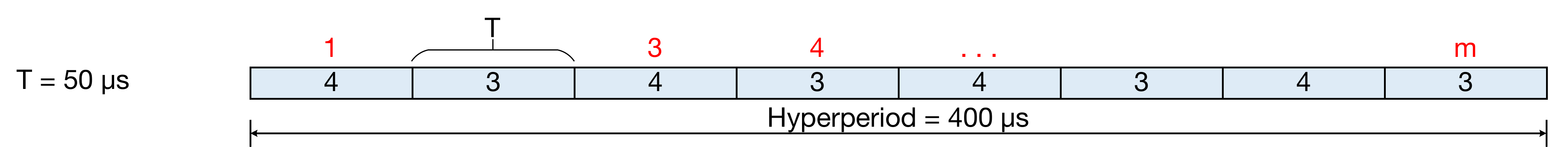}
  \caption{The Hypercycle also known as the scheduling cycle of the CQF (400 $\mu$s) with cycle duration (T) of 50$\mu$s. The different cycle slots are numbered as 1,2 $\cdots$ m in red.}
  \label{fig:hypercycle}
\end{figure}

In the CQF evaluation of TSNBench, we provide the cycle duration ($\mathrm{T}$) and the network-specific delays to the model as input through the prompt in all prompt variants. 

\begin{figure}[htbp]
  \centering
  \includegraphics[scale=0.2]{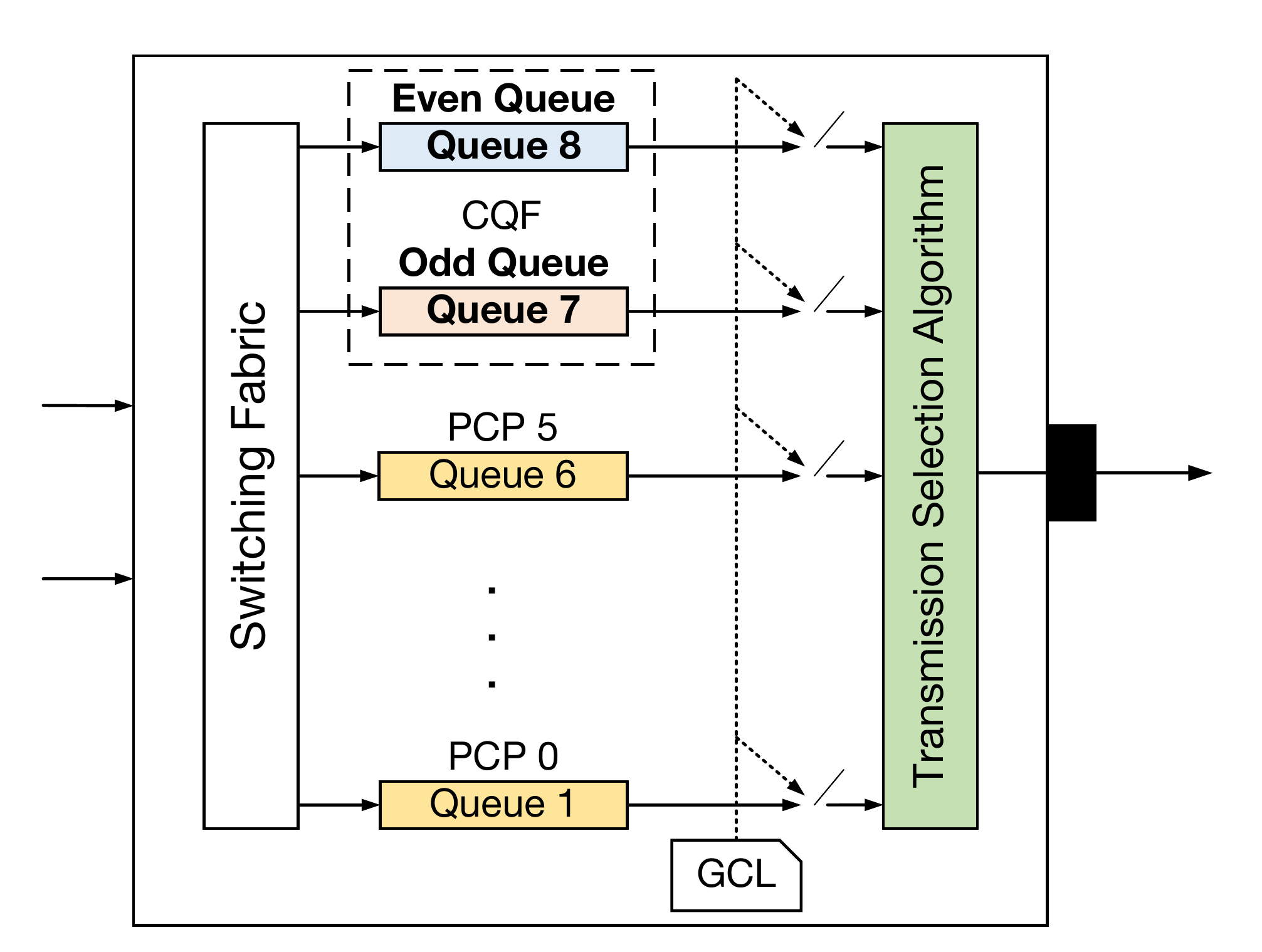}
  \caption{In this figure, we showcase the even and the odd queue in CQF architecture and during one cycle duration one queue receives the flows and another queue transmits the flows received in the previous cycle duration.}
  \label{fig:cqf}
\end{figure}

\textbf{WCD CQF:} The worst-case end-to-end delay of the TT flows in the CQF network is quantified as follows:
\begin{equation}
    \mathrm{WCD} = f_i.\phi + (\mathrm{SW_{num}}+1) \cdot \mathrm{T} + \xi,
    \label{eq:cqf_delay_appendix}
\end{equation} 
where $f_i.\phi$ is the offset of the flow $f_i$ in $\mu$s, $\mathrm{SW_{num}}$ is the total number of switches in the route of the TT flow, $\mathrm{T}$ is the cycle duration in $\mu$s, and $\xi$ denotes the network-specific delays: propagation delay ($d_\mathrm{prop}$), switching delay ($d_{\mathrm{sw}}$), and time synchronization error ($\mathrm{sync_{error}}$). In TSNBench, the switching delay $d_{\mathrm{sw}}$ is added once per switch, while the propagation delay and the time synchronization error are added once per flow:
\begin{equation}
    \xi = (\mathrm{SW_{num}} \cdot d_{\mathrm{sw}}) + d_{\mathrm{prop}} + \mathrm{sync_{error}}.
    \label{eq:cqf_xi}
\end{equation}

\section{More on TSNBench}
\label{appendix:more_on_tsnbench}

\subsection{MCQA keyword extraction and question generation}
\label{appendix:mcq-coverage}
Table~\ref{tab:gen_validator_model} lists the models used in the dataset pipeline. Claude Sonnet 4 extracts the TSN keywords, and Claude Sonnet 4, GPT-4o mini, and Llama 3.1 70B generate the MCQs, with each source paper assigned to exactly one generator in a round-robin manner. All models are used with default settings and were last accessed in April 2026.

\begin{table}[h]
  \caption{Models used in the TSNBench keyword extraction and question generation pipeline. All models are used with default settings and last accessed in April 2026. Claude Sonnet 4 serves two distinct roles: keyword extraction and question generation. These roles use identical model configurations but operate on different inputs and prompts. Full dataset generation details are provided in Appendix~\ref{appendix:more_on_tsnbench}.}
  \label{tab:gen_validator_model}
  \centering
  \small
  \resizebox{\textwidth}{!}{%
  \begin{tabular}{lllll}
  \toprule
    Model & API & Model ID & Organization & Usage \\
    \midrule
    Claude Sonnet 4 & Anthropic API & claude-sonnet-4-20250514 & Anthropic & Keyword extractor \\
    \midrule
    Claude Sonnet 4 & Anthropic API & claude-sonnet-4-20250514 & Anthropic & Generator \\
    GPT-4o mini & OpenAI API & gpt-4o-mini & OpenAI & Generator  \\
    Llama 3.1 70B & HF Router & Llama-3.1-70B-Instruct & Meta & Generator \\
    \bottomrule
  \end{tabular}
  }
\end{table}

\subsection{MCQ topic coverage by TSN mechanism}
\label{app:mcq-coverage}
Our 83 source papers were selected using explicit inclusion criteria, described in Sec. 3.1. A key part of this selection process was restricting our sources to published or accepted papers that are openly available. We chose to avoid paywalled versions of publications from IEEE and other publishers out of respect for their copyright terms. While we recognize this narrows the pool of available source material, we believe it reflects a responsible and ethical approach to constructing the dataset.

On the risk of under-representation, we provide a topic-coverage breakdown of the MCQs against different TSN mechanisms. By default, most of the research papers are published for TAS and CBS as these are the oldest mechanisms, and this is reflected in how much open-access material exists for each.

We used a 5\% representation threshold to check coverage, and all major TSN mechanisms are at or above this threshold. ATS sits almost exactly at the boundary (5.0\%), which we attribute to its comparatively recent standardization and the correspondingly smaller published literature base available for source-paper selection (Sec. 3.1). We do not consider this a critical gap. Instead, we highlight ATS as the mechanism most in need of attention in future dataset expansion.

"General background" category (29.2\%) consists of foundational TSN concepts, lexicon, keywords, and context spanning multiple mechanisms (TAS, CBS, ATS, CQF, GCL, reliability). We view this breadth as consistent with the benchmark's goal of assessing overall TSN domain knowledge, not solely mechanism-specific recall.

\begin{table}[htbp]
\caption{Topic coverage of the MCQ dataset, categorized by mechanism.}
\label{appendix:mcq_coverage}
\centering
\resizebox{\textwidth}{!}{
\small
\begin{tabular}{p{2.3cm} p{4.1cm} c c p{3.6cm}}
\toprule
\textbf{Category} & \textbf{TSN Mechanism} & \textbf{Count} & \textbf{\% of total} & \textbf{Comments} \\
\midrule
\multirow{4}{2.3cm}{Latency, schedule table, and configuration}
  & Time-Aware Shaper (TAS)              & 339 & 36.1\% & Most researched mechanism \\
  & Credit-Based Shaper (CBS)            & 94  & 10.0\% & Oldest mechanism \\
  & Asynchronous Traffic Shaper (ATS)    & 47  & 5.0\%  & Very recently standardized \\
  & Cyclic Queuing and Forwarding (CQF)  & 54  & 5.8\%  & Recently started gaining attention from researchers for its added benefits. \\
\midrule
\multirow{4}{2.3cm}{WCD analysis}
  & TAS            & \multirow{4}{*}{81} & \multirow{4}{*}{8.6\%} & \\
  & CBS            & & & \\
  & ATS    & & & \\
  & CQF  & & & \\
\midrule
\multirow{4}{2.3cm}{Reliability \& Resource Management}
  & Frame replication and elimination    & \multirow{4}{*}{50} & \multirow{4}{*}{5.3\%} & \\
  & Path control and reservation         & & & \\
  & Per stream filtering and policing    & & & \\
  & Stream reservation protocol          & & & \\
\midrule
\multirow{9}{2.3cm}{General background}
  & TAS & \multirow{9}{*}{274} & \multirow{9}{*}{29.2\%} & \multirow{9}{3.6cm}{General questions covering basic questions, keywords expansion covering all topics of TSN and legacy networks related to TSN.} \\
  & CBS & & & \\
  & ATS & & & \\
  & CQF  & & & \\
  & Frame replication and elimination & & & \\
  & Path control and reservation & & & \\
  & Stream reservation protocol  & & & \\
  & Time synchronization  & & & \\
  & Legacy system & & & \\
\midrule
\multicolumn{2}{l}{\textbf{Total}} & \textbf{939} & \textbf{100.0\%} & \\
\bottomrule
\end{tabular}}
\end{table}

\subsection{Human evaluation decision mechanism}
\label{appendix:sub_human_review}
To maintain the same standards across all human reviewers, we use the following rules to evaluate the MCQA dataset. There are four possible options for every question in the MCQA dataset.

\begin{enumerate}
    \item \textbf{Accept:}
    \begin{enumerate}[label=\roman*.]
        \item Technically correct.
        \item Clearly worded and self-contained.
        \item Unambiguous options.
        \item Accurate and sufficient explanation.
        \item The correct answer is actually the correct answer.
    \end{enumerate}
    \item \textbf{Reject:}
    \begin{enumerate}[label=\roman*.]
        \item Incorrect or misleading.
        \item Poorly constructed beyond revision.
        \item Irrelevant to TSN.
        \item Incomplete information.
        \item Too paper-dependent.
        \item Duplicate questions.
    \end{enumerate}
    \item \textbf{Revise:}
    \begin{enumerate}[label=\roman*.]
        \item Minor issues in grammar, clarity, or wording.
        \item Options need improvement.
        \item Explanation needs refinement.
    \end{enumerate}
    \item \textbf{Doubtful:}
    \begin{enumerate}[label=\roman*.]
        \item Paper-specific or uncertain about the correctness of the question.
        \item Explanation seems questionable.
        \item Needs further clarification.
    \end{enumerate}
    \textit{For a doubtful multiple-choice question, we read the research paper and re-evaluate the question. Afterward, the decision can be accept, reject, or revise; if it is still doubtful, we send it to another expert reviewer for a consensus-based group decision.}
\end{enumerate}

\textit{Key principles followed while reviewing the dataset:} We ensured that the MCQAs were technically accurate and aligned with TSN fundamentals. We avoided tricky questions and preferred clarity over complexity. The same set of rules was given to all expert reviewers who worked on this dataset and served as human judges. After the review, 185 questions were revised by the domain experts, as shown below in Table~\ref{tab:mcqa_human_review_stats}.

\begin{table}[htbp]
  \caption{Statistical data of the MCQA dataset after domain expert human review.}
  \label{tab:mcqa_human_review_stats}
  \centering
  \begin{tabular}{l|c}
    \toprule
    \textbf{Category} & \textbf{Count} \\
    \midrule
      Total questions revised by domain experts & 185 \\
    \bottomrule
  \end{tabular}
\end{table}

\subsection{Sample Questions}
\label{sub:sample_question_appendix}
We present three representative sample questions from our MCQA dataset below.

\vspace{6pt}
\noindent\fbox{%
\parbox{0.95\columnwidth}{%
\raggedright\small
\vspace{4pt}
\textbf{Q1} \hfill \textit{\small TSN Keyword}\\[6pt]
What does TAS stand for in TSN traffic management?\\[6pt]
\begin{tabular}{@{}ll@{}}
\textbf{A.} & Transmission Access Scheduler \\
\textbf{B.} & Traffic Analysis System \\
\textbf{C.} & Time-Aware Shaper \\
\textbf{D.} & Traffic Admission Service \\
[10pt]
Correct Answer: \textbf{C}
\end{tabular}
\vspace{4pt}
}}
\vspace{6pt}

\noindent\fbox{%
\parbox{0.95\columnwidth}{%
\raggedright\small
\vspace{4pt}
\textbf{Q2} \hfill \textit{\small Research Paper}\\[6pt]
In a Cyclic Queuing and Forwarding (CQF) network what fundamental 
limitation would prevent effective fault tolerance using Frame 
Replication and Elimination for Reliability (FRER) in a linear 
topology where each switch has maximum transmission unit (MTU) 
sized frames frequently queued?\\[6pt]
\begin{tabular}{@{}p{0.08\columnwidth}p{0.78\columnwidth}@{}}
\textbf{A.} & CQF's ping-pong queue switching would create timing conflicts with FRER's frame elimination mechanism.\\[3pt]
\textbf{B.} & EMI interference would corrupt both original and replicated frames equally, making spatial redundancy ineffective.\\[3pt]
\textbf{C.} & FRER cannot detect bit errors caused by EMI since it lacks Cyclic Redundancy Check (CRC) verification capabilities.\\[3pt]
\textbf{D.} & Linear topologies cannot provide the disjoint paths required for FRER's spatial redundancy approach, forcing expensive hardware additions. \\[20pt]
\multicolumn{2}{@{}l@{}}{Correct Answer: \textbf{D}}
\end{tabular}
\vspace{4pt}
}}
\vspace{6pt}

\noindent\fbox{%
\parbox{0.95\columnwidth}{%
\raggedright\small
\vspace{4pt}
\textbf{Q3} \hfill \textit{\small Research Paper}\\[6pt]
What fundamental challenge makes the Time Aware Shaper (TAS) 
implementation complex despite its ability to provide guaranteed 
end-to-end delays?\\[6pt]
\begin{tabular}{@{}p{0.08\columnwidth}p{0.78\columnwidth}@{}}
\textbf{A.} & The requirement to synchronize all network devices to a common time reference.\\[3pt]
\textbf{B.} & The need to maintain separate queues for each traffic class simultaneously.\\[3pt]
\textbf{C.} & The difficulty in estimating worst-case transmission times for variable-length frames.\\[3pt]
\textbf{D.} & The synthesis of the gate control list, which is an NP-hard problem.\\[10pt]
\multicolumn{2}{@{}l@{}}{Correct Answer: \textbf{D}}
\end{tabular}
}}
\vspace{6pt}

\subsection{Prompt design of open-ended questions}
\label{appendix:prompt_open_end}

\textbf{CB prompt:}
For the CBS and CQF mechanisms, two different approaches are used for WCD calculation. NC is used to calculate the CBS WCD, whereas an analytical mathematical calculation is used to find the WCD for the CQF mechanism. Since these two mechanisms work differently, we design prompts tailored to each mechanism.

\textit{Role:} We start by defining the role of the model: “You are an expert Time-Sensitive Networking (TSN) orchestrator.” We inject three network inputs: (i) network topology, (ii) TSN flow information, and (iii) the routes of the flows. We use the prompt-as-program~\citep{prompt_as_program} approach to separate the network topology, flow information, and flow routes. All of these are provided in text format. However, to evaluate different topologies, flows, and routes, we separate them from the prompt logic. This ensures that the prompt remains the same across different network topologies and parameters.

\textit{Constants:} To correctly calculate the WCD, information about the network parameters is required. To prevent the model from assuming these values and to keep the constant values consistent across all models, we provide this information in the prompt.

\vspace{6pt}
\noindent\fbox{%
\parbox{0.95\columnwidth}{%
\raggedright\small
\textbf{Constants for CBS open-ended questions:} \\[5pt]
$Bandwidth = 100$ Mbps,\; \\[3pt]
$Propagation \ delay = 1\,\mu$s,\; \\[3pt]
$Switching \ delay = 1\,\mu$s,\; \\[3pt]
$Time \ synchronization \ error = 1\,\mu$s,\;\\[3pt]
The switches of the network are cut-through switches,\; \\[3pt]
$IdleSlope$ $= 75\%$
}}
\vspace{6pt}

\noindent By controlling these network parameters, we directly mitigate hallucinations and assumptions about numerical values.

\textit{Architecture Restriction:}
TSN supports multiple architectures that affect the Quality of Service (QoS) and the WCD of the flows. The prompt restricts the model to using only one TSN mechanism through the following directive.

For the CBS mechanism, we use:

\vspace{6pt}
\noindent\fbox{%
\parbox{0.95\columnwidth}{%
\raggedright\small
\textbf{TSN Mechanism:} \\[10pt]
Only Credit-Based Shaper (CBS, IEEE 802.1Qav) is allowed; \\[10pt]
All flows are AVB Class A, PCP = 6, using queue 6 only. 
}}
\vspace{6pt}

For the CQF mechanism, we use:

\vspace{6pt}
\noindent\fbox{%
\parbox{0.95\columnwidth}{%
\raggedright\small
\textbf{TSN Mechanism:} \\[10pt]
Only Cyclic Queuing and Forwarding (CQF, IEEE 802.1Qch) is allowed; \\[10pt]
All flows are TT, PCP = 7, using queue 7 (odd) and 6 (even) only.
}}
\vspace{6pt}

Our reasoning is that letting the model select the TSN architecture or mechanism is a separate benchmarking problem, where the model is evaluated on architecture design performance. In TSNBench, our goal is to benchmark LLMs in TSN. Without an explicit restriction, the model may select an incorrect or inappropriate mechanism, producing a hallucinated architecture that does not satisfy the QoS requirements of the flows. This restriction forces the model to use a single solution space. It further ensures that the WCDs provided by different models are not caused by architectural faults or mechanism selection ambiguity, but rather by calculation and implementation errors within the specified mechanism.

\textit{Structured Output:}
We instruct the model through the prompt to provide the output strictly in JSON format~\citep{structeval}. 

\textbf{Formula-in-context (FC) prompt:}
The FC prompt is identical to the CB prompt above, except that the following description is provided in the prompt. 

\newcommand{\ptxt}[1]{\texttt{\detokenize{#1}}}
\vspace{6pt}
\noindent\fbox{%
\parbox{0.95\columnwidth}{%
\raggedright\small
Network Calculus Equations (use these directly; do not derive alternative formulations): \\[3pt]
Arrival curve, for flow \ptxt{f} with frame size \ptxt{l_f} and period \ptxt{P_f} at its source node \ptxt{h0}: \\
\ptxt{alpha_f^h0(t) = b_f^h0 + rho_f^h0 * t},   for \ptxt{t > 0}, else \ptxt{0}
where \ptxt{b_f^h0 = l_f},  \ptxt{rho_f^h0 = l_f / P_f} \\[4pt]
Arrival curve propagation to an intermediate node h (output arrival curve of \ptxt{f} departing server \ptxt{h-}): \\
\ptxt{alpha_f^h(t) = alpha_f^(h-)} (deconvolved with) \ptxt{delta_{D_f^(h-)}(t)} \\
where \ptxt{D_f^(h-)} is the latency upper bound of flow \ptxt{f} at server \ptxt{h-}, and \ptxt{delta_D(t)} is the pure-delay function. \\[4pt]
Aggregate arrival curve for all flows of priority class \ptxt{M_i} at port \ptxt{h}: \\
\ptxt{alpha_{M_i}^h(t) =} sum over incoming ports \ptxt{h-} and flows \ptxt{f} in class \ptxt{M_i} of \ptxt{[ alpha_{M_i}^{h-,h}(t) AND link-shaping-curve AND CBS-shaping-curve ]} \\
where \ptxt{AND} denotes \ptxt{min(x, y)}. \\[4pt]
Link shaping curve from \ptxt{h-} to \ptxt{h}: \\
\ptxt{sigma_link^{h-,h}(t) = C*t + l_{M_i}^{h-,h,max}}\\
where \ptxt{C} is the link rate and \ptxt{l_{M_i}^{h-,h,max}} is the max frame size of class \ptxt{M_i} flows from \ptxt{h-} to \ptxt{h}.\\[4pt]
CBS shaping curve of class \ptxt{M_i} from \ptxt{h-} to \ptxt{h}:\\
\ptxt{sigma_{M_i}^{h-,h}(t) = idleSlope_{M_i} * t + (c_{M_i}^{h-,max} - c_{M_i}^{h-,min}) + l_{M_i}^{h-,h,max}}\\[4pt]
Credit upper bound for class \ptxt{M_i} at port \ptxt{h}: \\
\ptxt{c_{M_i}^{h,max} = idleSlope_{M_i}^h * ( sum_{j<i} c_{M_j}^{h,min} - l_{>i}^{h,max} ) / ( sum_{j<i} idleSlope_{M_j}^h - C )} \\
where \ptxt{l_{>i}^{h,max}} is the max frame size of any class with priority lower than \ptxt{M_i} (or best-effort) at \ptxt{h}, \ptxt{j<i} means the AVB class \ptxt{j} with priority higher than \ptxt{i}. \\[4pt]

Credit lower bound for class \ptxt{M_i} at port \ptxt{h}: \\
\ptxt{c_{M_i}^{h,min} = sendSlope_{M_i}^h * l_{M_i}^{h,max} / C} \\
where \ptxt{sendSlope_{M_i}^h=idleSlope_{M_i}^h-C} \\[4pt]

CBS service curve (rate-latency form) for class \ptxt{M_i} at port \ptxt{h}: \\
\ptxt{beta_{M_i}^h(t) = idleSlope_{M_i}^h * [ t - c_{M_i}^{h,max} / idleSlope_{M_i}^h ]^+_{uparrow}}\\
where \ptxt{[f(t)]^+_{uparrow}=max_{0<=s<=t}\{f(s), 0\}}. \\[4pt]

Worst-case per-hop delay at port \ptxt{h} (horizontal deviation, NC Total Flow Analysis):
\ptxt{D_f^h = D_{M_i}^h = hDev(alpha_{M_i}^h, beta_{M_i}^h)}
\ptxt{= sup_{t>=0} { inf { tau >= 0 : alpha_{M_i}^h(t) <= beta_{M_i}^h(t + tau)}}} \\[4pt]

End-to-end WCD for flow \ptxt{f}:\\
\ptxt{WCD_f = sum of D_f^h} over every port \ptxt{h} along \ptxt{f}'s route (as given in Path input), \\
plus propagation delay, and switching delay per hop as specified in Constants.
}}
\vspace{6pt}

For CQF, we provide the following description:

\vspace{6pt}
\noindent\fbox{%
\parbox{0.95\columnwidth}{%
\raggedright\small
CQF worst case delay (WCD) equation (use these directly; do not derive alternative formulations):\\[3pt]
\ptxt{Hypercycle = LCM(all flow periods)}\\[3pt]
WCD of flow number \ptxt{i} is:\\[3pt]
\ptxt{WCD_f_i = offset_i + (n + 1) * cycle duration} \ptxt{+ n * switching delay + propagation delay} \ptxt{+ time synchronization error}\\[3pt]
\ptxt{offset_i} is the offset (start time) of flow \ptxt{i} at its source node.\\[3pt]
\ptxt{n} is the number of switches in the route of flow \ptxt{i} (\ptxt{n = hop count - 1}).\\[3pt]
Propagation delay and time synchronization error are added once per flow.
}}
\vspace{6pt}

\textit{Prompt variants:} For the prompt-sensitivity study (Section~\ref{sub:open_end_evaluation}), Variant A paraphrases the CB prompt, and Variant B replaces the JSON output instruction and schema with a fixed plain-text format. All inputs, constants, and the task itself are unchanged. The full text of the CB, FC, Variant A, and Variant B prompts for both mechanisms is available in the repository at \url{https://github.com/tum-esi/TSNBench/tree/main/open_end_evaluation} and in the dataset on Kaggle at \url{https://doi.org/10.34740/kaggle/ds/13019279}.

\subsection{TSNBench Open-Ended Question Details}
\label{appendix:open_end_question_details}
For the open-ended questions, there are three variable entries: network topology, flow information, and flow routing. We use the K-shortest path algorithm to determine the routes of the flows. The routes are then directly provided to the models as input for further evaluation.

\begin{figure}[htbp]
  \centering
  \includegraphics[scale=0.5, clip]{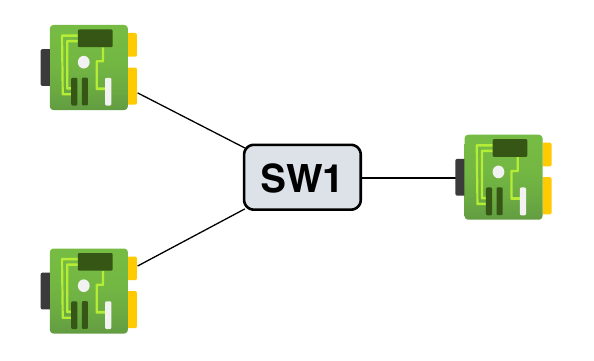}
  \caption{One-switch topology used to evaluate open-ended questions in TSNBench.}
  \label{fig:one_switch_topo}
\end{figure}

\begin{figure}[htbp]
  \centering
  \includegraphics[scale=0.3, trim=1cm 7cm 2cm 0cm, clip]{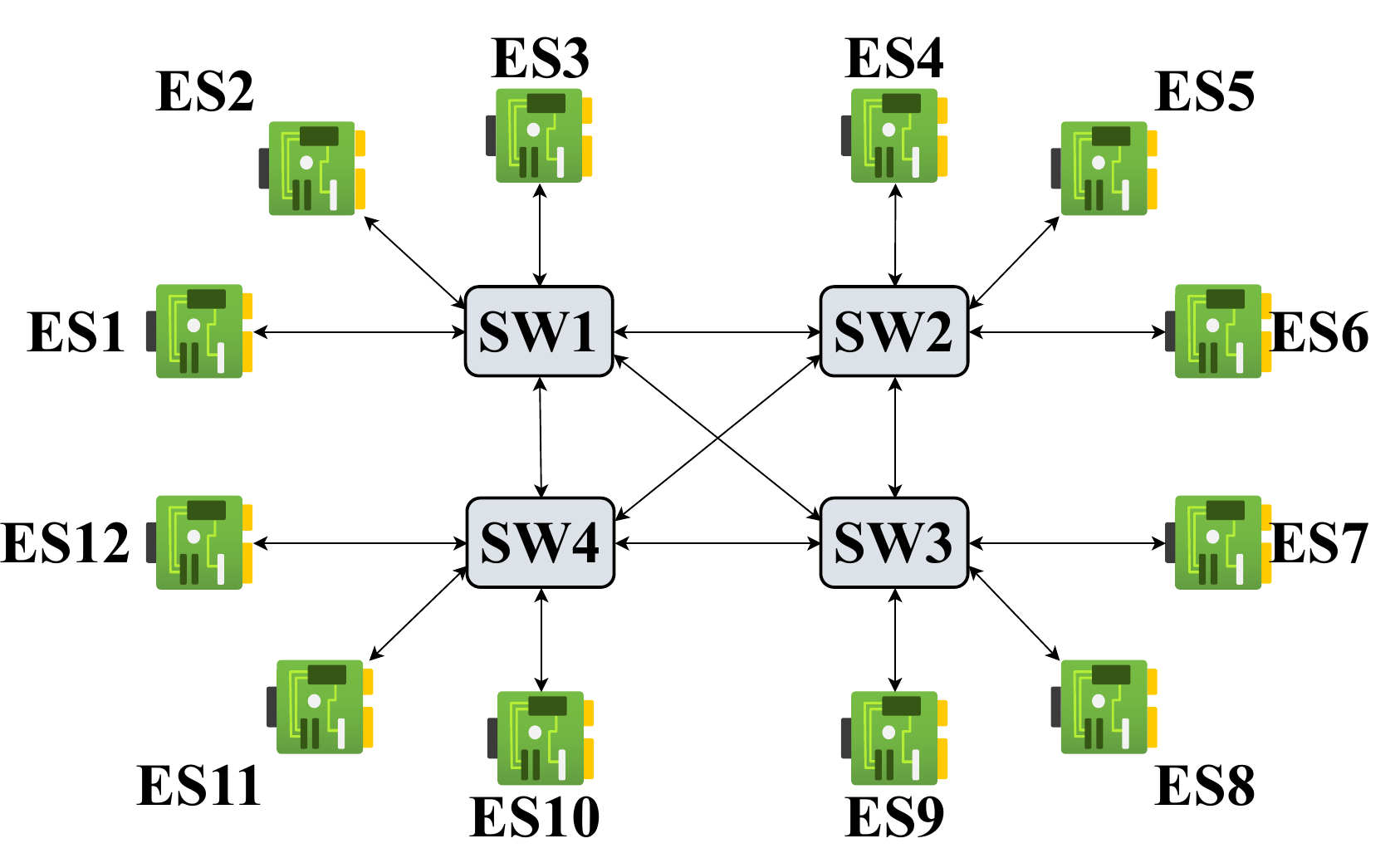}
  \caption{Medium-mesh topology used to evaluate open-ended questions in TSNBench.}
  \label{fig:mm_topo}
\end{figure}

\begin{figure}[htbp]
  \centering
  \includegraphics[scale=0.4]{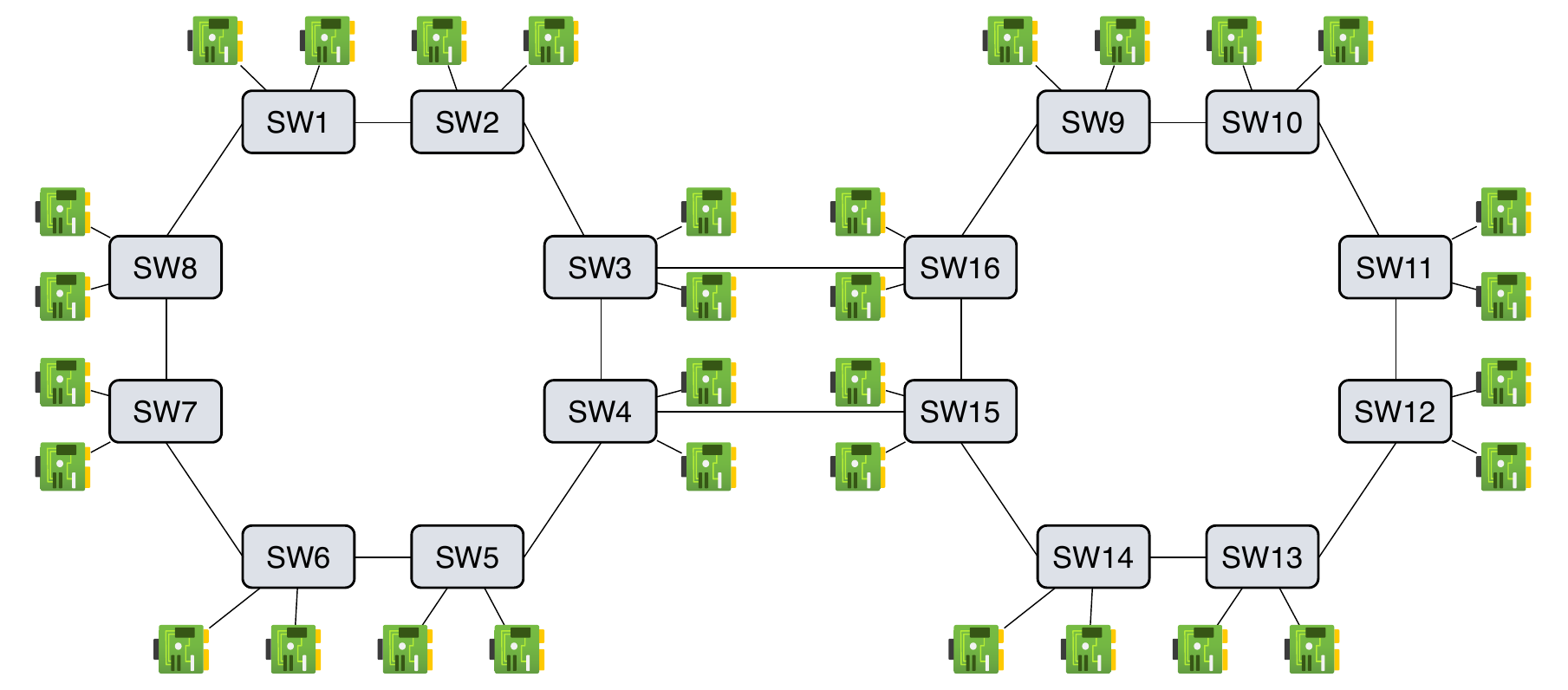}
  \caption{Ring topology representing the industrial ring network used to evaluate open-ended questions in TSNBench.}
  \label{fig:ring_topo}
\end{figure}

\textbf{Network Topologies Used:}
For the open-ended questions, we selected three different topologies to evaluate the models: a one-switch topology, a medium-mesh topology, and an industrial ring topology. Figures~\ref{fig:one_switch_topo}, \ref{fig:mm_topo}, and \ref{fig:ring_topo} represent the one-switch, medium-mesh, and ring topologies used in TSNBench, respectively. 

\begin{figure}[htbp]
  \centering
  \includegraphics[scale=0.4, trim=0cm 0cm 0cm 0cm, clip]{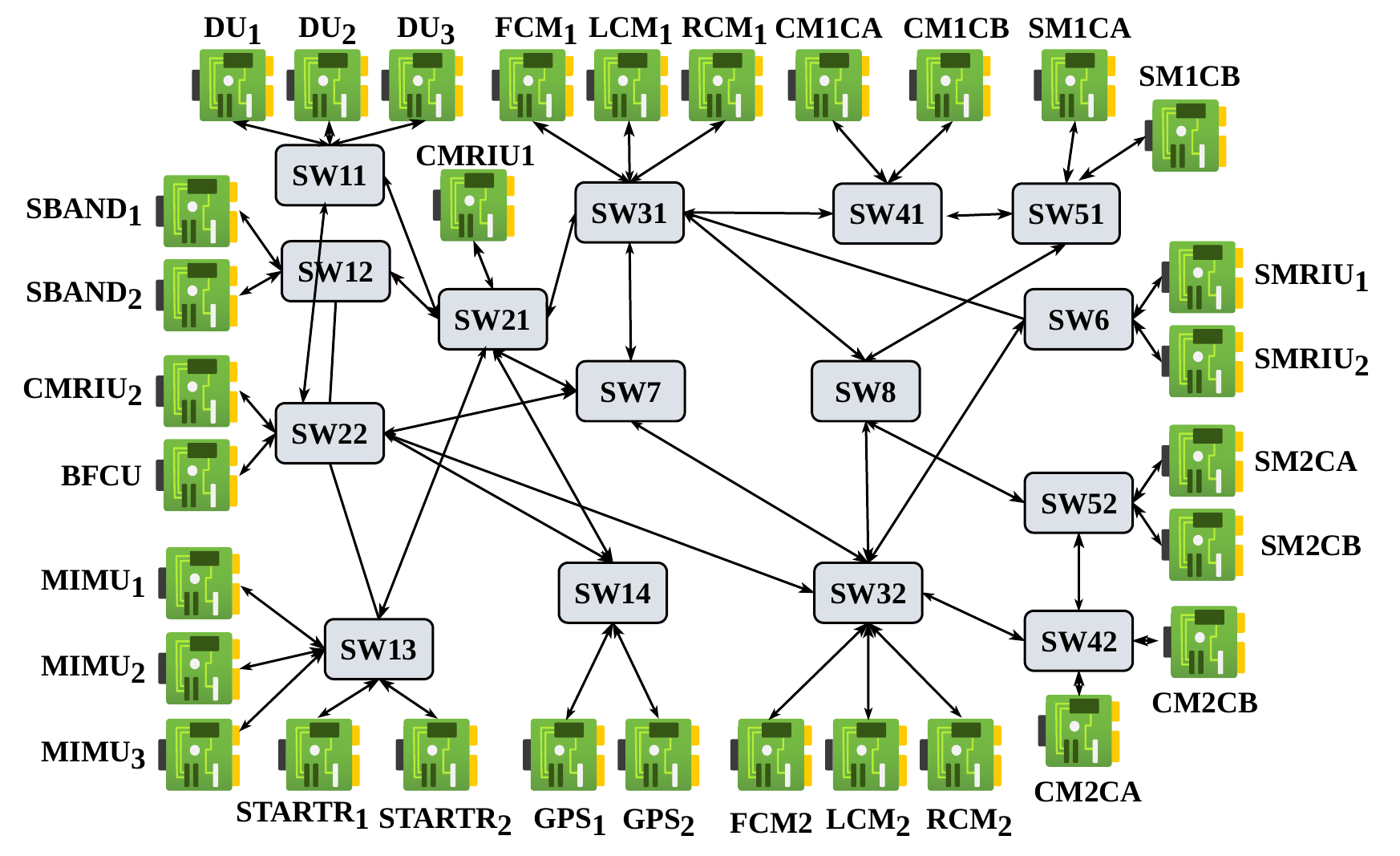}
  \caption{Orion Crew Exploration Vehicle (CEV) topology representing the realistic network topology used to evaluate open-ended questions in TSNBench.}
  \label{fig:orion_topo}
\end{figure}

\textbf{Flow parameters:}
We show the flow information used in TSNBench as follows.

\vspace{6pt}
\noindent\fbox{%
\parbox{0.95\columnwidth}{%
\raggedright\small
\vspace{4pt}
\textbf{Flow Information} \hfill \textit{\small TC1\_flows.txt}\\[6pt]
\texttt{0,node2\_1,node5\_2,2500,709,965}\\[2pt]
\texttt{1,node5\_4,node3\_2,2500,610,825}\\[2pt]
\texttt{2,node0\_4,node0\_1,1000,786,887}\\[2pt]
\texttt{3,node2\_3,node4\_3,2500,1088,1233}\\[2pt]
\texttt{4,node0\_4,node3\_3,1000,1015,488}\\[2pt]
\texttt{5,node0\_4,node0\_1,2500,926,501}\\[2pt]
\texttt{...}
\vspace{4pt}
}}
\vspace{6pt}

\textbf{Ground Truth WCD Values}
The ground-truth WCD values of the flows for all open-ended test cases for the CBS mechanism are calculated using a verified NC tool~\citep{luxi_avb, rubi_icc, voica_tta}. For the WCD of the CQF mechanism, we use the mathematical equation given in Eq.~\ref{eq:cqf_delay_appendix}.

\section{More on TSNBench Evaluation}
\label{appendix:more_on_evaluation}
We evaluate both open-source and closed-source state-of-the-art LLMs on TSNBench. A detailed list of the models, along with their model IDs and snapshots, is given in Table~\ref{tab:evaluation_models}. This ensures that the results are reproducible by the community.

\begin{table}[htbp]
  \caption{Details of the models used for the benchmarking on TSN. Both MCQA and open-ended questions are evaluated on these models. We provide the specific model ID and snapshot for reproducibility.} 
  \label{tab:evaluation_models}
  \centering
  \resizebox{\textwidth}{!}{%
  \begin{tabular}{lllll}
    \toprule
    \multicolumn{5}{c}{Chat Models} \\
    \midrule
    Model & Family & Model ID & Organization & Country \\
    \midrule
    Grok 4.1 Fast (Non-Reasoning) & Grok & grok-4-1-fast-non-reasoning & xAI & USA \\
    DeepSeek-V3.2 (Non-Thinking Mode) & DeepSeek & deepseek-chat & DeepSeek AI & China \\
    GPT‑4o & GPT & gpt-4o-2024-08-06 & OpenAI & USA \\
    GPT-4o mini & GPT & gpt-4o-mini-2024-07-18 & OpenAI & USA \\
    Llama 3.3 70B & Llama & Llama-3.3-70B-Instruct & Meta (via HF) & USA \\
    Mistral Medium 3.1 & Mistral & mistral-medium-2508 & Mistral AI & France \\
    Mistral Large 3 & Mistral & mistral-large-2512 & Mistral AI & France \\
    Claude Sonnet 4.5 & Claude & claude-sonnet-4-5-20250929 & Anthropic & USA  \\
    \midrule
    \multicolumn{5}{c}{Reasoning/Thinking Models} \\
    \midrule
     Grok 4.1 Fast & Grok & grok-4-1-fast-reasoning & xAI & USA \\
     o3 & GPT & o3-2025-04-16 & OpenAI & USA \\
     GPT-5 & GPT & gpt-5-2025-08-07 & OpenAI & USA \\
     DeepSeek-V3.2 (Thinking Mode) & DeepSeek & deepseek-reasoner & DeepSeek AI & China \\
     Gemini 2.5 Flash & Gemini & gemini-2.5-flash & Google & USA \\
    \midrule
    \multicolumn{5}{c}{Small Models} \\
    \midrule
     Llama 3.2 1B & Llama & llama-3.2-1B &  Meta (via HF) & USA  \\
     Qwen3 8B & Qwen & Qwen3-8B & Alibaba Cloud & China \\
    Ministral 3 8B & Ministral & ministral-8b-2512 & Mistral AI & France  \\
    \bottomrule
  \end{tabular}
}
\end{table}

\begin{table}[htbp]
\centering
\caption{Extended evaluation results of TSNBench MCQA dataset across different state-of-the-art models across different families. We provide the accuracy in percentage under different temperature settings (default, set to 0.0, and set to 0.7). The consistency shows the performance of the model in providing the same response across three runs. For those models which do not support temperature = 0.0, we use their default temperature and this is marked next to the model in the table.} 
\label{tab:more_tsnbench_evaluation_appendix_deafult_temp_temp_0}
\resizebox{\textwidth}{!}{%
\begin{tabular}{l|ccc|ccc}
\toprule
\multirow{2}{*}{\textbf{Model}}
    & \multicolumn{3}{c|}{\textbf{\shortstack{MCQA Accuracy (\%)}}}
    & \multicolumn{3}{c}{\textbf{Average Consistency}} \\
    \cmidrule(lr){2-4} \cmidrule(l){5-7}
    & \textbf{Default Temp.} & \textbf{Temp=0.0} & \textbf{Temp=0.7}
    & {\textbf{Default Temp.}} & {\textbf{Temp=0.0}} & \textbf{Temp=0.7}\\
\midrule
Grok 4.1 Fast$^\dagger$ & \cellcolor{red!20}\textbf{93.2} & \cellcolor{gray!20}-- & \cellcolor{gray!20}-- & 0.99 & \cellcolor{gray!20}-- & \cellcolor{gray!20}-- \\
Grok 4.1 Fast (Non-Reasoning) & \cellcolor{gray!20}-- & 91.7 & 91.6 & \cellcolor{gray!20}-- & 1.00 & 1.00 \\
DeepSeek-V3.2 (Non-Thinking) & \cellcolor{gray!20}-- & 94.0 & 93.4 & \cellcolor{gray!20}-- & 1.00 & 0.98 \\
GPT-4o & \cellcolor{gray!20}-- & 91.8 & 92.1 & \cellcolor{gray!20}-- & 1.00 & 0.98 \\
GPT-4o mini & \cellcolor{gray!20}-- & 88.3 & 88.2 & \cellcolor{gray!20}-- & 0.99 & 0.98 \\
Llama 3.3 70B & \cellcolor{gray!20}-- & 88.9 & 89.1 & \cellcolor{gray!20}-- & 0.99 & 0.99 \\
Mistral Medium 3.1 & \cellcolor{gray!20}-- & 92.1 & 92.3 & \cellcolor{gray!20}-- & 1.00 & 0.99 \\
Mistral Large 3 & \cellcolor{gray!20}-- & 92.8 & 92.9 & \cellcolor{gray!20}-- & 1.00 & 1.00 \\
Claude Sonnet 4.5 & \cellcolor{gray!20}-- & \cellcolor{green!20}\textbf{95.3} & \cellcolor{green!20}\textbf{95.3} & \cellcolor{gray!20}-- & 1.00 & 1.00 \\
o3$^\dagger$ & 94.7 & \cellcolor{gray!20}-- & \cellcolor{gray!20}-- & \cellcolor{red!20}\textbf{0.98} & \cellcolor{gray!20}-- & \cellcolor{gray!20}-- \\
GPT-5$^\dagger$ & \cellcolor{green!20}\textbf{95.0} & \cellcolor{gray!20}-- & \cellcolor{gray!20}-- & 0.99 & \cellcolor{gray!20}-- & \cellcolor{gray!20}-- \\
DeepSeek-V3.2 (Thinking)$^\dagger$ & 94.7 & \cellcolor{gray!20}-- & \cellcolor{gray!20}-- & \cellcolor{red!20}\textbf{0.98} & \cellcolor{gray!20}-- & \cellcolor{gray!20}-- \\
Gemini 2.5 Flash & \cellcolor{gray!20}-- & 90.1 & 90.8 & \cellcolor{gray!20}-- & \cellcolor{red!20}\textbf{0.98} & 0.97 \\
Llama 3.2 1B & \cellcolor{gray!20}-- & \cellcolor{red!20}\textbf{67.4} & \cellcolor{red!20}\textbf{67.0} & \cellcolor{gray!20}-- & 1.00 & \cellcolor{red!20}\textbf{0.93} \\
Qwen3 8B & \cellcolor{gray!20}-- & 83.7 & 82.8 & \cellcolor{gray!20}-- & 0.99 & 0.97 \\
Ministral 3 8B & \cellcolor{gray!20}-- & 86.9 & 86.5 & \cellcolor{gray!20}-- & 1.00 & 0.97 \\
\bottomrule
\end{tabular}%
}
\\[4pt]
\raggedright
\footnotesize
$^\dagger$ Temperature parameter not supported. Evaluated with default settings.
\end{table}

\subsection{Extended MCQA Experimental Evaluation}
We evaluate the models on the MCQA dataset under two different configurations: (i) temperature set to 0.7 and (ii) temperature set to 0.0 (or default temperature settings). The open-ended questions are evaluated at temperature 0.0 (or default temperature settings). As TSN is used in safety-critical networks, we want to ensure deterministic results, so we report results at temperature 0.0 wherever the model supports it. For models that do not support the temperature parameter, we use the default temperature for evaluation.

Table~\ref{tab:more_tsnbench_evaluation_appendix_deafult_temp_temp_0} provides the accuracy and average consistency of the models for the MCQA dataset under the default temperature, temperature set to 0.0, and temperature set to 0.7. Average consistency represents the ability of the model to provide the same results across three runs.

\subsection{MCQA and Open-Ended Cost and Latency}
\label{appendix:cost_latency}
The cost and latency of a model are important evaluation parameters for the research community. Spending a large amount of money on benchmark evaluation is a real bottleneck for research groups. Moreover, not all models can be evaluated locally. Table~\ref{tab:more_tsnbench_evaluation_cost_and_latency} presents the cost and latency of the TSNBench MCQA and open-ended questions. Evaluating MCQA is relatively much cheaper than evaluating open-ended questions. 

\begin{table}[htbp]
\centering
\caption{Extended results of cost and latency comparison for MCQA and CB open-ended evaluation in TSNBench. ``--'' indicates that cost and latency are not reported for this model, as it successfully evaluated fewer than 50 out of 100 TCs, where a TC is considered successfully evaluated only if the model provided WCD estimates for at least 80\% of the flows within that TC.}
\label{tab:more_tsnbench_evaluation_cost_and_latency}
\resizebox{\textwidth}{!}{%
\begin{tabular}{l|cc|cc|cc}
\toprule
\multirow{2}{*}{\textbf{Model}}
    & \multicolumn{2}{c|}{\textbf{\shortstack{MCQA}}}
    & \multicolumn{2}{c}{\textbf{CBS Open-ended questions}} & \multicolumn{2}{c}{\textbf{CQF Open-ended questions}} \\
    \cmidrule(lr){2-3} \cmidrule(l){4-5} \cmidrule(l){6-7}
    & \textbf{Cost (USD)} & \textbf{Latency (ms)}
    & \textbf{Cost (USD)} & \textbf{Latency (ms)} & \textbf{Cost (USD)} & \textbf{Latency (ms)} \\
\midrule
Grok 4.1 Fast$^\dagger$ & 0.2490 & 18,769,322 & \cellcolor{gray!20}-- & \cellcolor{gray!20}-- & 0.3047 & 49,367,058 \\
Grok 4.1 Fast (Non-Reasoning) & 0.2612 & 1,450,175 & 0.3256 & 3,200,483 & 0.3251 & 3,383,209 \\
DeepSeek-V3.2 (Non-Thinking) & \cellcolor{green!20}\textbf{0.0420} & 2,264,129 & \cellcolor{gray!20}-- & \cellcolor{gray!20}-- & 0.4816 & 13,211,941 \\
GPT-4o & 2.3661 & 2,053,601 & \cellcolor{gray!20}-- & \cellcolor{gray!20}-- & 4.3866 & \cellcolor{green!20}\textbf{2,122,167} \\
GPT-4o mini & 0.1417 & 2,251,786 & 0.3438 & 7,083,613 & 0.3642 & 7,092,033 \\
Llama 3.3 70B & 0.5224 & 1,028,334 & 0.7399 & \cellcolor{green!20}\textbf{1,264,847} & 0.7137 & 1,136,920 \\
Mistral Medium 3.1 & 0.4432 & 1,839,587 & 1.9918 & 6,335,803 & 2.2442 & 6,658,056 \\
Mistral Large 3 & 0.4868 & 15,487,495 & 1.5989 & 11,149,853 & 1.2298 & 8,226,564 \\
Claude Sonnet 4.5 & 3.5967 & 5,190,222 & 21.1719 & 14,342,470 & 18.3093 & 12,583,398 \\
o3$^\dagger$ & 7.6293 & 10,831,712 & 10.9954 & 10,127,407 & 10.0134 & 8,591,382 \\
GPT-5$^\dagger$ & \cellcolor{red!20}\textbf{12.8766} & 15,860,682 & \cellcolor{red!20}\textbf{57.5232} & \cellcolor{red!20}\textbf{74,473,434} & \cellcolor{red!20}\textbf{42.4470} & \cellcolor{red!20}\textbf{58,404,966} \\
DeepSeek-V3.2 (Thinking)$^\dagger$ & 0.4069 & 12,385,365 & \cellcolor{gray!20}-- & \cellcolor{gray!20}-- & \cellcolor{gray!20}-- & \cellcolor{gray!20}-- \\
Gemini 2.5 Flash & 0.4164 & 18,732,689 & 2.6471 & 24,160,941 & 2.7690 & 17,688,716 \\
Llama 3.2 1B & 0.0864 & 1,883,306 & \cellcolor{gray!20}-- & \cellcolor{gray!20}-- & \cellcolor{gray!20}-- & \cellcolor{gray!20}-- \\
Qwen3 8B & 0.3736 & \cellcolor{red!20}\textbf{42,272,121} & \cellcolor{gray!20}-- & \cellcolor{gray!20}-- & \cellcolor{gray!20}-- & \cellcolor{gray!20}-- \\
Ministral 3 8B & 0.1237 & \cellcolor{green!20}\textbf{972,292} & \cellcolor{green!20}\textbf{0.2425} & 10,122,749 & \cellcolor{green!20}\textbf{0.2434} & 9,603,062 \\
\bottomrule
\end{tabular}%
}
\\[4pt]
\raggedright
\footnotesize
$^\dagger$ Temperature parameter not supported. Evaluated with default settings.\\
\end{table}

\begin{figure}[htbp]
  \centering
  \includegraphics[width=\linewidth]{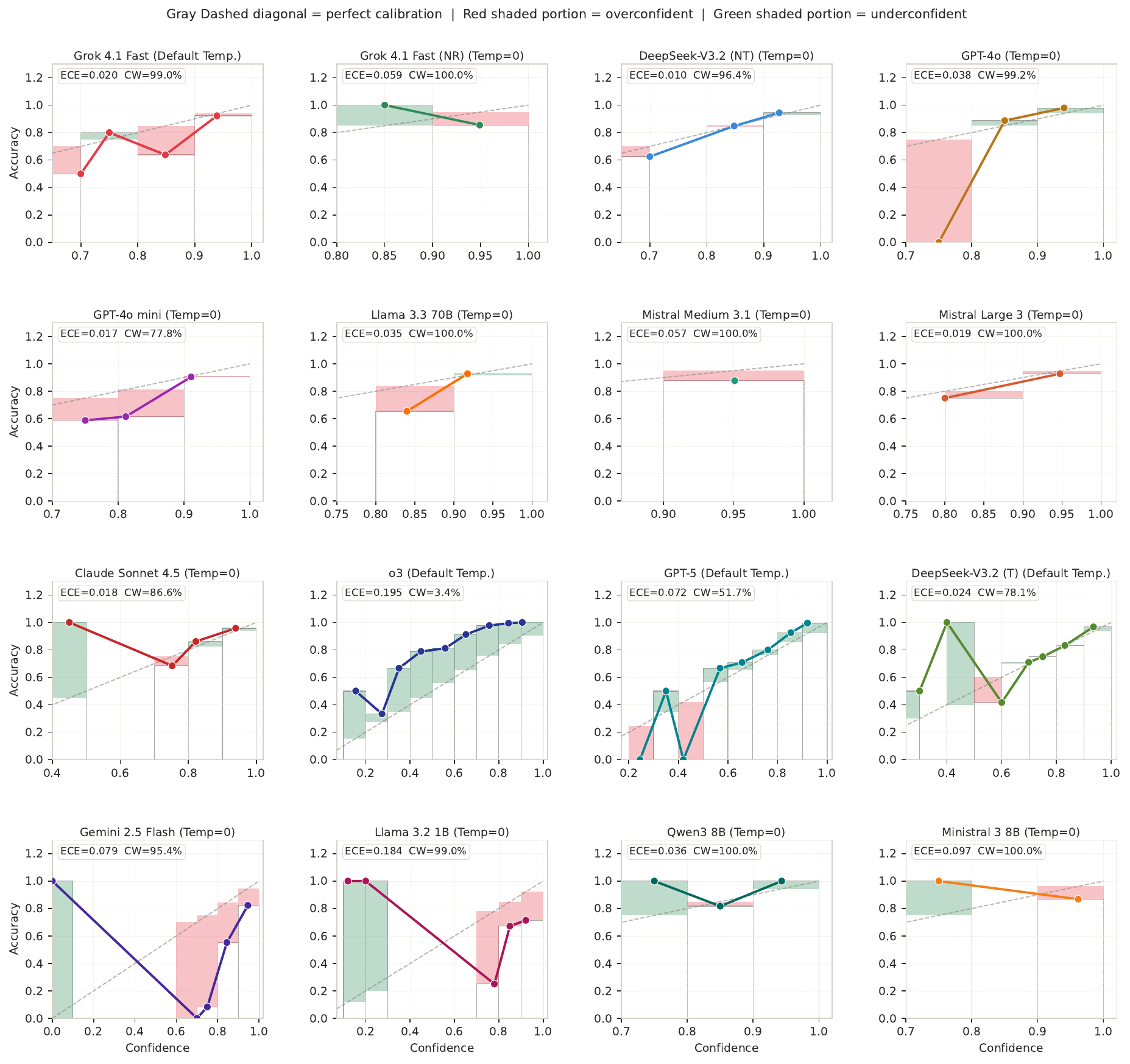}
  \caption{Reliability diagram representing the performance of all 16 state-of-the-art models evaluated on the MCQA dataset in TSNBench. The gray dashed line represents the perfect calibration where confidence is equal to the accuracy. A model which is 100\% confident and has 100\% accuracy will fall on this gray dashed line. The red shaded region represents the over-confidence of the model (model confidence exceeds the actual accuracy of the model), and the green shaded region represents the under-confidence of the model (actual accuracy of the model exceeds its given confidence).}
  \label{fig:reliability_diagram}
\end{figure}

\subsection{MCQA Reliability Analysis}
\label{sub:reliability_analysis}
Figure~\ref{fig:reliability_diagram} presents the reliability plot for all 16 evaluated models on the MCQA dataset. 

Each diagram shows the observed accuracy against the model's expressed confidence, binned across the confidence range. A perfectly calibrated model would fall on the gray dashed diagonal line. This means the model’s confidence would perfectly align with its actual accuracy. The red shaded region indicates overconfidence, meaning the model’s confidence exceeds its actual accuracy. The green shaded region indicates underconfidence, meaning the model is more accurate than its expressed confidence suggests.

In safety-critical TSN deployments, overconfidence is significantly more dangerous than underconfidence. A model that is incorrect but expresses high confidence may mislead a network engineer with an erroneous WCD estimate or misconfigured scheduling parameters. By contrast, an underconfident model that expresses uncertainty on correct answers prompts additional verification.

The majority of the evaluated models sit in the high-confidence region (0.8 to 1.0) regardless of their actual accuracy. This indicates that the models tend to exhibit overconfidence. 

Grok~4.1~Fast~(NR), Llama~3.3~70B, Mistral Medium~3.1, Mistral Large~3, Qwen3~8B, and Ministral~3~8B achieve CW rates of 100\%, meaning all incorrect answers fall in the high-confidence range. This represents the most critical calibration behavior for TSN deployment. Grok~4.1~Fast, GPT-4o, Gemini~2.5~Flash, and Llama~3.2~1B similarly exhibit CW rates exceeding 95\%. A notable exception is o3, which is the only model that falls predominantly in the green underconfident zone, with a CW rate of just 3.4\%. Despite having the highest ECE (0.1951) among all evaluated models, o3 is the safest model from a calibration perspective, as it rarely expresses high confidence on incorrect MCQA answers. This highlights an important distinction between aggregate calibration metrics and safety-relevant calibration behavior. DeepSeek-V3.2~(NT) achieves the lowest ECE (0.0104), suggesting strong overall calibration, yet maintains a CW rate of 96.4\%, demonstrating that a low ECE does not guarantee safe and realistic confidence behavior.

\subsection{More on TSNBench Open-Ended Questions Evaluation}
\label{appendix:open_end_question_evaluation}
We provide MAE evaluations for the open-ended questions. A sample calculation is given as follows:

\textbf{MAE calculation example:}
Consider a model evaluated on three test cases (TCs). These three TCs may have different topologies, different flows and flow parameters, and different routes. Table~\ref{tab:sample_mae_example} shows the GT WCD, the predicted WCD and the absolute error of each flow. As in the main evaluation, the predicted WCD of a flow is the average of its predictions over the valid runs of the TC.

\begin{table}[htbp]
\centering
\caption{Sample example of TCs with the GT, the predicted WCD (averaged over the valid runs) and the absolute error of each flow.}
\label{tab:sample_mae_example}
\small
\begin{tabular}{llccc}
\toprule
\textbf{TC} & \textbf{Flow} & 
\textbf{Ground Truth ($\mu$s)} & 
\textbf{Predicted ($\mu$s)} & 
\textbf{Abs. Error ($\mu$s)} \\
\midrule
TC1 & $f_0$ & 200 & 212 & 12 \\
TC1 & $f_1$ & 150 & 180 & 30 \\
TC1 & $f_2$ & 500 & 490 & 10 \\
\midrule
TC2 & $f_0$ & 100 & 108 & 8  \\
TC2 & $f_1$ & 300 & 255 & 45 \\
\midrule
TC3 & $f_0$ & 400 & 420 & 20 \\
TC3 & $f_1$ & 250 & 265 & 15 \\
TC3 & $f_2$ & 600 & 600 & 0  \\
\bottomrule
\end{tabular}
\end{table}

Per-TC MAE: Suppose TC1, TC2, and TC3 contain three, two, and three flows, respectively.
\begin{align*}
\{f_0, f_1, f_2\} \subseteq TC1;\\
\{f_0, f_1\} \subseteq TC2;\\
\{f_0, f_1, f_2\} \subseteq TC3;
\end{align*}
Let $\Gamma(f_0)$ denote the absolute error of flow $f_0$ in TC1, $\beta(f_0)$ denote the predicted WCD of flow $f_0$ given by the LLM model, and $\Omega(f_0)$ denote the GT of flow $f_0$. We calculate $\Gamma(f_0)$ as follows:
\begin{align*}
\Gamma(f_0) = |\beta(f_0) - \Omega(f_0)|
\end{align*}
In the given example, let $\Gamma(f_0)$ = 12, $\Gamma(f_1)$ = 30, and $\Gamma(f_2)$ = 10 for TC1. Similarly, for TC2, $\Gamma(f_0)$ = 8 and $\Gamma(f_1)$ = 45 and for TC3, $\Gamma(f_0)$ = 20, $\Gamma(f_1)$ = 15, and $\Gamma(f_2)$ = 0. We calculate the MAE for TC1, TC2, and TC3 represented as $\text{MAE}_{\text{TC1}}$, $\text{MAE}_{\text{TC2}}$, and $\text{MAE}_{\text{TC3}}$ as follows:
\begin{align*}
\text{MAE}_{\text{TC1}} &= (12 + 30 + 10) / 3 = 17.3~\mu\text{s} \\
\text{MAE}_{\text{TC2}} &= (8 + 45) / 2       = 26.5~\mu\text{s} \\
\text{MAE}_{\text{TC3}} &= (20 + 15 + 0) / 3  = 11.7~\mu\text{s}
\end{align*}

For every model, we have 100 TCs, and the final MAE is averaged across all valid TCs (in this example 3 TCs) and is represented as:
\begin{equation*}
\text{MAE} = (17.3 + 26.5 + 11.7) / 3 = 18.5~\mu\text{s}
\end{equation*}

The reported standard deviation is the population standard deviation of the per-TC MAE across the valid TCs (6.1~$\mu$s in this example), and the median MAE is the median of the per-TC MAE (17.3~$\mu$s). A TC is valid only if at least 80\% of its flows receive a prediction, and a model is reported only if it has at least 50 valid TCs.

In TSNBench, every TC contributes equally towards the model performance irrespective of the number of flows in the network. As per the network architecture, all flows are equally critical and need the same preference. This ensures that for each network scenario all the flows are weighted equally.

\subsection{Changes from the submitted version}
\label{app:changes}
Following the reviews, we made the following changes to the open-ended evaluation. First, model outputs were re-extracted with a stricter parser that reads the final WCD answer of each response. When a response contained malformed JSON, the earlier parser could read numbers from the model's reasoning, such as CBS credit values or frame sizes, instead of its answer. A calculation without a final result, a value in a unit other than $\mu$s, and values above $10^9~\mu$s now count as missing. Second, flow identifiers are matched to the GT regardless of their format (e.g., ``15'' and ``F15''), and a run that uses a flow identifier that does not exist in the TC is discarded. Third, each model is scored at one temperature (0.0, or the vendor default for models without temperature support) instead of pooling all temperatures. Fourth, MAPE is removed because it is unstable for small GT delays, and the median MAE and error-threshold rates are added. The MAE formula is unchanged and all values are recomputed. Models that return clean JSON with numeric values and the given flow identifiers are unaffected by the parser change, while the largest change is for Ministral 3 8B, whose earlier CBS error was dominated by extraction errors. Grok 4.1 Fast now has fewer than 50 valid CBS TCs and is no longer reported for CBS. Finally, we add the formula-in-context (FC) setting, a prompt-sensitivity study and a case study on the Orion CEV network topology.

\begin{table}[htbp]
\centering
\caption{MAE ($\mu$s) $\pm$ standard deviation and median MAE ($\mu$s) across the valid TCs for CBS under CB and FC settings (\textdownarrow{} lower is better). $n$ is the number of valid TCs out of 100, using one temperature per model. Models with fewer than 50 valid TCs are reported as ``--''.}
\label{appendix:cbs_mae_std_median}
\small
\resizebox{\textwidth}{!}{%
\begin{tabular}{l r r r r r r}
\toprule
& \multicolumn{3}{c}{\textbf{CB}} & \multicolumn{3}{c}{\textbf{FC}} \\
\cmidrule(lr){2-4} \cmidrule(lr){5-7}
\textbf{Model} & MAE $\pm$ std & Median & $n$ & MAE $\pm$ std & Median & $n$ \\
\midrule
Grok 4.1 Fast$^{\dagger,\P}$ & -- & -- & 48 & -- & -- & -- \\
Grok 4.1 Fast (Non-Reasoning)$^\P$ & $3637.3 \pm 3934.5$ & 2992.5 & 91 & -- & -- & -- \\
DeepSeek-V3.2 (Non-Thinking)$^{\P}$ & -- & -- & 28 & -- & -- & -- \\
GPT-4o$^\star$ & -- & -- & 49 & -- & -- & 17 \\
GPT-4o mini & $378.5 \pm 189.7$ & 337.7 & 89 & $461.6 \pm 999.5$ & 331.7 & 89 \\
Llama 3.3 70B & $301.2 \pm 171.0$ & 264.1 & 73 & $333.5 \pm 171.3$ & 308.5 & 63 \\
Mistral Medium 3.1 & $339.6 \pm 224.9$ & 260.3 & 100 & $374.2 \pm 232.7$ & 315.7 & 100 \\
Mistral Large 3 & $270.6 \pm 180.2$ & 231.2 & 100 & $273.5 \pm 179.9$ & 229.0 & 100 \\
Claude Sonnet 4.5 & $293.0 \pm 173.9$ & 264.4 & 100 & $307.7 \pm 178.3$ & 269.2 & 100 \\
o3$^\dagger$ & $255.8 \pm 315.4$ & 138.8 & 100 & $501.9 \pm 2341.3$ & 187.4 & 100 \\
GPT-5$^\dagger$ & $150.2 \pm 198.2$ & 92.4 & 99 & $161.4 \pm 188.0$ & 82.5 & 99 \\
DeepSeek-V3.2 (Thinking)$^{\dagger,\P}$ & -- & -- & 0 & -- & -- & -- \\
Gemini 2.5 Flash & $554.6 \pm 1831.6$ & 223.2 & 84 & $258.0 \pm 231.2$ & 180.4 & 91 \\
Llama 3.2 1B & -- & -- & 0 & -- & -- & 0 \\
Qwen3 8B & -- & -- & 24 & $51855.3 \pm 460034.8$ & 359.8 & 97 \\
Ministral 3 8B & $1643.3 \pm 2618.4$ & 853.7 & 100 & $1551.7 \pm 2735.6$ & 374.8 & 100 \\
\bottomrule
\end{tabular}
}
\\[4pt]
\raggedright
\footnotesize
$n$ is the number of valid TCs out of 100, using one temperature per model.\\ 
``--'' in the MAE columns: fewer than 50 valid TCs.\\
$n = 0$: evaluated, but no TC is valid. \\
``--'' in the $n$ column: not evaluated in this setting.\\
$^\dagger$, $^\star$ and $^\P$ are explained in Table~\ref{tab:open_end_std_dev}. The failure modes of models with few or no valid TCs are described in Appendix~\ref{appendix:failure_mode}.
\end{table}

\begin{table}[htbp]
\centering
\caption{MAE ($\mu$s) $\pm$ standard deviation and median MAE ($\mu$s) across the valid TCs for CQF under CB and FC settings (\textdownarrow{} lower is better). $n$ is the number of valid TCs out of 100, using one temperature per model. Models with fewer than 50 valid TCs are reported as ``--''.}
\label{appendix:cqf_mae_std_median}
\small
\resizebox{\textwidth}{!}{%
\begin{tabular}{l r r r r r r}
\toprule
& \multicolumn{3}{c}{\textbf{CB}} & \multicolumn{3}{c}{\textbf{FC}} \\
\cmidrule(lr){2-4} \cmidrule(lr){5-7}
\textbf{Model} & MAE $\pm$ std & Median & $n$ & MAE $\pm$ std & Median & $n$ \\
\midrule
Grok 4.1 Fast$^{\dagger,\P}$ & $130.1 \pm 85.0$ & 127.9 & 67 & -- & -- & -- \\
Grok 4.1 Fast (Non-Reasoning)$^\P$ & $163.9 \pm 37.8$ & 167.7 & 99 & -- & -- & -- \\
DeepSeek-V3.2 (Non-Thinking)$^{\P}$ & $168.1 \pm 66.3$ & 173.3 & 97 & -- & -- & -- \\
GPT-4o$^\star$ & $82.2 \pm 264.7$ & 1.2 & 100 & $11.3 \pm 15.8$ & 2.4 & 100 \\
GPT-4o mini & $180.5 \pm 82.0$ & 175.3 & 100 & $1377.1 \pm 2818.7$ & 323.4 & 88 \\
Llama 3.3 70B & $173.3 \pm 73.5$ & 173.7 & 93 & $5.1 \pm 15.0$ & 0.0 & 100 \\
Mistral Medium 3.1 & $150.3 \pm 42.2$ & 141.0 & 100 & $33.0 \pm 31.3$ & 31.5 & 100 \\
Mistral Large 3 & $54.2 \pm 28.2$ & 50.0 & 100 & $19.3 \pm 17.8$ & 15.7 & 100 \\
Claude Sonnet 4.5 & $213.2 \pm 1062.5$ & 65.8 & 99 & $30.3 \pm 37.2$ & 16.0 & 100 \\
o3$^\dagger$ & $101.6 \pm 75.9$ & 81.1 & 100 & $0.0 \pm 0.0$ & 0.0 & 100 \\
GPT-5$^\dagger$ & $107.0 \pm 69.0$ & 107.0 & 99 & $0.0 \pm 0.0$ & 0.0 & 100 \\
DeepSeek-V3.2 (Thinking)$^{\dagger,\P}$ & -- & -- & 1 & -- & -- & -- \\
Gemini 2.5 Flash & $112.0 \pm 90.1$ & 91.7 & 99 & $0.1 \pm 0.5$ & 0.0 & 99 \\
Llama 3.2 1B & -- & -- & 0 & -- & -- & 0 \\
Qwen3 8B & -- & -- & 0 & $2.0 \pm 6.2$ & 0.0 & 57 \\
Ministral 3 8B & $3057.4 \pm 4112.4$ & 1046.0 & 99 & $1354.6 \pm 2166.2$ & 776.2 & 100 \\
\bottomrule
\end{tabular}
}
\\[4pt]
\raggedright
\footnotesize
$n$ is the number of valid TCs out of 100, using one temperature per model.\\ 
``--'' in the MAE columns: fewer than 50 valid TCs.\\
$n = 0$: evaluated, but no TC is valid. \\
``--'' in the $n$ column: not evaluated in this setting.\\
$^\dagger$, $^\star$ and $^\P$ are explained in Table~\ref{tab:open_end_std_dev}. The failure modes of models with few or no valid TCs are described in Appendix~\ref{appendix:failure_mode}.
\end{table}

\begin{table}[htbp]
\centering
\caption{Percentage of flows whose absolute WCD error falls below 50, 100, 500, and 1000 $\mu$s in the FC setting for CBS and CQF. For each model, the percentage is computed per TC and averaged over all valid TCs.}
\label{appendix:fc_error_thresholds}
\small
\resizebox{\textwidth}{!}{%
\begin{tabular}{l|cccc|cccc}
\toprule
\multirow{2}{*}{\textbf{Model}}
& \multicolumn{4}{c|}{\textbf{CBS (\%)}}
& \multicolumn{4}{c}{\textbf{CQF (\%)}} \\
\cmidrule(lr){2-5} \cmidrule(lr){6-9}
& \textbf{${<}50\mu$s} & \textbf{${<}100\mu$s} & \textbf{${<}500\mu$s} & \textbf{${<}1000\mu$s}
& \textbf{${<}50\mu$s} & \textbf{${<}100\mu$s} & \textbf{${<}500\mu$s} & \textbf{${<}1000\mu$s} \\
\midrule
GPT-4o$^\star$ & -- & -- & -- & -- & 93.8 & 100.0 & 100.0 & 100.0 \\
GPT-4o mini & 0.9 & 8.8 & 75.8 & 96.4 & 8.5 & 20.4 & 65.0 & 82.5 \\
Llama 3.3 70B & 3.9 & 15.6 & 78.9 & 98.2 & 96.7 & 99.1 & 100.0 & 100.0 \\
Mistral Medium 3.1 & 2.2 & 10.1 & 75.5 & 96.7 & 84.7 & 96.6 & 100.0 & 100.0 \\
Mistral Large 3 & 13.5 & 25.0 & 87.9 & 98.0 & 89.3 & 100.0 & 100.0 & 100.0 \\
Claude Sonnet 4.5 & 5.8 & 16.9 & 84.4 & 97.9 & 84.2 & 94.3 & 100.0 & 100.0 \\
o3$^\dagger$ & 15.5 & 26.9 & 83.1 & 95.8 & 100.0 & 100.0 & 100.0 & 100.0 \\
GPT-5$^\dagger$ & 37.4 & 61.9 & 93.0 & 97.1 & 100.0 & 100.0 & 100.0 & 100.0 \\
Gemini 2.5 Flash & 18.0 & 34.0 & 85.2 & 96.3 & 100.0 & 100.0 & 100.0 & 100.0 \\
Qwen3 8B & 6.7 & 15.2 & 65.1 & 82.2 & 98.2 & 99.0 & 100.0 & 100.0 \\
Ministral 3 8B & 4.7 & 11.3 & 59.9 & 76.9 & 1.6 & 5.9 & 31.6 & 77.8 \\
\bottomrule
\end{tabular}
}
\\[4pt]
\raggedright
\footnotesize
``--'', $^\dagger$, $^\star$ and the models not shown are explained in Table~\ref{tab:open_end_std_dev}. All thresholds are in $\mu$s.
\end{table}

\begin{table}[htbp]
\centering
\caption{Sensitivity of the CB MAE ($\mu$s) for CBS to prompt wording and output format (100 TCs). CB is the original prompt of the main results, Variant A paraphrases the instructions and Variant B requests plain-text instead of JSON output. MAEs are computed per model on the TCs that are valid under all three prompts ($n$). Range, Std and CV (std/mean) are taken across the three prompts. Lower $\downarrow$ CV indicates a more stable model.} 
\label{tab:prompt_sensitivity_cbs_appendix}
\small
\begin{tabular}{l|ccc|ccc|r}
\toprule
\textbf{Model} & \textbf{CB} & \textbf{Variant A} & \textbf{Variant B} & \textbf{Range} & \textbf{Std} & \textbf{CV (\%)} $\downarrow$ & $n$ \\
\midrule
GPT-4o & -- & -- & -- & -- & -- & -- & 49 \\
GPT-4o mini & 371.4 & 357.9 & 117259.2 & 116901.3 & 55104.6 & \cellcolor{red!20}140.1 & 88 \\
Llama 3.3 70B & 278.3 & 268.9 & 768.3 & 499.4 & 233.2 & 53.2 & 60 \\
Mistral Medium 3.1 & 339.6 & 384.6 & 499.8 & 160.2 & 67.5 & 16.5 & 100 \\
Mistral Large 3 & 270.6 & 266.5 & 617.5 & 350.9 & 164.5 & 42.7 & 100 \\
Claude Sonnet 4.5 & 293.0 & 291.0 & 280.0 & 13.0 & 5.7 & 2.0 & 100 \\
o3$^\dagger$ & 255.8 & 253.6 & 248.8 & 7.0 & 2.9 & \cellcolor{green!20}1.2 & 100 \\
GPT-5$^\dagger$ & 142.4 & 121.6 & 122.5 & 20.8 & 9.6 & 7.4 & 93 \\
Gemini 2.5 Flash & 570.2 & 5435.5 & 388.6 & 5046.9 & 2337.5 & 109.7 & 73 \\
Llama 3.2 1B & -- & -- & -- & -- & -- & -- & 0 \\
Qwen3 8B & -- & -- & -- & -- & -- & -- & 8 \\
Ministral 3 8B & 1621.9 & 1056.1 & 7315.5 & 6259.5 & 2826.8 & 84.9 & 97 \\
\bottomrule
\end{tabular}
\\[4pt]
\raggedright
\footnotesize
$^\dagger$ Temperature parameter not supported. Evaluated with default settings.\\
``--'' Not reported: fewer than 50 valid TCs.\\
$n = 0$: evaluated, but no TC is valid (Llama 3.2 1B).\\
Not evaluated (retired before the prompt-sensitivity experiments): DeepSeek-V3.2 (Non-Thinking), DeepSeek-V3.2 (Thinking), Grok 4.1 Fast, Grok 4.1 Fast (Non-Reasoning).
\footnotesize
\end{table}

\begin{table}[htbp]
\centering
\caption{Sensitivity of the CB MAE ($\mu$s) for CQF to prompt wording and output format (100 TCs). CB is the original prompt of the main results, Variant A paraphrases the instructions and Variant B requests plain-text instead of JSON output. MAEs are computed per model on the TCs that are valid under all three prompts ($n$). Range, Std and CV (std/mean) are taken across the three prompts. Lower $\downarrow$ CV indicates a more stable model.}
\label{appendix:prompt_sensitivity_cv_cqf}
\small
\begin{tabular}{l|ccc|ccc|r}
\toprule
\textbf{Model} & \textbf{CB} & \textbf{Variant A} & \textbf{Variant B} & \textbf{Range} & \textbf{Std} & \textbf{CV (\%)} $\downarrow$ & $n$ \\
\midrule
GPT-4o & 85.3 & 35.5 & 48.6 & 49.8 & 21.1 & 37.3 & 96 \\
GPT-4o mini & 188.9 & 197.2 & 442.7 & 253.8 & 117.7 & 42.6 & 92 \\
Llama 3.3 70B & 172.4 & 154.3 & 77.7 & 94.6 & 41.0 & 30.4 & 91 \\
Mistral Medium 3.1 & 149.8 & 172.6 & 1952.9 & 1803.1 & 844.7 & \cellcolor{red!20}111.4 & 99 \\
Mistral Large 3 & 54.2 & 37.6 & 95.3 & 57.6 & 24.2 & 38.8 & 100 \\
Claude Sonnet 4.5 & 213.2 & 135.1 & 65.4 & 147.8 & 60.4 & 43.8 & 99 \\
o3$^\dagger$ & 101.6 & 84.9 & 89.4 & 16.7 & 7.1 & 7.7 & 100 \\
GPT-5$^\dagger$ & 107.0 & 74.6 & 101.5 & 32.3 & 14.1 & 15.0 & 99 \\
Gemini 2.5 Flash & 108.4 & 124.6 & 122.6 & 16.1 & 7.2 & \cellcolor{green!20}6.1 & 98 \\
Llama 3.2 1B & -- & -- & -- & -- & -- & -- & 0 \\
Qwen3 8B & -- & -- & -- & -- & -- & -- & 0 \\
Ministral 3 8B & 3057.4 & 260.0 & 995.2 & 2797.4 & 1184.1 & 82.4 & 99 \\
\bottomrule
\end{tabular}
\\[4pt]
\raggedright
\footnotesize
$^\dagger$ Temperature parameter not supported. Evaluated with default settings.\\
``--'' Not reported: fewer than 50 valid TCs.\\
$n = 0$: no TC is valid under all three prompts (Llama 3.2 1B, Qwen3 8B).\\
Not evaluated (retired before the prompt-sensitivity experiments): DeepSeek-V3.2 (Non-Thinking), DeepSeek-V3.2 (Thinking), Grok 4.1 Fast, Grok 4.1 Fast (Non-Reasoning). \\
\end{table}

\begin{table}[htbp]
\centering
\caption{MAE ($\mu$s) $\pm$ standard deviation and median MAE ($\mu$s) across the valid TCs for CBS under the CB (original), paraphrased (Variant A) and format-changed (Variant B) prompts (\textdownarrow{} lower is better). $n$ is the number of valid TCs out of 100, using one temperature per model.}
\label{appendix:prompt_sensitivity_cbs_appendix}
\small
\resizebox{\textwidth}{!}{%
\begin{tabular}{l r r r r r r r r r}
\toprule
& \multicolumn{3}{c}{\textbf{CB}} & \multicolumn{3}{c}{\textbf{Variant A}} & \multicolumn{3}{c}{\textbf{Variant B}} \\
\cmidrule(lr){2-4} \cmidrule(lr){5-7} \cmidrule(lr){8-10}
\textbf{Model} & MAE $\pm$ std & Median & $n$ & MAE $\pm$ std & Median & $n$ & MAE $\pm$ std & Median & $n$ \\
\midrule
GPT-4o & -- & -- & 49 & $312.3 \pm 263.4$ & 258.6 & 98 & $755.0 \pm 2178.2$ & 292.1 & 97 \\
GPT-4o mini & $378.5 \pm 189.7$ & 337.7 & 89 & $352.8 \pm 174.2$ & 329.0 & 99 & $103364.3 \pm 512945.1$ & 1414.3 & 100 \\
Llama 3.3 70B & $301.2 \pm 171.0$ & 264.1 & 73 & $285.4 \pm 186.3$ & 271.6 & 91 & $756.6 \pm 1343.0$ & 351.7 & 90 \\
Mistral Medium 3.1 & $339.6 \pm 224.9$ & 260.3 & 100 & $384.6 \pm 276.3$ & 303.5 & 100 & $499.8 \pm 908.4$ & 263.0 & 100 \\
Mistral Large 3 & $270.6 \pm 180.2$ & 231.2 & 100 & $266.5 \pm 181.3$ & 225.2 & 100 & $617.5 \pm 926.6$ & 294.8 & 100 \\
Claude Sonnet 4.5 & $293.0 \pm 173.9$ & 264.4 & 100 & $291.0 \pm 169.2$ & 259.5 & 100 & $280.0 \pm 378.1$ & 215.9 & 100 \\
o3$^\dagger$ & $255.8 \pm 315.4$ & 138.8 & 100 & $253.6 \pm 276.0$ & 169.0 & 100 & $248.8 \pm 263.5$ & 169.1 & 100 \\
GPT-5$^\dagger$ & $150.2 \pm 198.2$ & 92.4 & 99 & $129.1 \pm 131.5$ & 89.7 & 95 & $125.0 \pm 154.6$ & 77.0 & 99 \\
Gemini 2.5 Flash & $554.6 \pm 1831.6$ & 223.2 & 84 & $4318.4 \pm 36981.7$ & 250.7 & 94 & $375.5 \pm 415.7$ & 213.4 & 91 \\
Llama 3.2 1B & -- & -- & 0 & -- & -- & 0 & -- & -- & 0 \\
Qwen3 8B & -- & -- & 24 & -- & -- & 34 & $84655.7 \pm 606649.4$ & 370.7 & 55 \\ 
Ministral 3 8B & $1643.3 \pm 2618.4$ & 853.7 & 100 & $1083.2 \pm 1733.6$ & 454.1 & 100 & $7315.5 \pm 33679.4$ & 1873.7 & 97 \\
\bottomrule
\end{tabular}%
}
\\[4pt]
\raggedright
\footnotesize
``--'' Not reported: fewer than 50 valid TCs.\\
$n = 0$: evaluated, but no TC is valid (Llama 3.2 1B).\\
$^\dagger$ Temperature parameter not supported. Evaluated with default settings.\\
Not evaluated (retired before the prompt-sensitivity experiments): Grok 4.1 Fast, Grok 4.1 Fast (Non-Reasoning), DeepSeek-V3.2 (Non-Thinking), DeepSeek-V3.2 (Thinking).\\
Gemini 2.5 Flash's Variant A mean is dominated by TC72 (MAE 360,883.9~$\mu$s), where all three runs apply an incorrect CBS delay formula whose denominator is close to zero and report a WCD of 2,460,205.52~$\mu$s for one flow.
\end{table}

\begin{table}[htbp]
\centering
\caption{MAE ($\mu$s) $\pm$ standard deviation and median MAE ($\mu$s) across the valid TCs for CQF under the CB (original), paraphrased (Variant A), and format-changed (Variant B) prompts (\textdownarrow{} lower is better). $n$ is the number of valid TCs out of 100, using one temperature per model.} 
\label{appendix:prompt_sensitivity_cqf_appendix}
\small
\resizebox{\textwidth}{!}{%
\begin{tabular}{l r r r r r r r r r}
\toprule
& \multicolumn{3}{c}{\textbf{CB}} & \multicolumn{3}{c}{\textbf{Variant A}} & \multicolumn{3}{c}{\textbf{Variant B}} \\
\cmidrule(lr){2-4} \cmidrule(lr){5-7} \cmidrule(lr){8-10}
\textbf{Model} & MAE $\pm$ std & Median & $n$ & MAE $\pm$ std & Median & $n$ & MAE $\pm$ std & Median & $n$ \\
\midrule
GPT-4o & $82.2 \pm 264.7$ & 1.2 & 100 & $34.6 \pm 151.9$ & 1.0 & 100 & $48.6 \pm 46.5$ & 34.7 & 96 \\
GPT-4o mini & $180.5 \pm 82.0$ & 175.3 & 100 & $197.2 \pm 113.1$ & 177.3 & 92 & $491.8 \pm 1142.1$ & 165.6 & 100 \\
Llama 3.3 70B & $173.3 \pm 73.5$ & 173.7 & 93 & $159.1 \pm 59.8$ & 160.8 & 100 & $83.2 \pm 121.6$ & 40.9 & 98 \\
Mistral Medium 3.1 & $150.3 \pm 42.2$ & 141.0 & 100 & $173.1 \pm 56.0$ & 173.3 & 100 & $1952.9 \pm 11090.1$ & 84.3 & 99 \\
Mistral Large 3 & $54.2 \pm 28.2$ & 50.0 & 100 & $37.6 \pm 33.9$ & 33.3 & 100 & $95.3 \pm 30.4$ & 98.3 & 100 \\
Claude Sonnet 4.5 & $213.2 \pm 1062.5$ & 65.8 & 99 & $135.4 \pm 57.4$ & 143.1 & 100 & $65.1 \pm 46.8$ & 52.3 & 100 \\
o3$^\dagger$ & $101.6 \pm 75.9$ & 81.1 & 100 & $84.9 \pm 46.2$ & 76.1 & 100 & $89.4 \pm 51.0$ & 80.8 & 100 \\
GPT-5$^\dagger$ & $107.0 \pm 69.0$ & 107.0 & 99 & $75.0 \pm 58.6$ & 65.4 & 100 & $102.2 \pm 65.2$ & 92.9 & 100 \\
Gemini 2.5 Flash & $112.0 \pm 90.1$ & 91.7 & 99 & $125.6 \pm 94.1$ & 112.5 & 99 & $122.3 \pm 86.3$ & 120.2 & 100 \\
Llama 3.2 1B & -- & -- & 0 & -- & -- & 0 & -- & -- & 1 \\
Qwen3 8B & -- & -- & 0 & $183.4 \pm 85.7$ & 175.1 & 55 & $148.5 \pm 85.5$ & 147.5 & 100 \\
Ministral 3 8B & $3057.4 \pm 4112.4$ & 1046.0 & 99 & $258.3 \pm 408.0$ & 169.1 & 100 & $996.4 \pm 4916.2$ & 137.3 & 100 \\
\bottomrule
\end{tabular}%
}
\\[4pt]
\raggedright
\footnotesize
``--'' Not reported: fewer than 50 valid TCs.\\
$n = 0$: evaluated, but no TC is valid (Llama 3.2 1B under CB and Variant A, and Qwen3 8B under CB).\\
$^\dagger$ Temperature parameter not supported. Evaluated with default settings.\\
Not evaluated (retired before the prompt-sensitivity experiments): DeepSeek-V3.2 (Non-Thinking), DeepSeek-V3.2 (Thinking), Grok 4.1 Fast, Grok 4.1 Fast (Non-Reasoning).
\end{table}

\begin{table*}[htbp]
\centering
\caption{MAE ($\mu$s) for \textbf{CBS} open-ended evaluation across \textbf{One-Switch topology} TCs (TC41 to TC51) under the CB setting. The MAE of a TC is computed on the WCD averaged over the valid runs of the TC, using one temperature per model. ``--'' denotes an invalid, missing or partial response (fewer than 80\% of the flows of the TC predicted). ``0'' denotes a trivial failure (all-zero WCD values). Best result per TC shown in \textbf{bold}.}
\label{tab:cbs_per_tc_mae_one_switch}
\resizebox{\textwidth}{!}{%
\begin{tabular}{lccccccccccc}
\toprule
\textbf{Model} & \textbf{TC41} & \textbf{TC42} & \textbf{TC43} & \textbf{TC44} & \textbf{TC45} & \textbf{TC46} & \textbf{TC47} & \textbf{TC48} & \textbf{TC49} & \textbf{TC50} & \textbf{TC51} \\
\midrule
Grok 4.1 Fast & 127.77 & 91.48 & 127.06 & 32.98 & 24.74 & 202.69 & 132.81 & -- & 133.33 & 272.76 & 195.88 \\
Grok 4.1 Fast (Non-Reasoning) & -- & 5283.22 & -- & -- & 3916.63 & -- & -- & -- & 4248.83 & -- & 5159.01 \\
DeepSeek-V3.2 (Non-Thinking) & 410.75 & 484.56 & 291.06 & -- & 130.71 & -- & -- & -- & 237.33 & 508.76 & -- \\
GPT-4o & 505.94 & 486.53 & 294.39 & 85.41 & 124.37 & 265.02 & 204.64 & 63.72 & 237.22 & 510.61 & 415.26 \\
GPT-4o mini & 0 & 0 & 293.06 & 133.41 & 0 & 301.19 & 210.14 & 0 & 242.67 & 0 & 0 \\
Llama 3.3 70B & 516.89 & 322.23 & 301.96 & 140.34 & 43.20 & 303.73 & 213.14 & 58.87 & 250.67 & 391.42 & 427.25 \\
Mistral Medium 3.1 & 509.17 & 488.48 & 294.34 & 130.85 & 35.19 & 295.19 & 205.13 & 69.52 & 240.64 & 512.68 & 418.62 \\
Mistral Large 3 & 397.00 & 381.23 & 172.06 & 20.79 & 16.35 & 183.52 & 76.58 & 181.61 & 127.67 & 403.12 & 302.10 \\
Claude Sonnet 4.5 & 499.76 & 484.85 & 286.34 & 109.67 & 122.86 & 294.20 & 202.77 & 61.65 & 216.11 & 499.51 & 413.98 \\
o3 & 119.16 & \textbf{61.05} & 121.41 & \textbf{10.57} & 17.50 & 105.48 & 84.57 & 143.37 & \textbf{19.64} & \textbf{125.73} & 171.22 \\
GPT-5 & \textbf{60.23} & 177.15 & \textbf{86.44} & 30.66 & \textbf{15.86} & \textbf{50.59} & \textbf{40.84} & 14.55 & 23.66 & 136.91 & 225.17 \\
DeepSeek-V3.2 (Thinking) & -- & -- & -- & -- & -- & -- & -- & -- & -- & -- & -- \\
Gemini 2.5 Flash & -- & 225.61 & 220.86 & 61.78 & 57.02 & 135.30 & 84.65 & 39.35 & 164.93 & 334.01 & \textbf{158.82} \\
\midrule
\multicolumn{12}{c}{\textit{Small Models}} \\
\midrule
Llama 3.2 1B & -- & -- & -- & -- & -- & -- & -- & 0 & 0 & -- & -- \\
Qwen3 8B & -- & -- & -- & -- & -- & -- & -- & -- & -- & -- & -- \\
Ministral 3 8B & 828.44 & 1183.11 & 448.31 & 858.59 & 863.29 & 1360.14 & 1283.86 & \textbf{10.08} & 882.67 & 713.24 & 1183.01 \\
\bottomrule
\end{tabular}%
}
\end{table*}

\begin{table*}[htbp]
\centering
\caption{MAE ($\mu$s) for \textbf{CQF} open-ended evaluation across \textbf{One-Switch topology} TCs (TC41 to TC51) under the CB setting. The MAE of a TC is computed on the WCD averaged over the valid runs of the TC, using one temperature per model. ``--'' denotes an invalid, missing or partial response (fewer than 80\% of the flows of the TC predicted). ``0'' denotes a trivial failure (all-zero WCD values). Best result per TC shown in \textbf{bold}. “0.00” denotes an exact prediction (MAE of 0 µs).}
\label{tab:cqf_per_tc_mae_one_switch}
\resizebox{\textwidth}{!}{%
\begin{tabular}{lccccccccccc}
\toprule
\textbf{Model} & \textbf{TC41} & \textbf{TC42} & \textbf{TC43} & \textbf{TC44} & \textbf{TC45} & \textbf{TC46} & \textbf{TC47} & \textbf{TC48} & \textbf{TC49} & \textbf{TC50} & \textbf{TC51} \\
\midrule
Grok 4.1 Fast & 221.39 & 175.67 & 54.60 & 46.64 & 52.00 & 63.83 & -- & 61.55 & 212.00 & 409.67 & 128.18 \\
Grok 4.1 Fast (Non-Reasoning) & 95.00 & 93.67 & 89.00 & 91.00 & 93.67 & 95.50 & 91.00 & 95.00 & 93.67 & -- & 91.00 \\
DeepSeek-V3.2 (Non-Thinking) & 89.67 & 95.00 & 95.00 & 89.00 & 91.00 & 95.00 & 95.00 & 89.00 & 89.00 & -- & 37.89 \\
GPT-4o & 316.33 & 283.33 & 633.00 & \textbf{0.00} & \textbf{1.00} & 32.33 & 313.67 & 15.67 & 949.00 & 317.00 & 283.67 \\
GPT-4o mini & 88.33 & 85.67 & 93.00 & 91.00 & 97.00 & 98.33 & 97.00 & 29.67 & 93.00 & 91.00 & 95.44 \\
Llama 3.3 70B & 67.72 & 97.72 & 29.26 & 66.82 & 99.33 & 37.74 & 57.94 & 90.25 & 717.31 & 171.29 & 109.71 \\
Mistral Medium 3.1 & 92.67 & 93.00 & 91.00 & 90.33 & 91.00 & 93.00 & 91.17 & 90.00 & 93.00 & 92.67 & 93.00 \\
Mistral Large 3 & 100.67 & 84.00 & 101.00 & 82.67 & 101.00 & 83.33 & 101.00 & 101.67 & 84.00 & 50.00 & 101.00 \\
Claude Sonnet 4.5 & 87.43 & \textbf{44.48} & \textbf{11.68} & 46.29 & 47.44 & 43.38 & \textbf{30.28} & 48.89 & 46.54 & \textbf{13.69} & \textbf{31.28} \\
o3 & \textbf{34.33} & 159.11 & 146.89 & 68.99 & 32.78 & 69.33 & 32.40 & 70.37 & 47.00 & 126.11 & 81.15 \\
GPT-5 & 169.30 & 162.19 & 85.82 & 74.85 & 27.11 & 76.64 & 54.51 & 25.58 & 108.33 & 160.43 & 115.66 \\
DeepSeek-V3.2 (Thinking) & -- & -- & -- & -- & -- & -- & -- & -- & -- & -- & -- \\
Gemini 2.5 Flash & 80.68 & 91.67 & 79.79 & 5.34 & 15.68 & \textbf{21.28} & 49.73 & \textbf{8.42} & \textbf{44.36} & 149.51 & 80.49 \\
\midrule
\multicolumn{12}{c}{\textit{Small Models}} \\
\midrule
Llama 3.2 1B & 0 & -- & 0 & -- & -- & -- & -- & -- & -- & -- & 0 \\
Qwen3 8B & -- & -- & -- & -- & -- & -- & -- & -- & -- & -- & -- \\
Ministral 3 8B & 1297.00 & 1484.78 & 588.33 & 25.44 & 633.22 & 635.67 & 91.00 & 897.00 & 897.00 & -- & 1200.19 \\
\bottomrule
\end{tabular}%
}
\end{table*}

\begin{table*}[htbp]
\centering
\caption{MAE ($\mu$s) for \textbf{CBS} open-ended evaluation across 10  \textbf{medium-mesh topology} TCs (TC71 to TC80), under the CB setting. The MAE of a TC is computed on the WCD averaged over the valid runs of the TC, using one temperature per model. ``--'' denotes an invalid, missing or partial response (fewer than 80\% of the flows of the TC predicted). ``0'' denotes a trivial failure (all-zero WCD values). Best result per TC shown in \textbf{bold}.}
\label{tab:cbs_per_tc_mae_medium_mesh}
\resizebox{\textwidth}{!}{%
\begin{tabular}{lcccccccccc}
\toprule
\textbf{Model} & \textbf{TC71} & \textbf{TC72} & \textbf{TC73} & \textbf{TC74} & \textbf{TC75} & \textbf{TC76} & \textbf{TC77} & \textbf{TC78} & \textbf{TC79} & \textbf{TC80} \\
\midrule
Grok 4.1 Fast & \textbf{20.83} & 60.86 & \textbf{16.10} & 70.38 & 53.55 & \textbf{157.00} & -- & -- & 71.92 & 345.35 \\
Grok 4.1 Fast (Non-Reasoning) & 5545.59 & 4219.56 & 5320.84 & 2290.48 & 2848.59 & 4123.00 & 2749.35 & 4131.26 & 4171.83 & 2780.78 \\
DeepSeek-V3.2 (Non-Thinking) & 0 & 289.44 & 0 & 0 & 0 & 0 & 0 & 377.74 & 0 & 200.22 \\
GPT-4o & 2225.26 & 0 & 307.31 & 197.81 & 101.84 & 0 & 131.53 & 269.74 & 233.82 & 142.51 \\
GPT-4o mini & 166.41 & 295.44 & 415.31 & 331.66 & 209.84 & 387.00 & 237.65 & 379.74 & 340.17 & 206.22 \\
Llama 3.3 70B & 96.75 & 303.37 & 417.31 & 0 & 192.76 & 335.54 & 243.58 & 0 & 260.80 & 206.14 \\
Mistral Medium 3.1 & 86.15 & 283.44 & 326.31 & 327.59 & 120.24 & 303.20 & 154.67 & 290.24 & 256.82 & 123.68 \\
Mistral Large 3 & 73.25 & 192.89 & 295.31 & 231.69 & 99.22 & 289.40 & 138.89 & 271.74 & 239.77 & 107.56 \\
Claude Sonnet 4.5 & 124.07 & 170.41 & 336.11 & 284.87 & 142.77 & 332.72 & 186.06 & 230.38 & 300.15 & 149.62 \\
o3 & 690.12 & 55.96 & 109.72 & 381.87 & 62.76 & 192.38 & \textbf{47.62} & 179.95 & 194.04 & 78.94 \\
GPT-5 & 77.08 & \textbf{31.23} & 80.38 & \textbf{22.48} & \textbf{24.08} & 171.81 & 70.22 & \textbf{42.16} & \textbf{52.26} & 151.57 \\
DeepSeek-V3.2 (Thinking) & -- & -- & -- & -- & -- & -- & -- & -- & -- & -- \\
Gemini 2.5 Flash & 42.76 & 57.38 & 258.11 & -- & 3352.85 & 217.41 & 98.63 & 145.60 & 211.59 & \textbf{57.35} \\
\midrule
\multicolumn{11}{c}{\textit{Small Models}} \\
\midrule
Llama 3.2 1B & 0 & -- & -- & -- & -- & -- & -- & -- & -- & -- \\
Qwen3 8B & -- & -- & -- & -- & -- & -- & -- & -- & -- & 179.08 \\
Ministral 3 8B & 157.41 & 696.56 & 812.69 & 856.34 & 978.16 & 851.00 & 950.35 & 680.86 & 847.83 & 981.78 \\
\bottomrule
\end{tabular}%
}
\end{table*}

\begin{table*}[htbp]
\centering
\caption{MAE ($\mu$s) for \textbf{CQF} open-ended evaluation across 10 \textbf{medium-mesh topology} TCs (TC71 to TC80), under the CB setting. The MAE of a TC is computed on the WCD averaged over the valid runs of the TC, using one temperature per model. ``--'' denotes an invalid, missing or partial response (fewer than 80\% of the flows of the TC predicted). ``0'' denotes a trivial failure (all-zero WCD values). Best result per TC shown in \textbf{bold}. “0.00” denotes an exact prediction (MAE of 0 µs).}
\label{tab:cqf_per_tc_mae_medium_mesh}
\resizebox{\textwidth}{!}{%
\begin{tabular}{lcccccccccc}
\toprule
\textbf{Model} & \textbf{TC71} & \textbf{TC72} & \textbf{TC73} & \textbf{TC74} & \textbf{TC75} & \textbf{TC76} & \textbf{TC77} & \textbf{TC78} & \textbf{TC79} & \textbf{TC80} \\
\midrule
Grok 4.1 Fast & -- & 106.71 & 180.10 & 96.66 & 89.61 & 80.55 & 31.38 & -- & 84.00 & 55.00 \\
Grok 4.1 Fast (Non-Reasoning) & 139.00 & 142.00 & 142.00 & 139.00 & 136.00 & 139.00 & 139.00 & 139.00 & 139.00 & 139.00 \\
DeepSeek-V3.2 (Non-Thinking) & 142.00 & 140.00 & 142.00 & 98.00 & 142.00 & 142.00 & 142.00 & 140.00 & 142.00 & 142.00 \\
GPT-4o & \textbf{1.00} & \textbf{1.00} & \textbf{0.00} & \textbf{0.00} & \textbf{1.00} & \textbf{0.00} & \textbf{0.00} & \textbf{1.00} & \textbf{1.00} & \textbf{0.00} \\
GPT-4o mini & 146.00 & 142.00 & 142.00 & 142.00 & 142.00 & 142.00 & 142.00 & 148.00 & 146.00 & 142.00 \\
Llama 3.3 70B & 141.00 & 141.00 & 141.00 & 141.00 & 141.00 & 141.00 & 141.00 & 130.80 & 141.00 & 141.00 \\
Mistral Medium 3.1 & 141.00 & 143.00 & 141.00 & 141.00 & 141.00 & 141.00 & 144.00 & 141.00 & 141.00 & 141.00 \\
Mistral Large 3 & 50.00 & 50.00 & 52.00 & 50.00 & 50.00 & 50.00 & 49.00 & 49.00 & 50.00 & 50.00 \\
Claude Sonnet 4.5 & 74.19 & 104.53 & 34.18 & 3.51 & 19.02 & 2.48 & 2.13 & 28.42 & 13.70 & 2.17 \\
o3 & 2.00 & 153.00 & 2.00 & 28.07 & 41.03 & 48.56 & 129.11 & 137.80 & 44.57 & 113.05 \\
GPT-5 & 39.81 & 89.97 & 183.10 & 107.01 & 43.71 & 52.00 & 47.06 & 144.60 & 49.25 & 63.05 \\
DeepSeek-V3.2 (Thinking) & -- & -- & -- & -- & -- & -- & -- & -- & -- & -- \\
Gemini 2.5 Flash & 21.82 & 32.29 & 188.30 & 113.73 & 19.34 & 192.68 & 81.10 & 44.70 & 99.48 & 71.29 \\
\midrule
\multicolumn{11}{c}{\textit{Small Models}} \\
\midrule
Llama 3.2 1B & -- & -- & -- & -- & -- & -- & -- & 0 & -- & -- \\
Qwen3 8B & -- & -- & -- & -- & -- & -- & -- & -- & -- & -- \\
Ministral 3 8B & 1046.00 & 11846.00 & 1116.00 & 846.00 & 1096.00 & 1007.75 & 846.00 & 844.80 & 1046.00 & 846.00 \\
\bottomrule
\end{tabular}%
}
\end{table*}

\clearpage
\subsection{Case study: Orion CEV network topology}
To test whether our findings carry over from the synthetic topologies to a real-world topology, we evaluate the models on TCs derived from the Orion Crew Exploration Vehicle (CEV) network topology~\citep{luxi_tnsm}. This topology is a widely used reference in TSN research. It consists of 31 end systems and 15 switches connected by 100~Mbps links (Figure~\ref{fig:orion_topo}).
\paragraph{Test cases:} We generated 10 TCs per mechanism on this topology. Each TC follows the same format as the synthetic TCs. 
\paragraph{Evaluation:} All models are evaluated with the same prompts, the same CB and FC settings, up to three runs per TC and the same scoring as in Section~\ref{sub:open_end_evaluation}. The GT WCD is computed with the same NC analysis for CBS and the same closed-form bound (Eq.~\ref{eq:cqf_delay}) for CQF as for the main benchmark. Models retired by their providers before these experiments (Grok 4.1 Fast in reasoning and non-reasoning mode, and DeepSeek-V3.2 in thinking and non-thinking mode) are not included. With only 10 TCs, we replace the minimum of 50 valid TCs by a minimum of 5 valid TCs and report the number of valid TCs ($n$) for every model. For CBS, Table~\ref{appendix:orion_cbs_main} reports the mean and median MAE and $\Delta$ for each model under CB and FC and Table~\ref{appendix:orion_cbs_mae_std} the MAE with its standard deviation. Tables~\ref{appendix:orion_cqf_main} and \ref{appendix:orion_cqf_mae_std} report the same for CQF, and Tables~\ref{tab:orion_cbs_per_tc_cb_appendix}, \ref{tab:orion_cbs_per_tc_fc_appendix}, \ref{tab:orion_cqf_per_tc_cb_appendix}, and \ref{tab:orion_cqf_per_tc_fc_appendix} give the per-TC results.
\paragraph{CBS results:} As on the synthetic TCs, FC lowers the mean MAE for only a minority of models. Ten models have at least 5 valid TCs in both settings. The valid TCs of GPT-4o mini and Llama 3.3 70B differ between CB (Table~\ref{tab:orion_cbs_per_tc_cb_appendix}) and FC (Table~\ref{tab:orion_cbs_per_tc_fc_appendix}). As in Table 3, $\Delta$ compares the mean MAE over the TCs valid in each setting. FC lowers the mean MAE for four, namely GPT-4o mini ($+8.7\%$), o3 ($+4.3\%$), GPT-5 ($+43.8\%$), and Qwen3 8B ($+62.0\%$), and raises it for the other six ($\Delta$ between $-0.4\%$ and $-665.3\%$). The gain of GPT-5 rests on a single TC. Its CB mean is driven by TC110 (7,020.8~$\mu$s), while its other nine TCs stay between 43.5 and 145.4~$\mu$s, and FC lowers its error in only 1 of its 10 TCs (Tables~\ref{tab:orion_cbs_per_tc_cb_appendix} and~\ref{tab:orion_cbs_per_tc_fc_appendix}). Ministral 3 8B, which FC helps on the synthetic TCs, gets worse with FC in all 10 Orion TCs. In the CB setting, o3 has the lowest MAE (256.1~$\mu$s) and GPT-5 the lowest median MAE (99.5~$\mu$s).

\paragraph{CQF results:} Unlike CBS, FC lowers the mean MAE for most models on CQF (7 of 11, Table~\ref{tab:orion_cqf_summary}). o3, GPT-5, and Gemini 2.5 Flash reach an MAE of 0.0~$\mu$s in all 10 TCs, and Qwen3 8B reaches 4.7~$\mu$s (Table~\ref{appendix:orion_cqf_main}). FC raises the mean MAE for GPT-4o mini, Llama 3.3 70B, Claude Sonnet 4.5, and Ministral 3 8B. For Llama 3.3 70B, the increase comes from two TCs (TC104 and TC108), while its median falls from 164.1 to 26.3~$\mu$s. In the CB setting, GPT-4o has the lowest MAE (1.5~$\mu$s).

\begin{table}[htbp]
\centering
\caption{CBS WCD estimation on the Orion CEV topology (10 TCs) over the models with at least 5 valid TCs (\textdownarrow{} lower is better). Median MAE is the median of the models' mean MAEs. FC lowers the mean MAE for 4 of the 10 models that have at least 5 valid TCs in both settings.} 
\label{tab:orion_cbs_summary}
\small
\begin{tabular}{lccc}
\toprule
\textbf{Setting} & \textbf{Models} & \textbf{Median MAE ($\mu$s)} & \textbf{Best model (MAE, $\mu$s)} \\
\midrule
CB & 10 & 406.7 & o3 (256.1) \\
FC & 10 & 431.0 & o3 (245.1) \\
\bottomrule
\end{tabular}
\end{table}

\begin{table}[htbp]
\centering
\caption{CQF WCD estimation on the Orion CEV topology (10 TCs) over the models with at least 5 valid TCs (\textdownarrow{} lower is better). Median MAE is the median of the models' mean MAEs. FC lowers the mean MAE for 7 of the 11 models that have at least 5 valid TCs in both settings.} 
\label{tab:orion_cqf_summary}
\small
\begin{tabular}{lccc}
\toprule
\textbf{Setting} & \textbf{Models} & \textbf{Median MAE ($\mu$s)} & \textbf{Best model (MAE, $\mu$s)} \\
\midrule
CB & 11 & 136.3 & GPT-4o (1.5) \\
FC & 11 & 21.3 & o3$^{\ast}$ (0.0) \\
\bottomrule
\end{tabular}
\\[4pt]
\raggedright
\footnotesize
$^{\ast}$ Tied with GPT-5 and Gemini 2.5 Flash (0.0~$\mu$s).
\end{table}

\begin{table}[htbp]
\centering
\caption{Open-ended WCD estimation results for CBS on the Orion CEV topology (10 TCs), reported as mean and median MAE across the valid TCs ($\mu$s, \textdownarrow{} lower is better). $\Delta$ is the relative change of the mean MAE from CB to FC, with each mean taken over the valid TCs of its setting. Positive values mean the formula lowers the error. For GPT-4o mini and Llama 3.3 70B, the valid TCs differ between CB and FC (Tables~\ref{tab:orion_cbs_per_tc_cb_appendix} and \ref{tab:orion_cbs_per_tc_fc_appendix}), so their $\Delta$ compares means over different TCs.} 
\label{appendix:orion_cbs_main}
\small
\begin{tabular}{lrrrrr}
\toprule
& \multicolumn{2}{c}{\textbf{CB}} & \multicolumn{2}{c}{\textbf{FC}} & \\
\cmidrule(lr){2-3} \cmidrule(lr){4-5}
\textbf{Model} & \textbf{MAE} & \textbf{Median} & \textbf{MAE} & \textbf{Median} & \textbf{$\Delta$ (\%)} \\
\midrule
GPT-4o mini$^{\sharp}$ & 412.4 & 420.1 & 376.7 & 412.2 & $+8.7$ \\
Llama 3.3 70B$^{\sharp}$ & 324.2 & 293.8 & 487.4 & 251.6 & $-50.3$ \\
Mistral Medium 3.1 & 926.7 & 901.1 & 1304.5 & 872.4 & $-40.8$ \\
Mistral Large 3 & 339.6 & 335.4 & 340.9 & 337.5 & $-0.4$ \\
Claude Sonnet 4.5 & 358.2 & 350.3 & 366.9 & 352.7 & $-2.4$ \\
o3$^\dagger$ & 256.1 & 223.8 & 245.1 & 190.6 & $+4.3$ \\
GPT-5$^\dagger$ & 790.0 & 99.5 & 444.0 & 142.4 & $+43.8$ \\
Gemini 2.5 Flash & 400.9 & 312.2 & 418.1 & 435.6 & $-4.3$ \\
Qwen3 8B & 19058.4 & 7477.9 & 7249.0 & 2074.1 & $+62.0$ \\
Ministral 3 8B & 3031.6 & 2410.3 & 23200.1 & 9906.9 & $-665.3$ \\
\bottomrule
\end{tabular}
\\[4pt]
\raggedright
\footnotesize
$^\dagger$ Temperature parameter not supported, evaluated with default settings.\\
$^{\sharp}$ Valid TCs differ between CB and FC (Tables~\ref{tab:orion_cbs_per_tc_cb_appendix} and \ref{tab:orion_cbs_per_tc_fc_appendix}). As in Table~\ref{tab:open_end_std_dev}, each mean is taken over the TCs valid in that setting.\\
Not shown (fewer than 5 valid TCs in both settings): GPT-4o (1 valid TC), Llama 3.2 1B (no valid TC).\\
Not evaluated (retired before the Orion experiments): Grok 4.1 Fast, Grok 4.1 Fast (Non-Reasoning), DeepSeek-V3.2 (Non-Thinking), DeepSeek-V3.2 (Thinking).
\end{table}

\begin{table}[htbp]
\centering
\caption{MAE ($\mu$s) $\pm$ standard deviation and median MAE ($\mu$s) across the valid TCs for CBS on the Orion CEV topology under CB and FC (\textdownarrow{} lower is better). $n$ is the number of valid TCs out of 10.}
\label{appendix:orion_cbs_mae_std}
\small
\begin{tabular}{l r r r r r r}
\toprule
& \multicolumn{3}{c}{\textbf{CB}} & \multicolumn{3}{c}{\textbf{FC}} \\
\cmidrule(lr){2-4} \cmidrule(lr){5-7}
\textbf{Model} & MAE $\pm$ std & Median & $n$ & MAE $\pm$ std & Median & $n$ \\
\midrule
GPT-4o & -- & -- & 1 & -- & -- & 1 \\
GPT-4o mini & $412.4 \pm 116.0$ & 420.1 & 10 & $376.7 \pm 126.9$ & 412.2 & 7 \\
Llama 3.3 70B & $324.2 \pm 88.3$ & 293.8 & 8 & $487.4 \pm 622.4$ & 251.6 & 8 \\
Mistral Medium 3.1 & $926.7 \pm 322.6$ & 901.1 & 10 & $1304.5 \pm 1019.3$ & 872.4 & 10 \\
Mistral Large 3 & $339.6 \pm 95.2$ & 335.4 & 10 & $340.9 \pm 93.2$ & 337.5 & 10 \\
Claude Sonnet 4.5 & $358.2 \pm 97.7$ & 350.3 & 10 & $366.9 \pm 98.2$ & 352.7 & 10 \\
o3$^\dagger$ & $256.1 \pm 154.7$ & 223.8 & 10 & $245.1 \pm 159.7$ & 190.6 & 10 \\
GPT-5$^\dagger$ & $790.0 \pm 2077.1$ & 99.5 & 10 & $444.0 \pm 914.0$ & 142.4 & 10 \\
Gemini 2.5 Flash & $400.9 \pm 225.6$ & 312.2 & 9 & $418.1 \pm 96.7$ & 435.6 & 9 \\
Llama 3.2 1B & -- & -- & 0 & -- & -- & 0 \\
Qwen3 8B & $19058.4 \pm 31460.9$ & 7477.9 & 10 & $7249.0 \pm 11276.2$ & 2074.1 & 10 \\
Ministral 3 8B & $3031.6 \pm 2486.2$ & 2410.3 & 10 & $23200.1 \pm 27611.0$ & 9906.9 & 10 \\
\bottomrule
\end{tabular}
\\[4pt]
\raggedright
\footnotesize
``--'' Not reported: fewer than 5 valid TCs.\\
$n = 0$: evaluated, but no TC is valid. \\
$^\dagger$ Temperature parameter not supported, evaluated with default settings.\\
Not evaluated (retired before the Orion experiments): Grok 4.1 Fast, Grok 4.1 Fast (Non-Reasoning), DeepSeek-V3.2 (Non-Thinking), DeepSeek-V3.2 (Thinking).
\end{table}

\begin{table}[htbp]
\centering
\caption{MAE ($\mu$s) of each TC for CBS on the Orion CEV topology under the CB setting (\textdownarrow{} lower is better).}
\label{tab:orion_cbs_per_tc_cb_appendix}
\small
\resizebox{\textwidth}{!}{%
\begin{tabular}{l r r r r r r r r r r}
\toprule
\textbf{Model} & TC101 & TC102 & TC103 & TC104 & TC105 & TC106 & TC107 & TC108 & TC109 & TC110 \\
\midrule
GPT-4o & -- & -- & -- & -- & -- & -- & 361.1 & -- & -- & -- \\
GPT-4o mini & 209.1 & 548.4 & 538.3 & 547.9 & 239.5 & 425.5 & 472.1 & 331.5 & 414.7 & 396.8 \\
Llama 3.3 70B & 285.4 & -- & 485.3 & 399.8 & 176.9 & 276.8 & 300.9 & -- & 381.8 & 286.7 \\
Mistral Medium 3.1 & 1034.3 & 631.2 & 991.4 & 415.2 & 1293.9 & 672.6 & 1603.0 & 960.4 & 823.3 & 841.8 \\
Mistral Large 3 & 345.0 & 450.7 & 441.4 & 455.0 & 139.0 & 325.7 & 383.8 & 238.6 & 313.4 & 302.9 \\
Claude Sonnet 4.5 & 354.3 & 484.6 & 449.0 & 472.0 & 150.5 & 346.3 & 407.0 & 252.3 & 339.2 & 326.8 \\
o3$^\dagger$ & 209.5 & 361.5 & 183.1 & 359.2 & 88.9 & 93.9 & 274.4 & 238.1 & 627.6 & 124.8 \\
GPT-5$^\dagger$ & 88.2 & 103.4 & 120.8 & 145.4 & 43.5 & 81.1 & 94.6 & 95.7 & 106.1 & 7020.8 \\
Gemini 2.5 Flash & 442.5 & 472.7 & 285.4 & 299.2 & 162.8 & -- & 250.2 & 984.1 & 399.3 & 312.2 \\
Llama 3.2 1B & -- & -- & -- & -- & -- & -- & -- & -- & -- & -- \\
Qwen3 8B & 13989.3 & 677.6 & 25724.8 & 556.9 & 109081.0 & 12906.5 & 584.1 & 482.2 & 2049.3 & 24532.3 \\
Ministral 3 8B & 765.6 & 7630.4 & 3820.4 & 555.2 & 1000.3 & 4339.6 & 6428.3 & 778.5 & 4195.6 & 802.4 \\
\bottomrule
\end{tabular}
}
\\[4pt]
\raggedright
\footnotesize
``--'' No valid answer. \\
$^\dagger$ Temperature parameter not supported, evaluated with default settings.\\
Not evaluated (retired before the Orion experiments): Grok 4.1 Fast, Grok 4.1 Fast (Non-Reasoning), DeepSeek-V3.2 (Non-Thinking), DeepSeek-V3.2 (Thinking).
\end{table}

\begin{table}[htbp]
\centering
\caption{MAE ($\mu$s) of each TC for CBS on the Orion CEV topology under the FC setting (\textdownarrow{} lower is better).}
\label{tab:orion_cbs_per_tc_fc_appendix}
\small
\resizebox{\textwidth}{!}{%
\begin{tabular}{l r r r r r r r r r r}
\toprule
\textbf{Model} & TC101 & TC102 & TC103 & TC104 & TC105 & TC106 & TC107 & TC108 & TC109 & TC110 \\
\midrule
GPT-4o & -- & -- & -- & -- & -- & -- & 361.1 & -- & -- & -- \\
GPT-4o mini & 441.6 & -- & -- & 539.9 & 236.8 & -- & 465.1 & 146.6 & 412.2 & 394.4 \\
Llama 3.3 70B & 234.9 & 2124.4 & 297.5 & 251.9 & 149.5 & 195.6 & -- & -- & 251.3 & 393.8 \\
Mistral Medium 3.1 & 816.9 & 933.1 & 1489.5 & 415.2 & 852.2 & 409.4 & 3465.2 & 3040.6 & 730.2 & 892.6 \\
Mistral Large 3 & 349.7 & 448.3 & 442.6 & 452.9 & 146.4 & 325.2 & 381.9 & 237.0 & 321.1 & 303.9 \\
Claude Sonnet 4.5 & 321.6 & 492.9 & 492.6 & 487.1 & 181.8 & 367.1 & 396.4 & 261.9 & 338.2 & 329.5 \\
o3$^\dagger$ & 186.6 & 165.2 & 357.2 & 319.1 & 65.7 & 158.7 & 233.7 & 115.5 & 194.7 & 654.9 \\
GPT-5$^\dagger$ & 148.8 & 122.8 & 221.3 & 188.1 & 57.5 & 194.6 & 3182.2 & 104.4 & 136.1 & 84.3 \\
Gemini 2.5 Flash & 443.3 & 478.1 & 540.8 & 549.9 & 241.8 & -- & 435.6 & 334.4 & 415.9 & 322.6 \\
Llama 3.2 1B & -- & -- & -- & -- & -- & -- & -- & -- & -- & -- \\
Qwen3 8B & 3670.5 & 38235.5 & 11846.8 & 194.2 & 4697.0 & 121.2 & 477.8 & 12596.7 & 249.5 & 400.9 \\
Ministral 3 8B & 4483.5 & 61053.0 & 12836.5 & 36939.8 & 1473.7 & 86406.2 & 9704.0 & 1019.6 & 10109.8 & 7974.9 \\
\bottomrule
\end{tabular}
}
\\[4pt]
\raggedright
\footnotesize
``--'' No valid answer. \\
$^\dagger$ Temperature parameter not supported, evaluated with default settings.\\
Not evaluated (retired before the Orion experiments): Grok 4.1 Fast, Grok 4.1 Fast (Non-Reasoning), DeepSeek-V3.2 (Non-Thinking), DeepSeek-V3.2 (Thinking).
\end{table}

\begin{table}[htbp]
\centering
\caption{Open-ended WCD estimation results for CQF on the Orion CEV topology (10 TCs), reported as mean and median MAE across the valid TCs ($\mu$s, \textdownarrow{} lower is better). $\Delta$ is the relative change of the mean MAE from CB to FC, with each mean taken over the valid TCs of its setting. Positive values mean the formula lowers the error.}
\label{appendix:orion_cqf_main}
\small
\begin{tabular}{lrrrrr}
\toprule
& \multicolumn{2}{c}{\textbf{CB}} & \multicolumn{2}{c}{\textbf{FC}} & \\
\cmidrule(lr){2-3} \cmidrule(lr){4-5}
\textbf{Model} & \textbf{MAE} & \textbf{Median} & \textbf{MAE} & \textbf{Median} & \textbf{$\Delta$ (\%)} \\
\midrule
GPT-4o & 1.5 & 1.5 & 1.5 & 1.2 & $+0.6$ \\
GPT-4o mini & 190.9 & 174.0 & 401.0 & 405.1 & $-110.1$ \\
Llama 3.3 70B & 164.4 & 164.1 & 8029.5 & 26.3 & $-4785.1$ \\
Mistral Medium 3.1 & 189.7 & 121.3 & 21.3 & 9.7 & $+88.7$ \\
Mistral Large 3 & 45.4 & 45.4 & 22.7 & 30.2 & $+49.9$ \\
Claude Sonnet 4.5 & 24.7 & 21.1 & 43.8 & 50.3 & $-77.7$ \\
o3$^\dagger$ & 118.3 & 97.2 & 0.0 & 0.0 & $+100.0$ \\
GPT-5$^\dagger$ & 127.4 & 114.8 & 0.0 & 0.0 & $+100.0$ \\
Gemini 2.5 Flash & 136.3 & 131.9 & 0.0 & 0.0 & $+100.0$ \\
Qwen3 8B & 721.0 & 168.2 & 4.7 & 4.9 & $+99.3$ \\
Ministral 3 8B & 10079.0 & 867.5 & 26619.3 & 819.9 & $-164.1$ \\
\bottomrule
\end{tabular}
\\[4pt]
\raggedright
\footnotesize
$^\dagger$ Temperature parameter not supported, evaluated with default settings.\\
Not shown (fewer than 5 valid TCs in both settings): Llama 3.2 1B (no valid TC).\\
Not evaluated (retired before the Orion experiments): Grok 4.1 Fast, Grok 4.1 Fast (Non-Reasoning), DeepSeek-V3.2 (Non-Thinking), DeepSeek-V3.2 (Thinking).\\
FC raises Llama 3.3 70B’s mean MAE ($-4{,}785.1$\%) because of two TCs (TC104 and TC108, MAE 32,031.8 and 48,032.4 $\mu$s), while its median FC error falls from 164.1 to 26.3 $\mu$s.
\end{table}

\begin{table}[htbp]
\centering
\caption{MAE ($\mu$s) $\pm$ standard deviation and median MAE ($\mu$s) across the valid TCs for CQF on the Orion CEV topology under CB and FC (\textdownarrow{} lower is better). $n$ is the number of valid TCs out of 10.}
\label{appendix:orion_cqf_mae_std}
\small
\begin{tabular}{l r r r r r r}
\toprule
& \multicolumn{3}{c}{\textbf{CB}} & \multicolumn{3}{c}{\textbf{FC}} \\
\cmidrule(lr){2-4} \cmidrule(lr){5-7}
\textbf{Model} & MAE $\pm$ std & Median & $n$ & MAE $\pm$ std & Median & $n$ \\
\midrule
GPT-4o & $1.5 \pm 0.0$ & 1.5 & 10 & $1.5 \pm 0.8$ & 1.2 & 10 \\
GPT-4o mini & $190.9 \pm 52.3$ & 174.0 & 10 & $401.0 \pm 215.6$ & 405.1 & 7 \\
Llama 3.3 70B & $164.4 \pm 1.1$ & 164.1 & 10 & $8029.5 \pm 16396.4$ & 26.3 & 10 \\
Mistral Medium 3.1 & $189.7 \pm 201.3$ & 121.3 & 10 & $21.3 \pm 21.7$ & 9.7 & 10 \\
Mistral Large 3 & $45.4 \pm 0.0$ & 45.4 & 10 & $22.7 \pm 15.0$ & 30.2 & 10 \\
Claude Sonnet 4.5 & $24.7 \pm 18.2$ & 21.1 & 10 & $43.8 \pm 11.2$ & 50.3 & 10 \\
o3$^\dagger$ & $118.3 \pm 62.2$ & 97.2 & 10 & $0.0 \pm 0.0$ & 0.0 & 10 \\
GPT-5$^\dagger$ & $127.4 \pm 53.0$ & 114.8 & 10 & $0.0 \pm 0.0$ & 0.0 & 10 \\
Gemini 2.5 Flash & $136.3 \pm 55.3$ & 131.9 & 10 & $0.0 \pm 0.0$ & 0.0 & 10 \\
Llama 3.2 1B & -- & -- & 0 & -- & -- & 0 \\
Qwen3 8B & $721.0 \pm 1197.5$ & 168.2 & 10 & $4.7 \pm 3.7$ & 4.9 & 10 \\
Ministral 3 8B & $10079.0 \pm 17916.9$ & 867.5 & 10 & $26619.3 \pm 35065.0$ & 819.9 & 10 \\
\bottomrule
\end{tabular}
\\[4pt]
\raggedright
\footnotesize
``--'' Not reported: fewer than 5 valid TCs.\\
$n = 0$: evaluated, but no TC is valid. \\
$^\dagger$ Temperature parameter not supported, evaluated with default settings.\\
Not evaluated (retired before the Orion experiments): Grok 4.1 Fast, Grok 4.1 Fast (Non-Reasoning), DeepSeek-V3.2 (Non-Thinking), DeepSeek-V3.2 (Thinking).
\end{table}

\begin{table}[htbp]
\centering
\caption{MAE ($\mu$s) of each TC for CQF on the Orion CEV topology under the CB setting (\textdownarrow{} lower is better).}
\label{tab:orion_cqf_per_tc_cb_appendix}
\small
\resizebox{\textwidth}{!}{%
\begin{tabular}{l r r r r r r r r r r}
\toprule
\textbf{Model} & TC101 & TC102 & TC103 & TC104 & TC105 & TC106 & TC107 & TC108 & TC109 & TC110 \\
\midrule
GPT-4o & 1.5 & 1.5 & 1.5 & 1.5 & 1.5 & 1.5 & 1.5 & 1.5 & 1.5 & 1.5 \\
GPT-4o mini & 347.8 & 174.4 & 171.4 & 171.5 & 174.6 & 174.5 & 173.6 & 174.6 & 173.6 & 172.7 \\
Llama 3.3 70B & 164.1 & 164.0 & 164.1 & 163.9 & 164.1 & 163.6 & 164.0 & 163.9 & 164.1 & 167.7 \\
Mistral Medium 3.1 & 212.2 & 325.4 & 59.8 & 736.8 & 39.3 & 103.5 & 173.8 & 55.4 & 139.1 & 51.4 \\
Mistral Large 3 & 45.4 & 45.4 & 45.4 & 45.4 & 45.4 & 45.4 & 45.4 & 45.4 & 45.4 & 45.4 \\
Claude Sonnet 4.5 & 30.8 & 14.8 & 3.9 & 45.5 & 22.1 & 20.1 & 67.7 & 11.7 & 22.0 & 8.1 \\
o3$^\dagger$ & 87.7 & 53.2 & 260.0 & 148.9 & 33.5 & 96.2 & 94.2 & 98.2 & 179.5 & 131.2 \\
GPT-5$^\dagger$ & 120.1 & 104.9 & 125.0 & 197.1 & 73.6 & 77.7 & 142.3 & 76.9 & 246.5 & 109.5 \\
Gemini 2.5 Flash & 154.7 & 227.2 & 95.8 & 137.7 & 152.3 & 230.6 & 126.1 & 75.7 & 54.9 & 107.8 \\
Llama 3.2 1B & -- & -- & -- & -- & -- & -- & -- & -- & -- & -- \\
Qwen3 8B & 120.3 & 3814.5 & 178.3 & 128.5 & 166.3 & 170.1 & 136.5 & 2191.1 & 178.3 & 125.6 \\
Ministral 3 8B & 29503.5 & 57545.7 & 4268.8 & 4684.1 & 806.8 & 887.6 & 817.3 & 827.8 & 847.4 & 601.1 \\
\bottomrule
\end{tabular}
}
\\[4pt]
\raggedright
\footnotesize
``--'' No valid answer. \\
$^\dagger$ Temperature parameter not supported, evaluated with default settings.\\
Not evaluated (retired before the Orion experiments): Grok 4.1 Fast, Grok 4.1 Fast (Non-Reasoning), DeepSeek-V3.2 (Non-Thinking), DeepSeek-V3.2 (Thinking).
\end{table}

\begin{table}[htbp]
\centering
\caption{MAE ($\mu$s) of each TC for CQF on the Orion CEV topology under the FC setting (\textdownarrow{} lower is better).}
\label{tab:orion_cqf_per_tc_fc_appendix}
\small
\resizebox{\textwidth}{!}{%
\begin{tabular}{l r r r r r r r r r r}
\toprule
\textbf{Model} & TC101 & TC102 & TC103 & TC104 & TC105 & TC106 & TC107 & TC108 & TC109 & TC110 \\
\midrule
GPT-4o & 1.0 & 1.0 & 0.5 & 1.3 & 1.0 & 0.5 & 2.0 & 3.0 & 2.1 & 2.1 \\
GPT-4o mini & 560.7 & 730.0 & -- & 210.0 & 270.8 & 563.5 & 66.9 & 405.1 & -- & -- \\
Llama 3.3 70B & 59.9 & 27.3 & 14.3 & 32031.8 & 15.6 & 25.3 & 23.4 & 48032.4 & 18.4 & 46.4 \\
Mistral Medium 3.1 & 51.9 & 50.0 & 3.6 & 52.0 & 15.9 & 2.1 & 2.0 & 2.1 & 0.7 & 33.0 \\
Mistral Large 3 & 30.5 & 31.8 & 2.9 & 33.5 & 30.8 & 30.0 & 47.0 & 1.9 & 16.9 & 2.0 \\
Claude Sonnet 4.5 & 50.3 & 51.0 & 50.3 & 33.7 & 51.0 & 51.0 & 50.7 & 17.0 & 50.0 & 33.3 \\
o3$^\dagger$ & 0.0 & 0.0 & 0.0 & 0.0 & 0.0 & 0.0 & 0.0 & 0.0 & 0.0 & 0.0 \\
GPT-5$^\dagger$ & 0.0 & 0.0 & 0.0 & 0.0 & 0.0 & 0.0 & 0.0 & 0.0 & 0.0 & 0.0 \\
Gemini 2.5 Flash & 0.0 & 0.0 & 0.0 & 0.0 & 0.0 & 0.0 & 0.0 & 0.0 & 0.0 & 0.0 \\
Llama 3.2 1B & -- & -- & -- & -- & -- & -- & -- & -- & -- & -- \\
Qwen3 8B & 8.1 & 0.8 & 6.5 & 0.0 & 6.5 & 0.0 & 3.2 & 9.7 & 2.4 & 9.7 \\
Ministral 3 8B & 634.1 & 86861.6 & 814.4 & 699.0 & 438.8 & 86859.8 & 30067.2 & 814.4 & 825.4 & 58178.7 \\
\bottomrule
\end{tabular}
}
\\[4pt]
\raggedright
\footnotesize
``--'' No valid answer. \\
$^\dagger$ Temperature parameter not supported, evaluated with default settings.\\
Not evaluated (retired before the Orion experiments): Grok 4.1 Fast, Grok 4.1 Fast (Non-Reasoning), DeepSeek-V3.2 (Non-Thinking), DeepSeek-V3.2 (Thinking).
\end{table}

\clearpage
\begin{longtable}{p{0.96\textwidth}}
\caption{CBS Error Analysis Case 1: Lack of Specific Knowledge.}
\label{tab:cbs_error_case_study}\\
\toprule
\endfirsthead
\caption[]{CBS Error Analysis Case 1: Lack of Specific Knowledge. (continued)}\\
\toprule
\endhead
\bottomrule
\endfoot
\bottomrule
\endlastfoot
\begin{tcolorbox}[
    colback=violet!7,
    colframe=violet!55!black,
    boxrule=0.6pt,
    arc=2mm,
    left=2mm,
    right=2mm,
    top=1mm,
    bottom=1mm,
    width=\linewidth
]
\textbf{Test Case:} TC1 \\[2pt]
\textbf{TSN mechanism:} CBS \\
\tcblower
You are an expert Time-Sensitive Networking (TSN) orchestrator. Your task is to calculate the worst case delay (WCD) for each TSN flow.\\[3pt]
\textbf{Input:} \\
Network Topology ($TC1\_topo.txt$) \\[1pt]
Flow Information ($TC1\_flows.txt$) \\[1pt]
Routing of the Flow ($TC1\_route.txt$) \\[3pt]
\textbf{Constant:} \\
Bandwidth link = 100 Mbps; Propagation delay = 1~$\mu$s; Switching delay = 1~$\mu$s; \\
Time synchronization error = 1~$\mu$s; The switches of the network are cut-through switches. \\[4pt]
\textbf{TSN Mechanism:} \\
Only Credit-Based Shaper (CBS, IEEE 802.1Qav) is allowed; \\
All flows are AVB Class A, PCP = 6, using queue 6 only. \\[4pt]
\textbf{Task:}
\begin{enumerate}
    \item Map each egress port's queues and collect the set of flows traversing from that port, using the given topology, flows, and route of the flow.
    \item For each egress port, use the given IdleSlope and then compute the SendSlope.
    \item For each flow, construct an arrival curve from its frame size and periodicity.
    \item For each port, derive a lower-bounded CBS service curve.
    \item Calculate the worst case delay (WCD) in microseconds ($\mu$s) for each flow using Network Calculus method.
    \item Provide the confidence score between 0.0 and 1.0 from your answers. 1.0 means mathematically or procedurally provable from given info with zero ambiguity. 0.0 means zero confidence.
\end{enumerate}
\end{tcolorbox}
\\
\begin{tcolorbox}[
    colback=gray!7,
    colframe=gray!55!black,
    boxrule=0.6pt,
    arc=2mm,
    left=2mm,
    right=2mm,
    top=1mm,
    bottom=1mm,
    width=\linewidth
]
\begin{tcolorbox}[
    colback=teal!7,
    colframe=teal!55!black,
    boxrule=0.6pt,
    arc=2mm,
    left=2mm,
    right=2mm,
    top=1mm,
    bottom=1mm,
    width=\linewidth
]
\textbf{Grok 4.1 Fast (Non-Reasoning) output:}
\end{tcolorbox}
F0:~1452.0, F1:~1124.0, F2:~678.0, F3:~1234.0, F4:~1567.0, F5:~689.0, F6:~456.0, F7:~789.0, F8:~567.0, F9:~589.0, F10:~678.0, F11:~645.0, F12:~1123.0, F13:~789.0, F14:~456.0, F15:~678.0, F16:~789.0, F17:~1567.0, F18:~456.0, F19:~1234.0 \\
\tcblower
\textbf{Model Reasoning and Explanation Output:} \\[2pt]
Parameters used:\\
$\mathrm{send\_slope}$: $75\% \times 100~\mathrm{Mbps} = 75~\mathrm{Mbps} = 75000~\mathrm{bits}/\mu\mathrm{s}$\\
$\mathrm{frame\_overhead}$: 24 bytes (Ethernet header + FCS),\\
$\mathrm{mtu}$: 1522 bytes (max frame size),\\
$\mathrm{queue}$: 6 (PCP=6, AVB Class A),\\
$\mathrm{cbs\_credit\_max}$: $\mathrm{MTU} \times 8 = 12176$ bits\\[4pt]
\textbf{WCD equation given by the model:}
\[\alpha(t) = \frac{L_{\max}}{\mathrm{Period}} + \frac{t}{\mathrm{Period}} \quad \mathrm{(arrival\ curve)}\]
\[\beta(t) = R_{\mathrm{send}} \times t - C
\quad \mathrm{(CBS\ service\ curve)}\]
\[WCD = h \times\left(\mathrm{prop\_delay}+ \mathrm{switch\_delay}+\mathrm{sync\_error}\right)+ \max(\alpha \cap \beta)\]
\[\mathrm{SendSlope} = \mathrm{IdleSlope} \times \mathrm{Bandwidth} = 0.75 \times 100~\mathrm{Mbps}\]
\[ R_{\mathrm{send}} = \mathrm{SendSlope} = 75~\mathrm{Mbps} \]
\[ C = \mathrm{credit}_{\max} = \mathrm{MTU} \times 8~\mathrm{bits}\]
\[h = \mathrm{hop\ count\ from\ path}\]\\[3pt]
\textbf{Per-flow WCD:}\\
\[WCD_{F_i} = \sum_{h \in \mathrm{path}(F_i)} \left(d_{\mathrm{prop}}^{h} + d_{\mathrm{switch}}^{h} + d_{\mathrm{sync}}^{h} \right) + \max_{t \geq 0} \left(\alpha_{F_i}(t) \cap \beta_{\mathrm{port}}(t)\right)\]
\end{tcolorbox}
\\
\begin{tcolorbox}[
    colback=gray!7,
    colframe=gray!55!black,
    boxrule=0.6pt,
    arc=2mm,
    left=2mm,
    right=2mm,
    top=1mm,
    bottom=1mm,
    width=\linewidth
]
\begin{tcolorbox}[
    colback=red!7,
    colframe=red!55!black,
    boxrule=0.6pt,
    arc=2mm,
    left=2mm,
    right=2mm,
    top=1mm,
    bottom=1mm,
    width=\linewidth
]
\textbf{Run 2 of the model (Grok 4.1 Fast (Non-Reasoning)):}
\end{tcolorbox}
\textbf{WCD equation given by the model during run 2:} \\[4pt]
\[
\begin{aligned}
\text{Frame size:} \quad
L_i &= \mathrm{payload}\_i + 24~\mathrm{bytes},\\
\text{Transmission time:} \quad
T_{x_i} &= \frac{8L_i}{R}, \\
\text{Arrival curve:} \quad
\alpha_i(t) &= \frac{L_i}{p_i} \cdot t, \\
\text{CBS service curve:} \quad
\beta(t) &= \min \left\{ \alpha_{h_i}(t),\;
\mathrm{sendSlope} \cdot t + \mathrm{credit}_{\max} \right\}, \\
\text{Worst-case delay:} \quad
WCD_i &=
\max_{h \in \mathrm{path}}
\sum
\left(
T_{\mathrm{prop},h}
+
T_{\mathrm{switch},h}
+
T_{\mathrm{queue},h}
\right)
+
T_{\mathrm{sync}}, \\
\text{Queueing delay:} \quad
T_{\mathrm{queue},h}
&=
\frac{L_i}{\alpha C}
+
\mathrm{burst}\_{\mathrm{interference}}, \\
\text{Send slope:} \quad
\mathrm{sendSlope}
&=
\mathrm{idleSlope} \cdot R
-
\mathrm{idleSlope} \cdot R
=
\alpha R(1 - 2\alpha);
\end{aligned}
\]\\[4pt]
\textbf{Per-flow WCD:}\\
\[
\begin{aligned}
WCD_i &= \mathrm{hop\_count}_i \cdot (1+1) + \mathrm{max\_queue\_delay} + 1;
\end{aligned}
\]

\vspace{6pt}

\begin{tcolorbox}[
    colback=red!7,
    colframe=red!55!black,
    boxrule=0.6pt,
    arc=2mm,
    left=2mm,
    right=2mm,
    top=1mm,
    bottom=1mm,
    width=\linewidth
]
\textbf{Expert Explanation:}
\end{tcolorbox}
The model uses different equations across different runs even with temperature set to 0.0. When analyzing the first run, the model makes several fundamental errors in evaluating the interference/blocking experienced by the queue under analysis, i.e., both the arrival curve and CBS service curve.

First, it assumes the wrong maximum burst size and does not consider the variant output arrival curve at subsequent hops along the path.

Second, it assumes a non-zero maximum credit, which is possible in this scenario due to the non-preemption frame of lower priority of non-CBS traffic. However, the model applies the wrong maximum credit to the CBS service curve, incorrectly understanding the relation of the corresponding rate (75\%). 

Furthermore, in the final WCD calculation, the model considers only the arrival curve of the frame under analysis. This is incorrect: the proper approach requires using the aggregate arrival curve of all frames transmitted through the queue. It is also not clear whether the $max(\alpha \cap \beta)$ operation represents the maximum horizontal deviation between $\alpha$ and $\beta$. The output arrival curve is not considered in the following hops' calculation.

On the other hand, the second run exhibits even more severe hallucinations, producing calculations of the wrong arrival curve without considering any burst size, and the wrong CBS service curve considering the wrong service rate (sendSlope) and wrong relation between the corresponding rate and the maximum credit, as well as without considering the output arrival curve.

\vspace{6pt}

\begin{tcolorbox}[
    colback=violet!6,
    colframe=violet!55!black,
    boxrule=0.6pt,
    arc=2mm,
    left=2mm,
    right=2mm,
    top=1mm,
    bottom=1mm,
    width=\linewidth
]
\textbf{GPT-5 output:}
\end{tcolorbox}
F0:~714.65, F1:~821.79, F2:~171.48, F3:~687.17, F4:~907.26, F5:~171.48, F6:~183.9, F7:~453.91, F8:~611.08, F9:~181.3, F10:~611.08, F11:~321.35, F12:~1124.12, F13:~483.67, F14:~172.43, F15:~489.26, F16:~521.33, F17:~918.73, F18:~181.3, F19:~624.12
\tcblower
\end{tcolorbox}
\\
\begin{tcolorbox}[
    colback=gray!7,
    colframe=gray!55!black,
    boxrule=0.6pt,
    arc=2mm,
    left=2mm,
    right=2mm,
    top=1mm,
    bottom=1mm,
    width=\linewidth
]

\textbf{Model Reasoning and Explanation Output:} \\[2pt]
\[
\begin{aligned}
\mathrm{sendSlope} &= -25~\mathrm{Mbps}, \\[4pt]
\text{Arrival curve:} \quad
b &= \sum_{\mathrm{\forall F_i | {\ share\ output\ port}}}L_i~\mathrm{bits}, \\
r &= \sum_{\mathrm{\forall F_i | {\ share\ output\ port}}} \frac{L_i}{p_i}~\mathrm{bits}/\mu\mathrm{s}, \\
\alpha(t) &= b + r t,
\end{aligned}
\]
\[
\begin{aligned}
T_{\mathrm{tech,total}}
&=
T_{\mathrm{prop}}
+
T_{\mathrm{switching}}
+
T_{\mathrm{sync}} \\
&=
hops \times 1~\mu\mathrm{s}
+
(hops-1) \times 1~\mu\mathrm{s}
+
(hops-1) \times 1~\mu\mathrm{s}
=
13~\mu\mathrm{s}, \\[4pt]
WCD_{\mathrm{total}}
&=
WCD_{\mathrm{sum}}
+
T_{\mathrm{tech,total}} \\
&=
701.65~\mu\mathrm{s}
+
13~\mu\mathrm{s}
=
714.65~\mu\mathrm{s}.
\end{aligned}
\]

\vspace{6pt}

\begin{tcolorbox}[
    colback=red!7,
    colframe=red!55!black,
    boxrule=0.6pt,
    arc=2mm,
    left=2mm,
    right=2mm,
    top=1mm,
    bottom=1mm,
    width=\linewidth
]
\textbf{Expert Explanation:}
\end{tcolorbox}
GPT-5 uses the correct arrival and service curves. Its main error is that it fails to include the output port (output arrival curves) in the analysis.

\vspace{8pt}
\begin{tcolorbox}[
    colback=green!7,
    colframe=green!50!black,
    boxrule=0.6pt,
    arc=2mm,
    left=2mm,
    right=2mm,
    top=1mm,
    bottom=1mm,
    width=\linewidth
]
\textbf{Ministral 3 8B output:}
\end{tcolorbox}
F0:~1250.0, F1:~1100.0, F2:~500.0, F3:~1500.0, F4:~600.0, F5:~1100.0, F6:~2500.0, F7:~1200.0, F8:~500.0, F9:~500.0, F10:~2000.0, F11:~500.0, F12:~1000.0, F13:~1000.0, F14:~2000.0, F15:~500.0, F16:~1500.0, F17:~1000.0, F18:~500.0, F19:~2000.0
\tcblower
\textbf{Model Reasoning and Explanation Output:} \\ 
\[
\begin{aligned}
\text{Bandwidth:} \quad
R &= 100~\mathrm{Mbps}
= 12.5~\mathrm{MB/s}, \\[4pt]
\text{Propagation delay:} \quad
T_{\mathrm{prop}} &= 1~\mu\mathrm{s}
\quad \text{per hop}, \\[4pt]
\text{Switching delay:} \quad
T_{\mathrm{switch}} &= 1~\mu\mathrm{s}
\quad \text{per switch}, \\[4pt]
\text{Time synchronization error:} \quad
T_{\mathrm{sync}} &= 1~\mu\mathrm{s}, \\[4pt]
\text{IdleSlope:} \quad
\mathrm{IdleSlope}
&= 0.75R
= 75~\mathrm{Mbps}
= 9.375~\mathrm{MB/s}, \\[4pt]
\text{Cut-through switch:} \quad
T_{\mathrm{buffer}} &= 0, \\[4pt]
\text{Payload overhead:} \quad
L_{\mathrm{overhead}} &= 18~\mathrm{B} + 4~\mathrm{B} = 22~\mathrm{B}, \\[4pt]
\text{Maximum frame size:} \quad
L_{\max}&=1433~\mathrm{B}+22~\mathrm{B}=1455~\mathrm{B}, \\[4pt]
\text{Maximum transmission time:} \quad
T_{\mathrm{tx,max}}
&=
\frac{1455 \times 8}{100~\mathrm{Mbps}}
= 9.24~\mu\mathrm{s}.
\end{aligned}
\]\\

\end{tcolorbox}
\\
\begin{tcolorbox}[
    colback=gray!7,
    colframe=gray!55!black,
    boxrule=0.6pt,
    arc=2mm,
    left=2mm,
    right=2mm,
    top=1mm,
    bottom=1mm,
    width=\linewidth
]
\[
\begin{aligned}
\mathrm{SendSlope}&= \mathrm{IdleSlope}\,(1-\mathrm{IdleSlope}) \\
&=
9.375~\mathrm{MB/s} \times 0.25 \\
&=
2.34375~\mathrm{MB/s}, \\[6pt]
\text{Burst size:} \quad
B_{\mathrm{burst},i}
&=
\mathrm{SendSlope} \cdot p_i, \\[6pt]
\text{CBS service curve:} \quad
S(t)
&=
\mathrm{SendSlope}
\left(
t -
\frac{B_{\mathrm{burst}}}{\mathrm{SendSlope}}
\right),
\quad
t \geq
\frac{B_{\mathrm{burst}}}{\mathrm{SendSlope}}.
\end{aligned}
\]
\textbf{WCD Calculation:} \\
\[
\begin{aligned}
\text{Arrival curve for flow } F_x: \quad
R(t)
&=
\min \left\{
B_x,\;
S_x t
\right\}, \\[2pt]
B_x
&=
L_{\mathrm{payload},x}
+
L_{\mathrm{overhead}}, \\[2pt]
S_x
&=
\frac{8L_{\mathrm{payload},x}}{p_x}, \\[2pt]
\text{Service curve for each port:} \quad
S(t)
&=
\mathrm{SendSlope}
\left(
t -
\frac{B_{\mathrm{burst}}}{\mathrm{SendSlope}}
\right),
\quad
t \geq
\frac{B_{\mathrm{burst}}}{\mathrm{SendSlope}}, \\[2pt]
WCD
&=
\max \left(
0,\;
\frac{
B_{\mathrm{burst}}-\mathrm{SendSlope}\cdot Period
}{
\mathrm{SendSlope}
}
\right) \\
&\quad
+
T_{\mathrm{prop}}
+
T_{\mathrm{switch}}
+
T_{\mathrm{sync}}.
\end{aligned}
\]

\vspace{6pt}

\begin{tcolorbox}[
    colback=red!7,
    colframe=red!55!black,
    boxrule=0.6pt,
    arc=2mm,
    left=2mm,
    right=2mm,
    top=1mm,
    bottom=1mm,
    width=\linewidth
]
\textbf{Expert Explanation:}
\end{tcolorbox}
As a small model, Ministral 3 8B delivers a poor result. Similar to what was observed with Grok 4.1 Fast (Non-Reasoning), it fails to account for the arrival curves of flows and the wrong CBS service curve. 

In addition, it analyzes only a single queue, instead of computing the delay at the output port of each device along the end-to-end path. 

Moreover, it introduces unit inconsistencies such as dividing the frame size expressed in bytes by a link speed given in bits/$\mu$s, which leads to incorrect results.

\end{tcolorbox}
\end{longtable}

\begin{longtable}{p{0.96\textwidth}}
\caption{CQF Error Analysis Case 1: Lack of Specific Knowledge.}
\label{tab:cqs_error_case_study}\\
\toprule
\endfirsthead
\caption[]{CQF Error Analysis Case 1: Lack of Specific Knowledge. (continued)}\\
\toprule
\endhead
\bottomrule
\endfoot
\bottomrule
\endlastfoot
\begin{tcolorbox}[
    colback=violet!7,
    colframe=violet!55!black,
    boxrule=0.6pt,
    arc=2mm,
    left=2mm,
    right=2mm,
    top=1mm,
    bottom=1mm,
    width=\linewidth
]
\textbf{Test Case:} TC1 \\[2pt]
\textbf{TSN mechanism:} CQF \\
\tcblower
You are an expert Time-Sensitive Networking (TSN) orchestrator. Your task is to calculate the worst case delay (WCD) for each TSN flow.\\[3pt]
\textbf{Input:} \\
Network Topology ($TC1\_topo.txt$) \\[1pt]
Flow Information ($TC1\_flows.txt$) \\[1pt]
Routing of the Flow ($TC1\_route.txt$) \\[3pt]
\textbf{Constant:} \\
Bandwidth link = 100 Mbps; Propagation delay = 1~$\mu$s; Switching delay = 1~$\mu$s; \\
Time synchronization error = 1~$\mu$s; The switches of the network are cut-through switches; cycle duration is 50~$\mu$s. \\[4pt]
\textbf{TSN Mechanism:} \\
Only Cyclic Queuing and Forwarding (CQF, IEEE 802.1Qch) is allowed; \\
All flows are TT, PCP = 7, using queue 7 (odd) and 6 (even) only. \\[4pt]
\textbf{Task:}
\begin{enumerate}
    \item Map each egress port's queues and collect the set of flows traversing that port, using the given topology, flows, and route of the flow.
    \item For the entire network, use the given cycle duration and compute the Hypercycle.
    \item For each flow, set the offset or the start time of the flow from the sending node as 0.
    \item Calculate the worst case delay (WCD) in microseconds ($\mu$s) for each flow.
    \item Provide the confidence score between 0.0 and 1.0 from your answers. 1.0 means mathematically or procedurally provable from given info with zero ambiguity. 0.0 means zero confidence.
\end{enumerate}

\end{tcolorbox}
\\
\begin{tcolorbox}[
    colback=gray!7,
    colframe=gray!55!black,
    boxrule=0.6pt,
    arc=2mm,
    left=2mm,
    right=2mm,
    top=1mm,
    bottom=1mm,
    width=\linewidth
]
\begin{tcolorbox}[
    colback=teal!7,
    colframe=teal!55!black,
    boxrule=0.6pt,
    arc=2mm,
    left=2mm,
    right=2mm,
    top=1mm,
    bottom=1mm,
    width=\linewidth
]
\textbf{Claude Sonnet's output:}
\end{tcolorbox}
F0:~257.72, F1:~206.8, F2:~105.096, F3:~218.704, F4:~253.904, F5:~104.008, F6:~103.808, F7:~153.52, F8:~119.408, F9:~119.752, F10:~122.896, F11:~118.464, F12:~222.04, F13:~156.552, F14:~103.808, F15:~156.576, F16:~155.68, F17:~256.616, F18:~101.696, F19:~208.328
\tcblower
\textbf{Model Reasoning and Explanation Output:} \\[2pt]
\[
\begin{aligned}
\text{Cycle duration:} = 50~\mu\mathrm{s}, \\
\text{LCM of all periods and cycle:} \quad
\mathrm{LCM}(1000,2500,5000,50) &= 5000~\mu\mathrm{s}, \\
\text{Hypercycle:} = 5000~\mu\mathrm{s}.
\end{aligned}
\]
\[
\begin{aligned}
\text{WCD equation:} \quad
WCD&=\sum_{\mathrm{hops}}\left(T_{\mathrm{trans}}+T_{\mathrm{prop}}+T_{\mathrm{sw}}+T_{\mathrm{sync}}\right)+T_{\mathrm{buffering}}.\end{aligned}\]\\
\[
\begin{aligned}
T_{\mathrm{trans}}&=\frac{\left(\mathrm{Payload}+\mathrm{Ethernet\_overhead}\right)\cdot 8}{R}, \\[4pt]
T_{\mathrm{prop}}
&=1~\mu\mathrm{s}\quad \text{per link}, \\[4pt]
T_{\mathrm{sw}}&=
1~\mu\mathrm{s}\quad \text{per switch}, \\[4pt]
T_{\mathrm{sync}}&=1~\mu\mathrm{s}
\quad \text{per hop}, \\[4pt]
T_{\mathrm{buffering}}&=\left(\mathrm{hop\_count}-1
\right) \cdot T_{\mathrm{cycle}}
\end{aligned}
\]
In CQF, frames may wait up to one cycle duration at each intermediate switch if they arrive just after the transmission window closes. For cut-through with CQF, worst case adds $(N-1)*\mathrm{cycle\_duration}$ where $N$ is hop count.\\
\[
\begin{aligned}
\text{Hop count} &= 5, \\
\text{Payload\_bytes} &= 965~\mathrm{bytes}, \\
\text{Frame\_size\_bytes} &= 1003~\mathrm{bytes}, \\
\text{Transmission\_delay} &= 80.24~\mu\mathrm{s}, \\
\text{Total propagation delay} &= 5~\mu\mathrm{s}, \\
\text{Total switching delay} &= 4~\mu\mathrm{s}, \\
\text{Total synchronization error} &= 5~\mu\mathrm{s}, \\
\text{CQF buffering} &= 200~\mu\mathrm{s}, \\
\text{WCD} &= 257.72~\mu\mathrm{s}.
\end{aligned}
\]
\begin{tcolorbox}[
    colback=red!7,
    colframe=red!55!black,
    boxrule=0.6pt,
    arc=2mm,
    left=2mm,
    right=2mm,
    top=1mm,
    bottom=1mm,
    width=\linewidth
]
\textbf{Ground truth WCD:}
\end{tcolorbox}
\[
\begin{aligned}
\text{Route hop count of } f_0 &: 5, T = 50~\mu\mathrm{s}, \\
WCD_{f_0}
&= f_0.\phi + (SW_{\mathrm{num}} + 1) \cdot T + \xi \\
&= (4+1)\times 50 + (4 \times 1) + (1 + 1) \\
&= 256~\mu\mathrm{s}.
\end{aligned}
\]
\begin{tcolorbox}[
    colback=red!7,
    colframe=red!55!black,
    boxrule=0.6pt,
    arc=2mm,
    left=2mm,
    right=2mm,
    top=1mm,
    bottom=1mm,
    width=\linewidth
]

\textbf{Expert Explanation:}
\end{tcolorbox}
The model knows the working mechanism of CQF, but its WCD equation is a generic per-hop delay sum, not the CQF bound. It also calculates the wrong Hypercycle by including cycle duration with the periods. 
\end{tcolorbox}
\\
\begin{tcolorbox}[
    colback=gray!7,
    colframe=gray!55!black,
    boxrule=0.6pt,
    arc=2mm,
    left=2mm,
    right=2mm,
    top=1mm,
    bottom=1mm,
    width=\linewidth
]
\begin{tcolorbox}[
    colback=violet!6,
    colframe=violet!55!black,
    boxrule=0.6pt,
    arc=2mm,
    left=2mm,
    right=2mm,
    top=1mm,
    bottom=1mm,
    width=\linewidth
]
\textbf{GPT-4o output:}
\end{tcolorbox}
F0: 260.0, F1: 210.0, F2: 60.0, F3: 210.0, F4: 260.0, F5: 60.0, F6: 60.0, F7: 110.0, F8: 60.0, F9: 60.0, F10: 60.0, F11: 60.0, F12: 210.0, F13: 110.0, F14: 60.0, F15: 110.0, F16: 110.0, F17: 260.0, F18: 60.0, F19: 210.0
\tcblower
\textbf{Model Reasoning and Explanation Output:} \\[2pt]
\[
\begin{aligned}
\text{Hypercycle} = 100~\mu\mathrm{s}.
\end{aligned}
\]
\[ \mathrm{WCD} = \left(\mathrm{hop\_count}\times\left(\mathrm{cycle\_duration} + T_{\mathrm{switch}} + T_{\mathrm{prop}} + T_{\mathrm{sync}}\right)\right) + \frac{\mathrm{Payload}}{\mathrm{Bandwidth}}\]\\
Each flow's path, cycle duration, hypercycle, and CQF parameters are detailed, along with the equation used for WCD calculation. No missing inputs are identified, but assumptions are made based on typical network configurations.
\begin{tcolorbox}[
    colback=red!7,
    colframe=red!55!black,
    boxrule=0.6pt,
    arc=2mm,
    left=2mm,
    right=2mm,
    top=1mm,
    bottom=1mm,
    width=\linewidth
]
\textbf{Expert Explanation:}
\end{tcolorbox}
The Hypercycle calculated by the model is wrong. Moreover, the WCD equation used by the model has an extra wrong component $\frac{\mathrm{Payload}}{\mathrm{Bandwidth}}$.\\[2pt]

One of the most interesting findings is that all the models provide very different WCD equations across different runs and different test cases.\\
\begin{tcolorbox}[
    colback=green!6,
    colframe=green!55!black,
    boxrule=0.6pt,
    arc=2mm,
    left=2mm,
    right=2mm,
    top=1mm,
    bottom=1mm,
    width=\linewidth
]
\textbf{Ministral 3 8B output:}
\end{tcolorbox}
F0:~1160.0, F1:~1020.0, F2:~100.0, F3:~1160.0, F4:~1260.0, F5:~100.0, F6:~100.0, F7:~100.0, F8:~100.0, F9:~100.0, F10:~100.0, F11:~100.0, F12:~1160.0, F13:~100.0, F14:~100.0, F15:~100.0, F16:~100.0, F17:~1260.0, F18:~100.0, F19:~1160.0\\[4pt]
\textbf{Model Reasoning and Explanation Output:}\\[2pt]
\[
\begin{aligned}
\text{Hypercycle:} \quad
H &= 10000~\mu\mathrm{s}, \\[4pt]
\text{Max frame size} &= 1526~\mathrm{bytes}
= 1500 + 26~\mathrm{bytes}, \\[4pt]
\text{Max transmission time} &= 12.208~\mu\mathrm{s}, \\[4pt]
\text{Max queuing delay} &= 48~\mu\mathrm{s}, \\[6pt]
WCD
&=
h\left(
T_{\mathrm{prop}}
+
T_{\mathrm{switch}}
+
T_{\mathrm{sync}}
\right) \\[2pt]
&\quad
+
T_{\mathrm{tx,max}}
\left\lceil
\frac{\mathrm{payload}}{\mathrm{max\_frame\_size}}
\right\rceil \\[2pt]
&\quad
+ T_{\mathrm{queue,max}}(h-1), \\[6pt]
\end{aligned}
\]
\end{tcolorbox}
\\
\begin{tcolorbox}[
    colback=gray!7,
    colframe=gray!55!black,
    boxrule=0.6pt,
    arc=2mm,
    left=2mm,
    right=2mm,
    top=1mm,
    bottom=1mm,
    width=\linewidth
]
\[
\begin{aligned}
\text{Hop count:} \quad
h &= 5, \\[4pt]
\text{Propagation + switching + sync} &= 5 \times (1+1+1) = 15~\mu\mathrm{s}, \\[4pt]
\text{Transmission time} &= 12.208 \times
\left\lceil
\frac{2500}{1526}
\right\rceil
=
12.208 \times 2
=
24.416~\mu\mathrm{s}, \\[4pt]
\text{Queueing delay} &= 48 \times (5-1) = 192~\mu\mathrm{s}, \\[4pt]
\text{Total} =
&15 + 24.416 + 192
=
231.416~\mu\mathrm{s}, \\[4pt]
\text{Adjusted worst-case} &=
1160~\mu\mathrm{s}.
\end{aligned}
\]
\vspace{4pt}
\begin{tcolorbox}[
    colback=red!7,
    colframe=red!55!black,
    boxrule=0.6pt,
    arc=2mm,
    left=2mm,
    right=2mm,
    top=1mm,
    bottom=1mm,
    width=\linewidth
]
\textbf{Expert Explanation:}
\end{tcolorbox}
The WCD equation provided by the model is wrong. Although the model considers the number of hops in the route and the delays accumulated at each hop, and also calculates the hop count, it misses the most crucial term of the WCD equation, which is the cycle duration. Furthermore, the two components of the WCD equation ($T_{\mathrm{tx,max}}\left\lceil \frac{\mathrm{payload}}{\mathrm{max\_frame\_size}}\right\rceil$) and ($T_{\mathrm{queue,max}}(h-1)$) considered by the model are entirely hallucinated. These two components are mainly contributing to the large WCD values of this model.
\end{tcolorbox}
\end{longtable}

\section{Failure Mode Analysis}
\label{appendix:failure_mode}
To understand the nature of WCD computation failures, we identify three distinct failure modes observed across models and mechanisms. Some models show more than one of them. Tables~\ref{tab:cbs_error_case_study} and~\ref{tab:cqs_error_case_study} show example CBS and CQF outputs with expert explanations.

\paragraph{Trivial Zero Failure:} The model returns WCD~=~0 for all flows, producing a structurally valid JSON response but with no computational content. This failure mode affects GPT-4o for 51 of the 100 CBS TCs in the CB setting, leaving 49 valid TCs, DeepSeek-V3.2 (Non-Thinking) on CBS, and Llama 3.2 1B in about half of its CB runs for CBS and CQF. It also occurs in 18\% of the CB runs of GPT-4o mini and 30\% of those of Llama 3.3 70B on CBS. This suggests these models recognize the output format requirement but do not carry out the WCD computation in these responses.
\paragraph{Invalid Flow Identifiers:} The model returns WCD values for flows that do not exist in the TC. Such runs are discarded by the flow-ID rule (Appendix~\ref{app:changes}). Llama 3.2 1B lists up to 1,013 invented flows (F0 to F1012) on CBS and CQF instead of the flows of the TC and returns WCD~=~0 for all of them. For Qwen3 8B, the rule discards 209 CB runs on CBS, which leaves 24 valid CBS TCs. On CQF, all 36 of its answered CB runs use flow IDs that are not in the TC. For Grok 4.1 Fast, the rule discards 36 CB runs on CBS and 45 on CQF. 
\paragraph{Missing or Truncated Output:} The model returns no answer, or an answer that ends before the WCD values. DeepSeek-V3.2 (Thinking), evaluated in the CB setting only, returns empty responses for all CBS TCs and all but one CQF TC. Grok 4.1 Fast returns truncated JSON with flow profile metadata but no WCD values, and has only 48 valid CBS TCs. Llama 3.2 1B returns no parsable WCD value in 56 of its 123 CB runs on CBS. For Qwen3 8B, 16 of its 300 CB runs on CBS and 87 of its 123 CB runs on CQF failed at the inference endpoint, most of them with a gateway timeout. As a result, neither Qwen3 8B nor Llama 3.2 1B reaches the 50 valid TCs required for reporting on the synthetic topologies in the CB setting with the original prompts. Qwen3 8B is therefore reported in the FC setting (97 valid CBS TCs and 57 valid CQF TCs), for the CBS Variant B prompt (55 valid TCs), and for the CQF Variant A and B prompts (55 and 100 valid TCs), but not in the CB setting.

\end{document}